 \documentclass[final,5p,times,twocolumn]{elsarticle}
 \date{}
 \nopreprintlinetrue

\usepackage{amssymb}
\usepackage{microtype}
\usepackage{subfigure}
\usepackage{booktabs} % for professional tables
\usepackage{multirow, rotating} 
\usepackage{tabularx}
\usepackage{longtable}
\usepackage{hyperref}
\usepackage{bigstrut}
 \usepackage{xurl}
\hypersetup{breaklinks=true}
\usepackage{amsmath}
\usepackage{mathtools}
\usepackage{amsthm}
\usepackage[capitalize,noabbrev]{cleveref}
\usepackage{subfigure} 
\usepackage{subcaption}
\usepackage{pdflscape}
\usepackage{algorithm,algorithmic}
\begin{document}

\begin{frontmatter}



\title{Iterative Trace Minimization for the Reconciliation of Very Short Hierarchical Time Series}
% Converging on Perfection: the Iterative Trace Minimization for Hierarchical Time Series Reconciliation


\author[TUD,GSofLog]{Louis Steinmeister\corref{cor1}}
\ead{louis.steinmeister@tu-dortmund.de}
\cortext[cor1]{Corresponding author: Louis Steinmeister}

\author[TUD,GSofLog,RC]{Markus Pauly}

\affiliation[TUD]{organization={Department of Statistics, TU Dortmund University},%Department and Organization
            addressline={Vogelpothsweg 78}, 
            city={Dortmund},
            postcode={44227}, 
            %state={},
            country={Germany}}
\affiliation[GSofLog]{organization={Graduate School of Logistics},%Department and Organization
            addressline={Leonhard-Euler-Straße 5}, 
            city={Dortmund},
            postcode={44227}, 
            %state={},
            country={Germany}}
\affiliation[RC]{organization={Research Center Trustworthy Data Science and Security},%Department and Organization
            addressline={Joseph-von-Fraunhofer-Straße 25}, 
            city={Dortmund},
            postcode={44227}, 
            %state={},
            country={Germany}}

\begin{abstract}
Time series often appear in an additive hierarchical structure. In such cases, higher-level time series are the sums of their subordinate time series. This hierarchical structure places a natural constraint on forecasts. However, univariate forecasting techniques are incapable of ensuring this forecast coherence. An obvious solution is to forecast only bottom time series and obtain higher level forecasts through aggregation. This approach is also known as the bottom-up approach. In their seminal paper, \citep{Wickramasuriya2019} propose an optimal reconciliation approach 
named MinT. It tries to minimize the trace of the underlying covariance matrix of all forecast errors. The MinT algorithm has demonstrated superior performance to the bottom-up and other approaches and enjoys great popularity. 
This paper provides a simulation study examining the performance of MinT for very short time series and larger hierarchical structures. This scenario makes the covariance estimation required by MinT difficult. A novel iterative approach is introduced which significantly reduces the number of estimated parameters. This approach is capable of improving forecast accuracy further.
The application of MinTit is also demonstrated with a case study at the hand of a semiconductor dataset based on data provided by the World Semiconductor Trade Statistics (WSTS), a premier provider of semiconductor market data. 
\end{abstract}

%\includegraphics[width=\columnwidth]{Humans_vs_Machines_2k.png}


\end{frontmatter}


\section{Executive Summary}
\label{ExecutiveSummary}

The main \textbf{objective} of this paper is to introduce iterative versions (MinTit) of the Trace Minimization (MinT) algorithm, which was introduced by \citet{Wickramasuriya2019}. The goal of the iterative algorithms is to simplify covariance estimation for large hierarchical structures, especially in the case of time series with very few observations.

The \textbf{motivation} stems from the difficulty of estimating large residual covariance matrices with very few observations. In the case study presented in \cref{sec:CaseStudy}, a total of 110 dimensions implies the estimation of 6160 parameters with only 24 observations in the most extreme case. \citet{athanasopoulos2023forecast} point out that this challenge of dimensionality can lead to ``volatile performance of MinT, especially for short time series''.

\textbf{Methods:} A simulation study resembling \citet{Wickramasuriya2019} 
for extremely short time series is conducted to assess the performance of MinT in this case and determine whether an iterative approach can improve upon the performance of MinT and existing alternatives such as structural scaling. Here, performance is measured by a reduction in root mean squared error (RMSE) compared to the base forecasts. For the latter we considered three common forecast models: Gaussian process regression (GPR), Error, Trend, and Seasonality (ETS) as well as auto.ARIMA models. The latter automatically seeks to find an optimal parameter combination including potential seasonal components as described in \citet{Hyndman2008AutomaticForecasting}. 
In addition, a case study is presented which showcases the application of various reconciliation methods on a real-world dataset.

\textbf{Results:} 
In many of the simulated scenarios, MinTit, the iterative adaptations of MinT, achieved a higher reduction in average RMSE than any other reconciliation method. Nevertheless, MinT proved remarkably robust -- even in the presence of extremely short time series. It was generally the best performing reconciliation method for forecasts based on GPR. However, the forecasts based on the ETS and ARIMA models were much more accurate in most of the simulated cases. For these two base forecasts, the new MinTit approaches were often more favorable. 

\textbf{Conclusion:}  
MinTit reconciliation has the potential to improve forecast accuracy for extremely short time series and when time series lengths differ across the hierarchical structure. Ideally, the reconciliation method should be chosen based on a cross validation when the length of the considered time series permits, as performance may differ depending on the structure of the data and the employed forecasting algorithm. 

As a rule of thumb, MinT may be the reconciliation method of choice when there is reason to assume strong seasonal effects: the ETS and auto.ARIMA models (both seek an optimal model, including models incorporating seasonal components) struggled to adequately identify seasonality with short time series, which appeared to have negatively affected the performance of the iterative reconciliation methods. Otherwise, MinTit may be a good choice with the potential to achieve a lower RMSE. MinTit also allows for efficient use of information when time series lengths differ, which can provide an additional edge.

\section{Brief Introduction}
\label{sec:Introduction}

Accurate forecasts play an increasingly important role in a more and more interconnected and digitized world. In a corporate setting, forecasts routinely inform investment decisions, capital allocation, production, strategy, and help companies anticipate technological change \citep{Agrawal.2020,  modis1994lifeCycles, Petropoulos2022ForecastingTP,Steinmeister2023, Pauly2023}. In many settings in practice, time series occur in a hierarchical context. A textbook example is the Australian tourism data set \citep{Athanasopoulos2009tourism}. Here, quarterly tourism demand is broken down into three levels: by purpose of travel, states, and nights spent in the regional capital or other cities.
\citet{Kuhlmann2024CPSL} provide a corporate example of the occurrence of hierarchical time series where sales numbers are broken down by product and sales channel. Furthermore, the case study in \cref{sec:CaseStudy} consists of time series in a hierarchical structure with market revenues aggregated by semiconductor product categories. Forecasts at different levels of detail are crucial for operational planning. However, ensuring these forecasts and plans are coherent is essential for a unified operational strategy. 

The paper is structured as follows: 
The above brief introduction motivates the discussion of hierarchical time series reconciliation methods, including the popular trace minimization algorithm \citet{Wickramasuriya2019}. The latter is explained in \cref{sec:HTS}, which also introduces new iterative versions of trace minimization (\cref{sec:ItMinT}).
Similar to \citet{Wickramasuriya2019}, a simulation study to compare the performance of various reconciliation algorithms %, briefly presented in \cref{sec:sim_rec},
is conducted in \cref{sec:sim}. There, we explore the effect of correlation, the impact of the smoothing effect of aggregation, the effects of seasonality, different time series lengths, or hierarchical degeneracy, as well as the effect of a 'larger' hierarchy. \cref{sec:CaseStudy} presents a case study of a semiconductor market data set: The World Semiconductor Trade Statistics (WSTS).
Finally, \cref{sec:Discussion} provides a discussion of the overall findings.

\section{Hierarchical Time Series Reconciliation}
\label{sec:HTS}

\begin{figure}[ht]
%\vskip 0.2in
\begin{center}
\centerline{\includegraphics[width=.8\columnwidth]{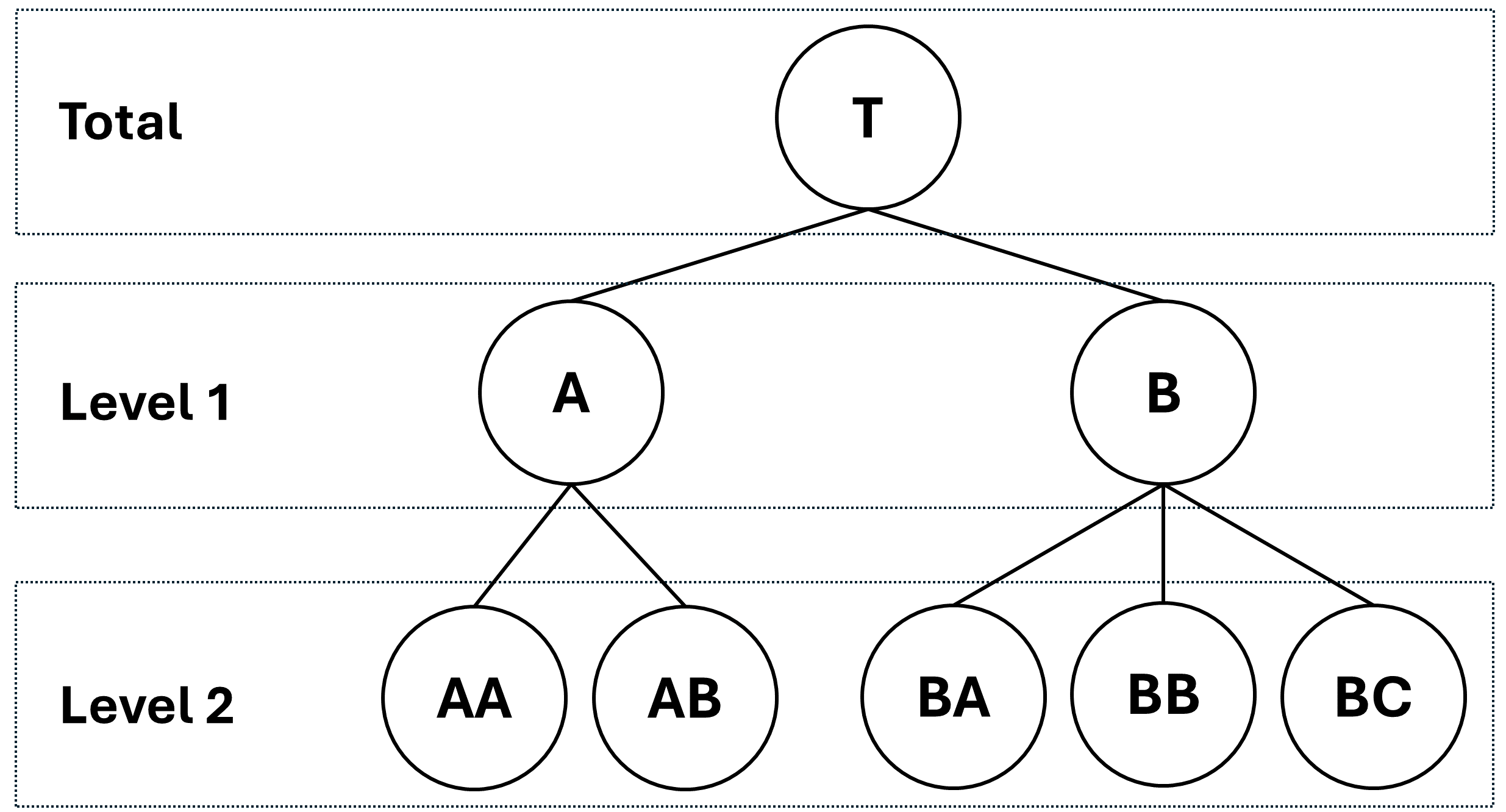}}
\caption{\small Illustration of a simple time series hierarchy with two levels.}
\label{fig:hts_simple}
\end{center}
\vskip -0.3in
\end{figure}

A hierarchical time series is defined as a set of (bottom) time series which can be summed to higher level time series. An illustration of a simple such hierarchy is provided in \cref{fig:hts_simple}. In this example, the bottom time series encompass 
$$X^{(bottom)} = \left(X^{(AA)}, X^{(AB)}, X^{(BA)}, X^{(BB)}, X^{(BC)}\right)^\intercal.$$
These are progressively summed according to the overall hierarchical structure:
\begin{align*}
    X^{(A)} & = X^{(AA)} + X^{(AB)} \\
    X^{(B)} & = X^{(BA)} + X^{(BB)} + X^{(BC)} \\
    X^{(T)} & = X^{(A)} + X^{(B)} = X^{(AA)} + X^{(AB)} + X^{(BA)} + X^{(BB)} + X^{(BC)}.
\end{align*}

Note that this set of equations can be summarized in a simple matrix multiplication: $X = \mathbf{S}X^{(bottom)}$, where $X = (X^{(T)}, X^{(A)}, X^{(B)}, \dots,X^{(BC)})^\intercal$ are the pooled time series from all levels in hierarchical order and $\mathbf{S}$ is an appropriate structure matrix. In the example of \cref{fig:hts_simple}, clearly

\begin{align*}
    \mathbf{S} &=  
       \begin{bmatrix}
           1 & 1 & 1 & 1 & 1 \\           
           1 & 1 & 0 & 0 & 0 \\
           0 & 0 & 1 & 1 & 1 \\
           1 & 0 & 0 & 0 & 0 \\
           0 & 1 & 0 & 0 & 0 \\
           0 & 0 & 1 & 0 & 0 \\
           0 & 0 & 0 & 1 & 0 \\
           0 & 0 & 0 & 0 & 1 
      \end{bmatrix}.
\end{align*}

The goal of hierarchical time series reconciliation is to ensure the coherence of base forecasts, which are obtained through appropriate forecasting models like ARIMA, ETS, or other forecasting approaches. This means that forecasts should be constrained by the underlying hierarchical structure, i.e. the appropriate bottom time series forecasts should add up to the corresponding higher level forecasts.

Probably the most intuitive method to achieve this goal is called the bottom-up approach (BU). Here, forecasts are only generated on the bottom time series. Higher level forecasts are derived by summing the respective bottom forecasts. However, this approach has a significant drawback: information on higher aggregation levels is omitted. To make matters worse, higher aggregation levels are observed to often feature a preferable signal-to-noise-ratio compared to lower levels, potentially making the forecasting process more difficult and less accurate.

The opposite approach is called the top-down approach (TD). As the name suggests, it consists of disaggregating a single, top-level forecast to arrive at lower level forecasts. However, this implies that all information of lower level time series is omitted in the process. Furthermore, \citet{Hyndman2011} and \citet{Panagiotelis2021} showed that a top-down approach necessarily introduces bias. 

\citet{Hyndman2011} showed that any of the existing methods can be expressed as 
\begin{align}
\label{eqn:recFcst_G}
    \Tilde{X} = \mathbf{SG} \hat{X},
\end{align}
for an appropriate matrix $\mathbf{G}$, where $\Tilde{X}$ are the reconciled forecasts based on the base forecasts $\hat{X}$.
For instance, the bottom-up approach for the example hierarchy from \cref{fig:hts_simple} can be expressed with 
$$\mathbf{G} = \left[ \mathbf{0}_{5 \times 3}, \mathbf{I}_5 \right].$$

Ideally, all available forecasts should be utilized in a way that optimally combines the information on all hierarchical levels. This motivated the introduction of the trace minimization algorithm (MinT) by \citet{Wickramasuriya2019}. Denote by $\mathbf{V}$ the covariance matrix of all forecast errors (from all levels). As the name suggests, MinT aims to minimize the trace of $\mathbf{V}$. In other words, MinT is a solution of the minimization problem that minimizes the total sum of forecast errors under the constraint that the forecasts have to obey the hierarchical structure. Applying \cref{eqn:recFcst_G}, this minimization problem can be written as
\begin{align}
\label{eqn:MinT_min_prob}
    \min_\mathbf{G} \left\{\mathbf{SGW}\mathbf{G}^\intercal\mathbf{S}^\intercal\right\}
\end{align}
under the constraint that the reconciled forecasts are unbiased or, as \citet{Hyndman2011} showed, under the condition that $\mathbf{SGS} = \mathbf{S}$. Here, $\mathbf{W}$ is defined as the covariance matrix of the base forecast errors. \citet{Wickramasuriya2019} further show that the solution to the minimization problem in \cref{eqn:MinT_min_prob} has the closed form solution
\begin{align}
\label{eqn:MinT_sol}
    \mathbf{G} = \left(\mathbf{S}^\intercal \mathbf{W}^{-1}\mathbf{S} \right)^{-1}\mathbf{S}^\intercal\mathbf{W}^{-1}.
\end{align}
This algorithm is accessible through the \verb|hts|\citep{Hyndman2021htsLib} and \verb|fable|\citep{OHara-Wild2023} packages in R or the \verb|pyhts| and \verb|hierarchicalforecast| libraries in Python \citep{Olivares.772022}. \citet{Steinmeister2024} provide an adapted version of the \verb|hts| implementation of MinT, allowing for degenerate hierarchical structures.

Note that the solution in \cref{eqn:MinT_sol} requires the inversion of the covariance matrix of base forecast errors $\mathbf{W}$. With a total of $p$ time series, the estimation of $\mathbf{W}$ thus requires the estimation of $\frac{p(p+1)}{2}$ variables. The inversion of this matrix can cause further issues, particularly when $\mathbf{W}$ is badly conditioned. This may amplify estimation errors and lead to an unstable solution. Therefore, in the case of hierarchies with a large number of time series and a relatively small number of observations, \citet{Wickramasuriya2019} propose the use of a shrinkage estimator introduced by \citet{Schafer2005}. This shrinkage estimator finds a %n optimal 
combination of the standard covariance estimator and a diagonal matrix containing only the base forecast variances. The result is a covariance estimator where the off-diagonal entries are shrunk towards zero. 
While this approach significantly stabilizes the algorithms, we propose an iterative approach in the following section which additionally reduces the number of estimated parameters significantly. This approach is compared against common reconciliation methods in an extensive simulation study in \cref{sec:sim}.

\subsection{Iterative Trace Minimization}
\label{sec:ItMinT}

\begin{figure*}[ht]
%\vskip 0.2in
\begin{center}
\centerline{\includegraphics[width=0.9\linewidth]{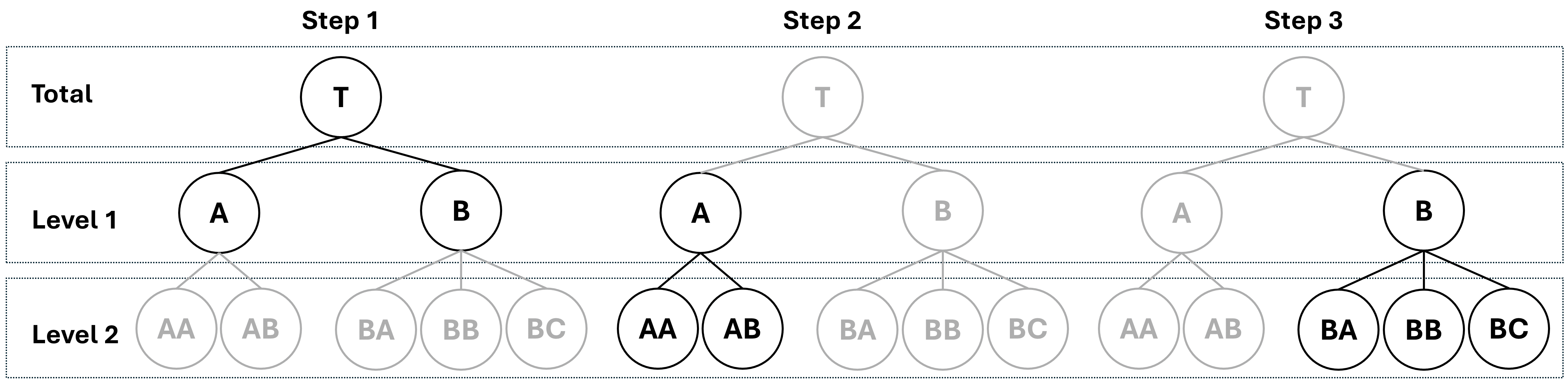}}
\caption{The steps with their respective sub-hierarchies in each iteration.}
\label{fig:MinTit_steps}
\end{center}
\vskip -0.3in
\end{figure*}

In the spirit of divide and conquer, we propose breaking a large hierarchy down into smaller sub-hierarchies to mitigate the stability issues with MinT. An illustration of the steps that the algorithm passes through in each iteration for the example hierarchy in \cref{fig:hts_simple} is provided in \cref{fig:MinTit_steps}. The iterative process views each 1-level sub-hierarchy as an optimization problem in itself. The intuition is that this allows for the estimation of several small covariance matrices instead of one large one. Iterating this process allows information from the higher levels to be propagated to lower levels and vice versa, similar to back-propagation in neural networks \citep{Rumelhart1986backprop}.

The total number of parameters is greatly reduced by this method. For illustration, assume a hierarchy with $D$ levels and $w>1$ sub-nodes each. Then the total number of nodes, or time series, is given by 
\begin{align}
\label{eq:cont_cov}
    p = 
    \sum_{k=0}^D w^k = 
    \frac{w^{D+1}-1}{w-1}.
\end{align}
This implies that $\frac{w^{D+1}-1}{w-1}\frac{w^{D+1}-w-2}{w-1}\frac{1}{2} = \mathcal{O}\left(w^{2D}\right)$ parameters have to be estimated to obtain the covariance matrix. Thus, the complexity is polynomial of order $2D$ in terms of the hierarchy's width and of exponential order in terms of the hierarchy's depth.

In contrast, the iterative approach goes through $\frac{w^{D}-1}{w-1}$ steps, each requiring the estimation of $w(w+1)/2$ parameters. Thus, yielding a polynomial order of $D+1$ degrees in terms of the hierarchy's width. The order of complexity for the depth remains exponential, but with a (significantly) smaller base. Overall, the complexity is given by $\mathcal{O}\left(w^{D+1}\right)$. \cref{tab:MinT_Iterative_compex} illustrates the impact that this difference in complexity can have in practice. Particularly when hierarchies are large, it can be seen that the iterative algorithm, as detailed in \cref{algo:MinTit}, has a much smaller complexity in terms of internal parameters that need to be estimated. However, this advantage usually does not translate into faster runtime. The iterative approach requires several iterations, usually ranging from high double digits to several hundreds. Furthermore, more iterations may be required when the hierarchy is large, as it may take longer for the information to propagate.

\begin{table}[htbp]
  \footnotesize\centering
  \caption{\small Comparison of the number of parameters required to estimate the covariance matrix for simple hierarchies each with constant widths of 2 and 3 and depths ranging from 2 to 6.}
    \begin{tabular}{r|rrr|rrr}
    \multicolumn{1}{l|}{Width} & \multicolumn{3}{c|}{2} & \multicolumn{3}{c}{3} \\
    \hline
    \multicolumn{1}{l|}{Depth} & \multicolumn{1}{l}{MinT} & \multicolumn{1}{l}{Iterative} & \multicolumn{1}{l|}{Fraction} & \multicolumn{1}{l}{MinT} & \multicolumn{1}{l}{Iterative} & \multicolumn{1}{l}{Fraction} \\
    \hline
    2     & 28    & 9     &           0.32  & 91    & 24    &           0.26  \\
    3     & 120   & 21    &           0.18  & 820   & 78    &           0.10  \\
    4     & 496   & 45    &           0.09  & 7381  & 240   &           0.03  \\
    5     & 2016  & 93    &           0.05  & 66430 & 726   &           0.01  \\
    6     & 8128  & 189   &           0.02  & 597871 & 2184  &           0.00  \\
     % 7     & 32640 & 381   &           0.01  & 5380840 & 6558  &           0.00  \\
    % 8     & 130816 & 765   &           0.01  & 48427561 & 19680 &           0.00  \\
    % 9     & 523776 & 1533  &           0.00  & 4.36E+08 & 59046 &           0.00  \\
    % 10    & 2096128 & 3069  &           0.00  & 3.92E+09 & 177144 &           0.00  \\
    \end{tabular}%
  \label{tab:MinT_Iterative_compex}%
  \vskip -0.3in
\end{table}%

\begin{algorithm}[ht]
\caption{Iterative Trace Minimization}
\label{algo:MinTit}
\begin{algorithmic}[1]
\STATE \textbf{Input:} A hierarchical structure with levels $L=\{L_0,...,L_N\}$ each containing the time series $TS_{L_i} = \{TS_{L_i}^0, ...,TS_{L_i}^{K_{L_i}}\}$ with respective constituents $C(TS_{L_i}^{k_{L_i}})$. Forecasts are denoted $\widehat{TS}$ and $\widehat{C(TS)}$ and residuals are denoted $R(TS)$ and $R(C(TS))$. Define $\epsilon>0$ as the convergence threshold.
\FOR{$k \in \{1,..., maxit\}$}
    \STATE Save vector of all forecasts in $forecasts_{old}$ 
    \FOR{$l \in L$}
        \FOR{$TS \in l$}
            \STATE $\mathbf{S} \gets$ structure matrix containing $TS$ and $C(TS)$
            \STATE $\mathbf{\Sigma} \gets$ covariance matrix$^\ast$ of $[TS, C(TS)]$
            \STATE $\left[\widehat{TS}, \widehat{C(TS)}\right] \gets$ MinT$\left(\left[\widehat{TS}, \widehat{C(TS)}\right], \mathbf{\Sigma}, \mathbf{S}\right)$
        \ENDFOR
    \ENDFOR
    \STATE Save vector of all updated forecasts in $forecasts_{new}$
    \IF{$\left\lVert forecasts_{new} - forecasts_{old}\right\lVert < \epsilon$}
        \STATE return $forecasts_{new}$
    \ENDIF
\ENDFOR
\end{algorithmic}
$^\ast$ The covariance matrix can either be locally estimated (using only the sub-hierarchy) or globally by using the global shrinkage estimator proposed by \citet{Schafer2005} and subsetting the global matrix appropriately. The former will be denoted as MinTit$_l$ and the latter as MinTit$_g$
\end{algorithm}

\section{Monte Carlo Simulation}
\label{sec:sim}
To evaluate the performance of the iterative algorithm and compare it against established methods, we conduct Monte Carlo experiments with $5000$ iterations in \cref{sec:sim_exp}, unless stated otherwise.

The compared reconciliation methods are summarized in \cref{sec:sim_rec} and an overview over the utilized forecasting methods is given in \cref{sec:sim_fcst}. Results are presented in root mean squared error (RMSE) relative to the baseline forecasts.

\subsection{Reconciliation Methods}
\label{sec:sim_rec}

The following reconciliation methods are compared: the bottom-up approach (BU), trace minimization (MinT) and two iterative versions: MinTit$_l$ and MinTit$_g$. They are all briefly explained below:

The \textbf{bottom-up} (BU) approach requires forecasts only for the bottom time series. These forecasts are then aggregated according to the specified hierarchical structure. This is the simplest approach in the comparison -- but nevertheless commonly used as it presents a canonical way of ensuring coherence.

\textbf{Trace minimization} (MinT) functions by minimizing the sum of expected forecasting errors. 
A closed form solution is given by \cref{eqn:MinT_sol}. This approach has been shown to excel in many situations \citep{Wickramasuriya2019}. 

\textbf{Iterative trace minimizations} (MinTit) is introduced in \cref{sec:HTS} and iteratively reconciles sub-hierarchies until forecasts converge, as illustrated in \cref{fig:MinTit_steps} and defined in \cref{algo:MinTit}. It was inspired by the idea that it may not be necessary to specify the complete covariance matrix as in MinT, potentially leading to higher performance in settings with large hierarchical structures and very short time series or time series of varying lengths. 
Two options for estimation of the covariance matrix in the sub-hierarchy are compared: either the shrinkage estimator \citep{Schafer2005} can be computed globally and subdivided according to the respective sub-hierarchies (MinTit$_g$), or the shrinkage estimator can be computed locally, leading to locally differing shrinkage factors (MinTit$_l$). The latter has the advantage that more data can be incorporated when available on the sub-hierarchy-level.

Additionally, two weighted least squares methods are compared, which differ from MinT mainly in assumptions regarding the form of the covariance matrix $\mathbf{W}$:

\textbf{Structural Scaling} (WLS$_s$) as in \citet{ATHAN2017TempHier} assumes equal and uncorrelated residual variances of the reconciled bottom time series. This results in higher variances on higher levels in the hierarchy, proportional to the number of subsumed bottom time series. This assumption can be expressed as
$$\mathbf{W} = k \mathbf{\Lambda},$$
where $\mathbf{\Lambda} = diag(\mathbf{S1})$ is a diagonal matrix with the row sums of the structure matrix $\mathbf{S}$ on the diagonal.

\textbf{Variance Scaling} (WLS$_v$) as in \citet{ATHAN2017TempHier} assumes that reconciled errors of all time series (also including higher hierarchical levels) are uncorrelated. Thus, it uses
$$\mathbf{W} = diag\left(\hat{\mathbf{W}}\right),$$
where $\hat{\mathbf{W}}$ is the sample covariance matrix of the observed forecast errors.

\subsection{Forecasting Methods}
\label{sec:sim_fcst}
Base forecasts are generated with three common models: Error Trend and Seasonality (ETS), ARIMA, and Gaussian Process Regression (GPR). ETS and ARIMA are standard time series models, which are, e.g., explained in more detail in \citet{Hyndman.2018Book}.

This study uses the R implementations of the models available in the \verb|forecast| package \citep{Hyndman.2008forecastLibrary}. For ARIMA, we utilize the \verb|auto.arima| function, which searches for an optimal model, including models with seasonality. The exact process is explained in \citet{Hyndman2008AutomaticForecasting}. In all evaluations and result discussions, we use the abbreviation ARIMA for ease of presentation. 

GPR is explained in more detail in \citet{Wang.2023}. It was one of the strongest performing forecasting models on studies with the M3-Competition dataset \citet{Ahmed2010, Makridakis2018} and performed very well on a previous study involving the WSTS dataset \citep{steinmeister2024human}. Here, the GPR implementation in the \verb+kernlab+ R package \citep{Karatzoglou.2004} is used.

\subsection{Experiments}
\label{sec:sim_exp}
The design of our numeric experiments is inspired by \citet{Wickramasuriya2019}, where similar Monte Carlo simulations were run to assess the performance of the MinT algorithm. 
In contrast to \citet{Wickramasuriya2019}, which examined time series of length $T = 60$, $180$, and $300$ observations, the simulation study in this paper focuses the challenging situation of shorter time series with $T = 15$, $30$, and $60$ observations with the final $h = 4$, $4$, and $8$ observations withheld respectively for out-of-sample evaluation.

To give detailed recommendations for such shorter time series, we study several important scenarios. In particular, we investigate  
\begin{enumerate}
    \item the impact of correlated time series (\cref{sec:sim_impactOfCorr}) in a setup similar to Section 3.2 of \citet{Wickramasuriya2019}, 
    \item the effect of smoothing, which is supposed to replicate higher signal-to-noise-ratios often observed in practice (\cref{sec:sim_effOfAggr}) based on the experimental setup in Section 3.3 of \citet{Wickramasuriya2019},
    \item reconciliation in the presence of seasonality (\cref{sec:sim_season}), similar to Section 3.4 in \citet{Wickramasuriya2019},
    \item the effect of differing time series lengths (\cref{sec:sim_len}),
    \item the effect of degeneracy on the choice of the reconciliation algorithm (\cref{sec:sim_deg}), and
    \item reconciliation for a large hierarchy (\cref{sec:sim_largeHier}) in a setup similar to Section 3.5 of \citet{Wickramasuriya2019}.
\end{enumerate}
The concrete scenario settings and the resulting performances of all approaches are discussed below.

\subsubsection{Impact of Correlation}
\label{sec:sim_impactOfCorr}
\begin{figure}[ht]
%\vskip 0.2in
\begin{center}
\centerline{\includegraphics[width=1.0\linewidth]{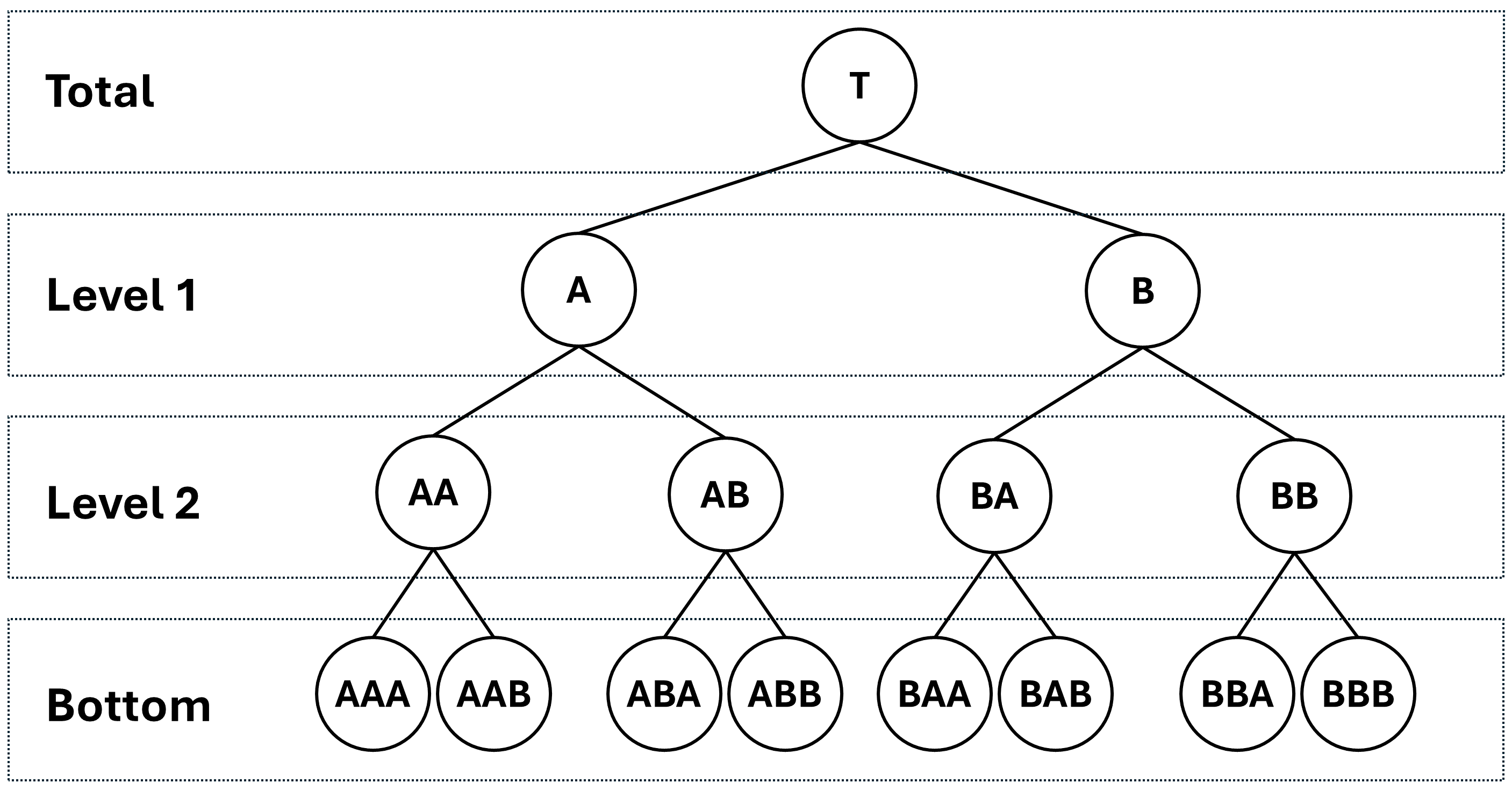}}
\caption{\small Hierarchical structure for the simulations in \cref{sec:sim_impactOfCorr}.}
\label{fig:sim_corr_hier}
\end{center}
\vskip -0.3in
\end{figure}

\citet{Wickramasuriya2019} considered a two level hierarchy with $7$ total time series. Because the iterative version proposed in \cref{sec:ItMinT} is intended for larger hierarchies with short histories, a slightly larger hierarchy of level $3$ with $15$ total time series, which is illustrated in \cref{fig:sim_corr_hier}, is considered.

The bottom time series follow a similar data generating process as in \citet{Wickramasuriya2019}: Each time series on the bottom level were generated from an $\text{ARIMA}(p,d,q)$ process, where $p$ and $q$ were uniformly drawn from $\{0,1,2\}$ and $d$ from $\{0,1\}$. For each time series, the ARIMA coefficients were drawn from different uniform distributions as detailed in \cref{tab:ARIMA_coef}. The given boundaries are the same as in \citet{Wickramasuriya2019} to ensure better comparability. 
\begin{table}[t]
  \centering
  \caption{\small Overview of the sampling ranges of the ARIMA parameters, analogous to \citet{Wickramasuriya2019}.}
    \begin{tabular}{rll}
    \multicolumn{3}{c}{AR coefficients}  \\
    \hline
    \multicolumn{1}{l}{$p$} & coefficient & sampling range  \\
    \hline
    1     & $\phi_1$ & $[0.5,0.7]$  \\
    2     & $\phi_1$ & $[\phi_2-0.9,0.9-\phi_2]$ \\
          & $\phi_2$ & $[0.5,0.7]$ \\
    \multicolumn{3}{c}{MA coefficients}  \\
    \hline
    \multicolumn{1}{l}{$q$} & coefficient & sampling range  \\
    \hline
    1     & $\theta_1$ & $[0.5,0.7]$  \\
    2     & $\theta_1$ & $[-\frac{0.9+\theta_2}{3.2},\frac{0.9+\theta_2}{3.2}]$ \\
          & $\theta_2$ & $[0.5,0.7]$ \\
    \end{tabular}%
  \label{tab:ARIMA_coef}%
\end{table}%

Moreover, the bottom time series were generated with contemporaneous errors following a predefined covariance matrix to introduce correlation of time series throughout the hierarchy.
However, since the larger hierarchy calls for more bottom time series, the covariance matrix used here is a larger version of the one used in \citet{Wickramasuriya2019} and given by $\mathbf{\Sigma}$ as defined in \cref{eqn:contempErr}.

\begin{align}
\label{eqn:contempErr}
\small
\mathbf{\Sigma} = 
    \begin{bmatrix}
        5 & 3 & 2 & 1 & 1 & 1 & 1 & 1 \\ 
        3 & 4 & 2 & 1 & 1 & 1 & 1 & 1 \\
        2 & 2 & 5 & 3 & 2 & 1 & 1 & 1 \\ 
        1 & 1 & 3 & 4 & 3 & 2 & 1 & 1 \\
        1 & 1 & 2 & 3 & 5 & 3 & 2 & 1 \\ 
        1 & 1 & 1 & 2 & 3 & 4 & 2 & 1 \\
        1 & 1 & 1 & 1 & 2 & 2 & 5 & 3 \\
        1 & 1 & 1 & 1 & 1 & 1 & 3 & 4
    \end{bmatrix}
\end{align}
Similarly to the simulation in Section 3.2 of \citet{Wickramasuriya2019}, this structure creates more strongly correlated errors for time series under the same parent node in the hierarchical structure and generally more moderately correlated errors for time series of more distant parent nodes. The thus generated bottom time series were then summed according to the defined hierarchical structure as illustrated in \cref{sec:sim_impactOfCorr}.

\textit{Results for ETS forecasts.} \cref{tab:sim_cor_ETS} summarizes the simulation results for the ETS-generated forecasts. The table is divided into three column groups containing the results for time series of total lengths $T = 15$, $T = 30$, and $T= 60$, respectively. Each of them contains columns of relative RMSE changes for forecasting horizons of $h = 1$, $h = 1:2$ (the average over the 1-step and 2-step ahead forecasts), and $h = 1:4$ in the first two cases ($T=15$ and $T=30$) . In the case of $T = 60$, the forecast horizons are $h = 1$, $h = 1:4$, and $h = 1:8$. The last column provides row-wise averages.

The rows contain groups relating to the different hierarchical levels. Each of these groups provides details of the relative RMSE of the compared reconciliation method. Negative values indicate a \%-improvement over the unreconciled base forecasts and positive values indicate a \%-increase in RMSE. Hence, smaller values indicate better forecast performance. 

The introduced contemporaneous covariance structure gives rise to lower SNRs on higher hierarchical levels. This explains how the simple BU approach on average achieves relatively large error reductions. 

However, on average, the MinTit reconciliation methods lead to the highest reduction in RMSE for forecasts generated by 
%ARIMA (see \cref{tab:sim_cor_ARIMA}) and 
ETS. This largely stems from its superior performance on higher hierarchical levels, with substantial improvements over the forecasts reconciled with MinT or the WLS approaches.
%-- particularly for the ARIMA forecasts. 
However, the performance of MinT and MinTit was much closer on the bottom-level: both MinT and MinTit$_g$ achieving a $0.8\%$ improvement for the ETS forecasts.
%and MinT achieving a $5.3\%$ improvement (against $5.1\%$) for ARIMA forecasts. 

The \textit{results for ARIMA forecasts} (see \cref{tab:sim_cor_ARIMA}) are similar to those for ETS. MinTit performed best on average -- leading to average improvements in RMSE of 10\% (compared to 9.1\% for MinT). As was observed with the ETS forecasts, this strong performance largely stemmed from the larger RMSE reductions on higher hierarchical levels (up to 11.9\%). On the bottom level, MinT outperformed MinTit, achieving a $5.3\%$ improvement against $5.1\%$.

\begin{figure}[t]
%\vskip 0.2in
\begin{center}
\centerline{\includegraphics[width=\columnwidth]{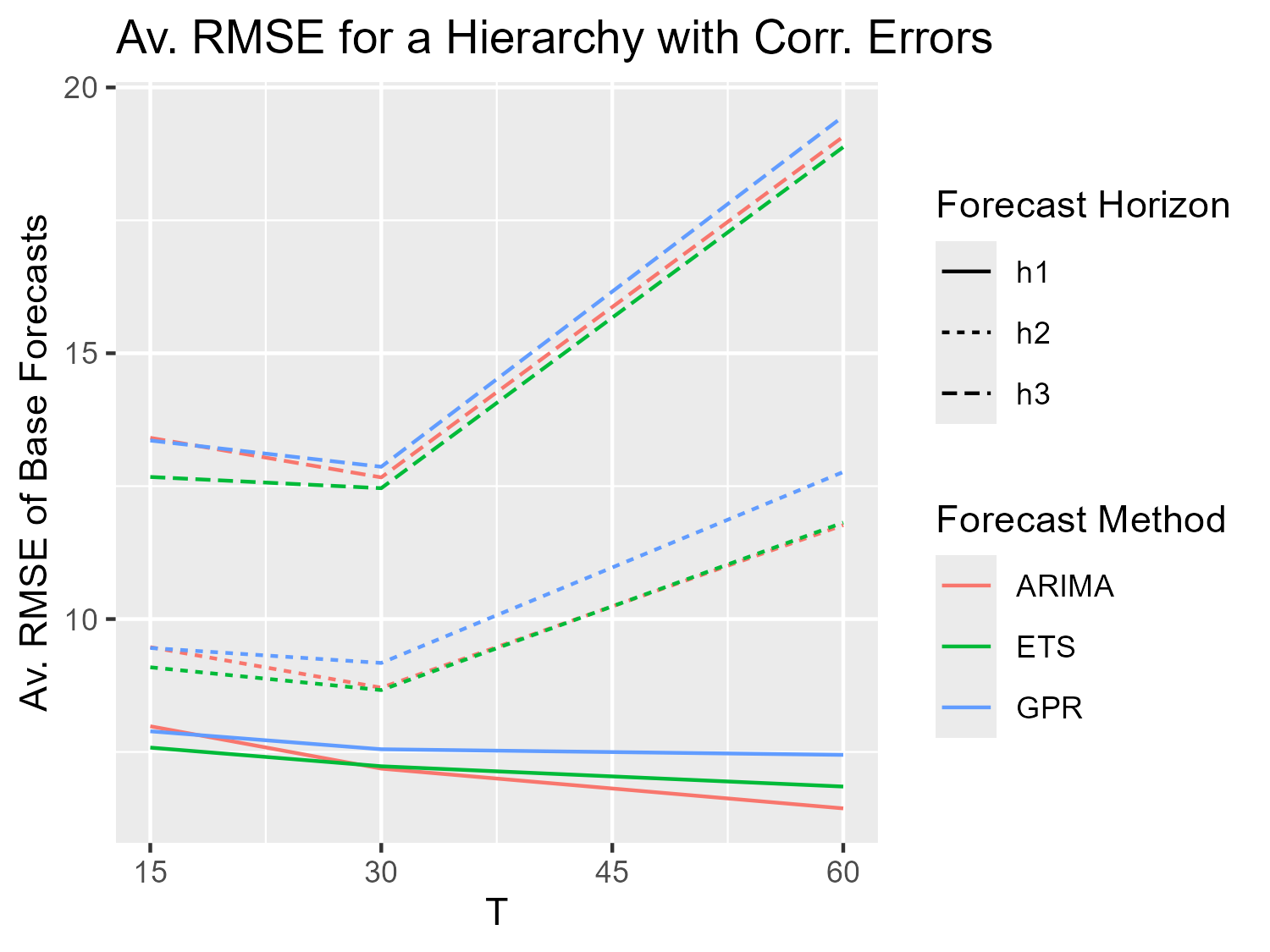}}
\caption{Average RMSE for the various base forecasts in \cref{sec:sim_impactOfCorr} (ARIMA in red, ETS in green, and GPR in blue) by total length of the time series. A forecast horizon of $h1$ corresponds to $h = 1$, $h2$ to $h = 1:2$ in the case of $T = 15, 30$ and $h = 1:4$ for $T = 60$. Similarly, $h3$ corresponds to the longer forecasting horizons of $h = 1:4$ and $h = 1:8$, respectively.}
\label{fig:sim_cor_bfe}
\end{center}
\vskip -0.3in
\end{figure}

\textit{Results for GPR forecasts.} MinT and MinTit result in similar improvements for the GPR-generated forecasts (\cref{tab:sim_cor_GPR}) with MinT leading to slightly more accurate reconciled forecasts. However, the base forecast performance of GPR was worse than for ETS and ARIMA, see \cref{fig:sim_cor_bfe}. Reconciliation resulted in a maximum average error reduction of $7.3\%$ using the MinT reconciliation method as compared to a $10\%$ reduction in RMSE for ARIMA forecasts using MinTit$_g$ reconciliation. Thus, these GPR forecasts are less accurate, putting the comparative advantage of MinT against MinTit into perspective.

\begin{table}[t]
  \centering
    \footnotesize\tabcolsep2.5pt
  \renewcommand{\arraystretch}{.9}
  \caption{\small Out-of-sample forecast performance for the hierarchy with contemporaneously correlated time series
  in \cref{sec:sim_impactOfCorr} for ETS-generated forecasts relative to the ETS-generated base forecasts.}
\begin{tabular}{lrrrlrrrlrrrr|r}
      & \multicolumn{13}{c}{ETS} \bigstrut[b]\\
\cline{2-14}      & \multicolumn{3}{c}{T = 15} &       & \multicolumn{3}{c}{T = 30} &       & \multicolumn{3}{c}{T = 60} & \multicolumn{1}{c}{} &  \bigstrut\\
\cline{2-4}\cline{6-8}\cline{10-12}      & \multicolumn{1}{c}{h=1} & \multicolumn{1}{c}{1:2} & \multicolumn{1}{c}{1:4} &       & \multicolumn{1}{c}{h=1} & \multicolumn{1}{c}{1:2} & \multicolumn{1}{c}{1:4} &       & \multicolumn{1}{c}{h=1} & \multicolumn{1}{c}{1:4} & \multicolumn{1}{c}{1:8} & \multicolumn{1}{c}{} & \multicolumn{1}{c}{Av.} \bigstrut\\
\cline{2-12}\cline{14-14}      & \multicolumn{13}{c}{Top-Level} \bigstrut\\
\cline{2-14}MinT  & -6.8  & -5.3  & -3.5  &       & -9.1  & -8.5  & -7.6  &       & -8.9  & -6.2  & -4    &       & -6.7 \bigstrut[t]\\
WLS$_s$ & -7.4  & \textbf{-6.7} & \textbf{-6.5} &       & -8.8  & -9.3  & \textbf{-9.7} &       & -7.3  & -7.2  & \textbf{-6.6} &       & -7.7 \\
WLS$_v$ & -7.4  & \textbf{-6.7} & \textbf{-6.5} &       & -8.8  & -9.3  & \textbf{-9.7} &       & -7.3  & -7.2  & \textbf{-6.6} &       & -7.7 \\
BU    & -7.3  & -5.6  & -4.8  &       & \textbf{-10.5} & -9.6  & -8.3  &       & -9.5  & -6.4  & -3.6  &       & -7.3 \\
MinTit$_g$ & \textbf{-7.7} & -6.6  & -5.8  &       & -10.1 & \textbf{-10} & \textbf{-9.7} &       & -9.3  & \textbf{-7.7} & -6.2  &       & \textbf{-8.1} \\
MinTit$_l$ & \textbf{-7.7} & -6.5  & -5.6  &       & -10   & -9.8  & -9.3  &       & \textbf{-9.6} & -7.6  & -6    &       & -8.0 \\
      & \multicolumn{13}{c}{Level 1} \bigstrut[b]\\
\cline{2-14}MinT  & -6.7  & -5.6  & -3.8  &       & -8.5  & -7.8  & -7    &       & -8    & -6.1  & -4.2  &       & -6.4 \bigstrut[t]\\
WLS$_s$ & -6.9  & \textbf{-6.5} & \textbf{-6.2} &       & -7.5  & -8    & -8.3  &       & -6.1  & -6.5  & \textbf{-6.2} &       & -6.9 \\
WLS$_v$ & -6.8  & -6.4  & -6.1  &       & -7.5  & -8    & -8.2  &       & -6.1  & -6.5  & -6.1  &       & -6.9 \\
BU    & -6.9  & -5.4  & -4.5  &       & \textbf{-9.5} & -8.3  & -7    &       & -8.4  & -5.9  & -3.3  &       & -6.6 \\
MinTit$_g$ & \textbf{-7.1} & -6.2  & -5.2  &       & -9.1  & \textbf{-8.8} & \textbf{-8.4} &       & -8.3  & \textbf{-7.2} & -5.9  &       & \textbf{-7.4} \\
MinTit$_l$ & \textbf{-7.1} & -6    & -5    &       & -9    & -8.5  & -8    &       & \textbf{-8.5} & -7.1  & -5.7  &       & -7.2 \\
      & \multicolumn{13}{c}{Level 2} \bigstrut[b]\\
\cline{2-14}MinT  & \textbf{-4.3} & -3.6  & -2.4  &       & -5.2  & -4.9  & -4.2  &       & -4.6  & -3.1  & -2.4  &       & -3.9 \bigstrut[t]\\
WLS$_s$ & -4    & \textbf{-3.8} & \textbf{-3.9} &       & -4    & -4.6  & -4.9  &       & -2.8  & -3.3  & \textbf{-3.7} &       & -3.9 \\
WLS$_v$ & -3.7  & -3.6  & -3.6  &       & -3.8  & -4.5  & -4.8  &       & -2.6  & -3.2  & -3.6  &       & -3.7 \\
BU    & -4    & -2.8  & -2.2  &       & \textbf{-5.8} & -4.9  & -3.8  &       & -4.8  & -2.7  & -1.1  &       & -3.6 \\
MinTit$_g$ & -4.1  & -3.5  & -2.9  &       & -5.3  & \textbf{-5.3} & \textbf{-5} &       & -4.7  & \textbf{-3.9} & -3.5  &       & \textbf{-4.2} \\
MinTit$_l$ & -4.1  & -3.3  & -2.6  &       & -5.2  & -4.9  & -4.5  &       & \textbf{-4.9} & -3.6  & -3.2  &       & -4.0 \\
      & \multicolumn{13}{c}{Bottom-Level} \bigstrut[b]\\
\cline{2-14}MinT  & \textbf{-0.8} & \textbf{-1.3} & -0.8  &       & \textbf{-0.1} & -0.6  & -1.1  &       & \textbf{-0.5} & -0.8  & -1.6  &       & \textbf{-0.8} \bigstrut[t]\\
WLS$_s$ & 0     & -0.7  & \textbf{-1.1} &       & 1.4   & 0.2   & -0.8  &       & 1.5   & -0.4  & -1.7  &       & -0.2 \\
WLS$_v$ & 0.8   & -0.1  & -0.5  &       & 2.1   & 0.6   & -0.5  &       & 2.2   & -0.2  & -1.5  &       & 0.3 \\
BU    & 0     & 0     & 0     &       & 0     & 0     & 0     &       & 0     & 0     & 0     &       & 0.0 \\
MinTit$_g$ & -0.2  & -0.9  & -0.9  &       & 0.1   & \textbf{-0.7} & \textbf{-1.4} &       & -0.2  & \textbf{-1.2} & \textbf{-2.2} &       & \textbf{-0.8} \\
MinTit$_l$ & -0.4  & -0.8  & -0.8  &       & 0.2   & -0.4  & -1.1  &       & -0.4  & -1    & -2    &       & -0.7 \\
      & \multicolumn{13}{c}{Average} \bigstrut[b]\\
\cline{2-14}MinT  & -5.4  & -4.5  & -2.9  &       & -7    & -6.6  & -5.8  &       & -6.8  & -4.8  & -3.3  &       & -5.2 \bigstrut[t]\\
WLS$_s$ & -5.6  & \textbf{-5.2} & \textbf{-5} &       & -6.2  & -6.7  & -7.1  &       & -5    & -5.2  & \textbf{-5.1} &       & -5.7 \\
WLS$_v$ & -5.3  & -5    & -4.9  &       & -6.1  & -6.7  & -7    &       & -4.8  & -5.2  & \textbf{-5.1} &       & -5.6 \\
BU    & -5.5  & -4.1  & -3.4  &       & \textbf{-8} & -7    & -5.8  &       & -7.1  & -4.6  & -2.4  &       & -5.3 \\
MinTit$_g$ & -5.7  & -5    & -4.2  &       & -7.6  & \textbf{-7.5} & \textbf{-7.2} &       & -7    & \textbf{-5.9} & -4.9  &       & \textbf{-6.1} \\
MinTit$_l$ & \textbf{-5.8} & -4.9  & -4    &       & -7.5  & -7.2  & -6.8  &       & \textbf{-7.2} & -5.7  & -4.7  &       & -6.0 \\
    \end{tabular}%
  \label{tab:sim_cor_ETS}%
\end{table}%

\begin{table}[t]
  \centering
  \footnotesize\tabcolsep2.5pt
  \renewcommand{\arraystretch}{.9}
  \caption{\small Out-of-sample forecast performance for the hierarchy with contemporaneously correlated time series
  in \cref{sec:sim_impactOfCorr} for ARIMA-generated forecasts relative to the ARIMA-generated base forecasts.}
    % Table generated by Excel2LaTeX from sheet 'paper'
% Table generated by Excel2LaTeX from sheet 'paper'
\begin{tabular}{lrrrlrrrlrrrr|r}
      & \multicolumn{13}{c}{ARIMA} \bigstrut[b]\\
\cline{2-14}      & \multicolumn{3}{c}{T = 15} &       & \multicolumn{3}{c}{T = 30} &       & \multicolumn{3}{c}{T = 60} & \multicolumn{1}{c}{} &  \bigstrut\\
\cline{2-4}\cline{6-8}\cline{10-12}      & \multicolumn{1}{c}{h=1} & \multicolumn{1}{c}{1:2} & \multicolumn{1}{c}{1:4} &       & \multicolumn{1}{c}{h=1} & \multicolumn{1}{c}{1:2} & \multicolumn{1}{c}{1:4} &       & \multicolumn{1}{c}{h=1} & \multicolumn{1}{c}{1:4} & \multicolumn{1}{c}{1:8} & \multicolumn{1}{c}{} & \multicolumn{1}{c}{Av.} \bigstrut\\
\cline{2-12}\cline{14-14}      & \multicolumn{13}{c}{Top-Level} \bigstrut\\
\cline{2-14}MinT  & -16.2 & -13.7 & -10.1 &       & -11.6 & -10.8 & -9.6  &       & -8.6  & -7.2  & -6.7  &       & -10.5 \bigstrut[t]\\
WLS$_s$ & -13.9 & -12.5 & -10.8 &       & -11.2 & -10.6 & -10.5 &       & -8.9  & -8.2  & -8.5  &       & -10.6 \\
WLS$_v$ & -13.9 & -12.5 & -10.8 &       & -11.2 & -10.6 & -10.5 &       & -8.9  & -8.2  & -8.5  &       & -10.6 \\
BU    & -14.6 & -12.1 & -9.4  &       & -8.3  & -8.5  & -9.3  &       & -9.8  & -8.5  & -7.7  &       & -9.8 \\
MinTit$_g$ & -16.4 & \textbf{-13.9} & \textbf{-10.9} &       & \textbf{-13.5} & \textbf{-12.5} & \textbf{-11.6} &       & \textbf{-10.7} & -9.2  & \textbf{-8.8} &       & \textbf{-11.9} \\
MinTit$_l$ & \textbf{-16.5} & \textbf{-13.9} & -10.7 &       & \textbf{-13.5} & \textbf{-12.5} & -11.5 &       & -10.6 & \textbf{-9.3} & -8.7  &       & \textbf{-11.9} \\
      & \multicolumn{13}{c}{Level 1} \bigstrut[b]\\
\cline{2-14}MinT  & \textbf{-15.3} & \textbf{-12.7} & -9.2  &       & -13   & -12.3 & -9.6  &       & -7.8  & -6.6  & -5.7  &       & -10.2 \bigstrut[t]\\
WLS$_s$ & -12.3 & -11   & -9.4  &       & -10.9 & -10.7 & -9.5  &       & -7.3  & -6.8  & -6.8  &       & -9.4 \\
WLS$_v$ & -12.3 & -11   & -9.3  &       & -11   & -10.7 & -9.5  &       & -7.4  & -6.9  & -6.8  &       & -9.4 \\
BU    & -12.8 & -10.4 & -7.8  &       & -7.3  & -7.9  & -7.8  &       & -7.9  & -7    & -5.7  &       & -8.3 \\
MinTit$_g$ & -15.1 & -12.5 & \textbf{-9.5} &       & \textbf{-14.4} & \textbf{-13.5} & \textbf{-11} &       & \textbf{-9.4} & \textbf{-8} & \textbf{-7} &       & \textbf{-11.2} \\
MinTit$_l$ & -15.1 & -12.4 & -9.3  &       & \textbf{-14.4} & \textbf{-13.5} & -10.8 &       & \textbf{-9.4} & \textbf{-8} & -6.8  &       & -11.1 \\
      & \multicolumn{13}{c}{Level 2} \bigstrut[b]\\
\cline{2-14}MinT  & \textbf{-11.2} & \textbf{-9} & \textbf{-6.3} &       & -11.4 & -9.8  & -7.3  &       & -5.5  & -4.4  & -3.7  &       & -7.6 \bigstrut[t]\\
WLS$_s$ & -7.4  & -6.5  & -5.9  &       & -7.1  & -6.5  & -6    &       & -4.3  & -4.1  & -4.1  &       & -5.8 \\
WLS$_v$ & -7.3  & -6.4  & -5.7  &       & -7.8  & -6.8  & -6.2  &       & -4.6  & -4.3  & -4.2  &       & -5.9 \\
BU    & -7.8  & -5.9  & -4.4  &       & -2.7  & -2.8  & -3.9  &       & -4.3  & -4.2  & -3.3  &       & -4.4 \\
MinTit$_g$ & -10.3 & -8.3  & -6.2  &       & -12   & \textbf{-10.2} & \textbf{-7.9} &       & \textbf{-6.6} & \textbf{-5.3} & \textbf{-4.5} &       & \textbf{-7.9} \\
MinTit$_l$ & -10.2 & -8.1  & -5.9  &       & \textbf{-12.1} & -10.1 & -7.6  &       & -6.5  & -5.2  & -4.1  &       & -7.8 \\
      & \multicolumn{13}{c}{Bottom-Level} \bigstrut[b]\\
\cline{2-14}MinT  & \textbf{-4.7} & \textbf{-3.9} & \textbf{-2.3} &       & -13.9 & -11   & -5.9  &       & -3.5  & -1.3  & -1.1  &       & \textbf{-5.3} \bigstrut[t]\\
WLS$_s$ & 0.3   & -0.4  & -1    &       & -3.1  & -2.5  & -1.5  &       & 0     & 0.1   & -0.5  &       & -1.0 \\
WLS$_v$ & 0.2   & -0.3  & -0.5  &       & -8.8  & -6.7  & -3.6  &       & -2.2  & -0.7  & -0.9  &       & -2.6 \\
BU    & 0     & 0     & 0     &       & 0     & 0     & 0     &       & 0     & 0     & 0     &       & 0.0 \\
MinTit$_g$ & -3.3  & -2.7  & -1.6  &       & -14.1 & -10.9 & \textbf{-6} &       & \textbf{-4.4} & \textbf{-1.8} & \textbf{-1.4} &       & -5.1 \\
MinTit$_l$ & -3.3  & -2.6  & -1.5  &       & \textbf{-14.4} & \textbf{-11.1} & -5.9  &       & -4.2  & -1.7  & -1.2  &       & -5.1 \\
      & \multicolumn{13}{c}{Average} \bigstrut[b]\\
\cline{2-14}MinT  & \textbf{-13.5} & \textbf{-11.1} & -7.9  &       & -12.3 & -11.1 & -8.6  &       & -7.2  & -5.7  & -4.9  &       & -9.1 \bigstrut[t]\\
WLS$_s$ & -10.3 & -9.2  & -7.9  &       & -9.1  & -8.6  & -7.9  &       & -6.5  & -5.9  & -5.9  &       & -7.9 \\
WLS$_v$ & -10.3 & -9.2  & -7.8  &       & -10.1 & -9.3  & -8.3  &       & -6.9  & -6    & -6    &       & -8.2 \\
BU    & -10.9 & -8.7  & -6.6  &       & -5.7  & -5.9  & -6.3  &       & -7    & -6    & -5.1  &       & -6.9 \\
MinTit$_g$ & -13.1 & -10.9 & \textbf{-8.2} &       & -13.5 & \textbf{-12.1} & \textbf{-9.8} &       & \textbf{-8.8} & \textbf{-7.1} & \textbf{-6.3} &       & \textbf{-10.0} \\
MinTit$_l$ & -13.2 & -10.8 & -8    &       & \textbf{-13.6} & \textbf{-12.1} & -9.7  &       & -8.7  & \textbf{-7.1} & -6.1  &       & -9.9 \\
\end{tabular}%

  \label{tab:sim_cor_ARIMA}%
\end{table}%

\begin{table}[t]
  \centering
    \footnotesize\tabcolsep2.5pt
  \renewcommand{\arraystretch}{.9}
  \caption{\small Out-of-sample forecast performance for the hierarchy with contemporaneously correlated time series
  in \cref{sec:sim_impactOfCorr} for GPR-generated forecasts relative to the GPR-generated base forecasts.}
    % Table generated by Excel2LaTeX from sheet 'paper'
\begin{tabular}{lrrrlrrrlrrrr|r}
      & \multicolumn{13}{c}{GPR} \bigstrut[b]\\
\cline{2-14}      & \multicolumn{3}{c}{T = 15} &       & \multicolumn{3}{c}{T = 30} &       & \multicolumn{3}{c}{T = 60} & \multicolumn{1}{c}{} &  \bigstrut\\
\cline{2-4}\cline{6-8}\cline{10-12}      & \multicolumn{1}{c}{h=1} & \multicolumn{1}{c}{1:2} & \multicolumn{1}{c}{1:4} &       & \multicolumn{1}{c}{h=1} & \multicolumn{1}{c}{1:2} & \multicolumn{1}{c}{1:4} &       & \multicolumn{1}{c}{h=1} & \multicolumn{1}{c}{1:4} & \multicolumn{1}{c}{1:8} & \multicolumn{1}{c}{} & \multicolumn{1}{c}{Av.} \bigstrut\\
\cline{2-12}\cline{14-14}      & \multicolumn{13}{c}{Top-Level} \bigstrut\\
\cline{2-14}MinT  & -13.6 & \textbf{-12.7} & -9.4  &       & -10.1 & \textbf{-9.2} & -8    &       & \textbf{-7.9} & \textbf{-6.3} & -5.7  &       & -9.2 \bigstrut[t]\\
WLS$_s$ & -10.9 & -10   & -9.3  &       & -7.7  & -6.9  & -6.9  &       & -5.3  & -4.6  & -4.9  &       & -7.4 \\
WLS$_v$ & -10.9 & -10   & -9.3  &       & -7.7  & -6.9  & -6.9  &       & -5.3  & -4.6  & -4.9  &       & -7.4 \\
BU    & \textbf{-13.7} & -12.1 & -10.1 &       & -8.9  & -7.1  & -6.7  &       & -5.9  & -4    & -4.3  &       & -8.1 \\
MinTit$_g$ & -13.4 & -12.4 & \textbf{-10.5} &       & -10   & -8.8  & -8.1  &       & -7.5  & -5.8  & -5.9  &       & -9.2 \\
MinTit$_l$ & -13.5 & -12.4 & -10.2 &       & \textbf{-10.2} & -9    & \textbf{-8.3} &       & -7.5  & -6.2  & \textbf{-6.4} &       & \textbf{-9.3} \\
      & \multicolumn{13}{c}{Level 1} \bigstrut[b]\\
\cline{2-14}MinT  & \textbf{-12.3} & \textbf{-11} & -8.9  &       & \textbf{-10.3} & \textbf{-8.8} & \textbf{-6.8} &       & \textbf{-9.1} & \textbf{-5.8} & -4.5  &       & \textbf{-8.6} \bigstrut[t]\\
WLS$_s$ & -8.9  & -7.9  & -8.5  &       & -7.1  & -6    & -5.3  &       & -5.7  & -3.8  & -3.5  &       & -6.3 \\
WLS$_v$ & -8.8  & -7.8  & -8.5  &       & -7.1  & -5.9  & -5.2  &       & -5.8  & -3.8  & -3.4  &       & -6.3 \\
BU    & -11.9 & -10   & -9    &       & -9    & -6.7  & -5.4  &       & -7.2  & -3.7  & -3.2  &       & -7.3 \\
MinTit$_g$ & -11.5 & -10.1 & \textbf{-9.4} &       & -9.8  & -8.1  & -6.6  &       & -8.4  & -5.2  & -4.3  &       & -8.2 \\
MinTit$_l$ & -11.5 & -10.1 & -9    &       & -9.8  & -8.2  & -6.7  &       & -8.4  & -5.4  & \textbf{-4.7} &       & -8.2 \\
      & \multicolumn{13}{c}{Level 2} \bigstrut[b]\\
\cline{2-14}MinT  & \textbf{-8.2} & \textbf{-7.1} & \textbf{-4.9} &       & \textbf{-7.4} & \textbf{-6.2} & \textbf{-4.4} &       & \textbf{-7} & \textbf{-4.1} & \textbf{-2.7} &       & \textbf{-5.8} \bigstrut[t]\\
WLS$_s$ & -4.5  & -3.9  & -4.2  &       & -4    & -3.2  & -2.6  &       & -3.6  & -1.9  & -1.6  &       & -3.3 \\
WLS$_v$ & -4.2  & -3.6  & -4    &       & -3.8  & -3.1  & -2.6  &       & -3.5  & -1.9  & -1.6  &       & -3.1 \\
BU    & -7.3  & -5.8  & -4.6  &       & -5.9  & -4.1  & -2.9  &       & -5.2  & -2.1  & -1.5  &       & -4.4 \\
MinTit$_g$ & -6.8  & -5.8  & -4.8  &       & -6.4  & -5.1  & -3.7  &       & -6.1  & -3.3  & -2.4  &       & -4.9 \\
MinTit$_l$ & -6.8  & -5.8  & -4.4  &       & -6.4  & -5.1  & -3.6  &       & -6    & -3.4  & \textbf{-2.7} &       & -4.9 \\
      & \multicolumn{13}{c}{Bottom-Level} \bigstrut[b]\\
\cline{2-14}MinT  & \textbf{-1.9} & \textbf{-2.1} & \textbf{-1} &       & \textbf{-2.2} & \textbf{-2.7} & \textbf{-2} &       & \textbf{-2.4} & \textbf{-2.4} & \textbf{-1.6} &       & \textbf{-2.0} \bigstrut[t]\\
WLS$_s$ & 2.2   & 1.4   & 0.3   &       & 1.5   & 0.6   & 0.2   &       & 1.2   & 0.1   & -0.1  &       & 0.8 \\
WLS$_v$ & 3     & 2     & 0.7   &       & 1.9   & 0.8   & 0.2   &       & 1.4   & -0.1  & -0.2  &       & 1.1 \\
BU    & 0     & 0     & 0     &       & 0     & 0     & 0     &       & 0     & 0     & 0     &       & 0.0 \\
MinTit$_g$ & 0.2   & -0.3  & -0.4  &       & -0.7  & -1.1  & -0.9  &       & -1.2  & -1.4  & -1.1  &       & -0.8 \\
MinTit$_l$ & 0.1   & -0.4  & -0.3  &       & -0.8  & -1.2  & -1    &       & -1.1  & -1.6  & -1.3  &       & -0.8 \\
      & \multicolumn{13}{c}{Average} \bigstrut[b]\\
\cline{2-14}MinT  & \textbf{-10.7} & \textbf{-9.6} & -7.1  &       & \textbf{-8.6} & \textbf{-7.6} & \textbf{-6} &       & \textbf{-7.3} & \textbf{-5.1} & -4    &       & \textbf{-7.3} \bigstrut[t]\\
WLS$_s$ & -7.4  & -6.6  & -6.6  &       & -5.6  & -4.8  & -4.5  &       & -4.2  & -3.1  & -3    &       & -5.1 \\
WLS$_v$ & -7.2  & -6.5  & -6.5  &       & -5.5  & -4.8  & -4.5  &       & -4.1  & -3.1  & -3    &       & -5.0 \\
BU    & -10.2 & -8.6  & -7.2  &       & -7.2  & -5.4  & -4.5  &       & -5.3  & -2.9  & -2.7  &       & -6.0 \\
MinTit$_g$ & -9.8  & -8.8  & \textbf{-7.5} &       & -8    & -6.7  & -5.7  &       & -6.6  & -4.4  & -3.9  &       & -6.8 \\
MinTit$_l$ & -9.9  & -8.8  & -7.2  &       & -8.1  & -6.9  & -5.8  &       & -6.6  & -4.7  & \textbf{-4.3} &       & -6.9 \\
\end{tabular}%

  \label{tab:sim_cor_GPR}%
\end{table}%

\subsubsection{Impact of the Smoothing Effect of Aggregation}
\label{sec:sim_effOfAggr}
As \citet{Wickramasuriya2019} note, aggregated time series display a higher signal-to-noise-ratio (SNR) compared to their component time series. Borrowing the same design, which was based on the work of \citet{VanErven2015}, this section studies the performance of the different reconciliation methods when the SNR decreases with the hierarchical level on the 3-level hierarchy used in \cref{sec:sim_impactOfCorr} and illustrated in \cref{fig:sim_corr_hier}. \citet{Wickramasuriya2019} used the following design to generate a total of four bottom time series for their 2-level hierarchy:
\begin{align*}
    y_{AA, t} & = z_{AA,t}-\eta_t-0.5\omega_t, \\
    y_{AB, t} & = z_{AB,t}+\eta_t-0.5\omega_t, \\
    y_{BA, t} & = z_{BA,t}-\eta_t+0.5\omega_t, \\
    y_{BB, t} & = z_{BB,t}+\eta_t+0.5\omega_t, 
\end{align*}
where $z_{AA}, z_{Ab}, z_{bA}$ and $z_{BB}$ are simulated as independent ARIMA processes, otherwise as specified in \cref{sec:sim_impactOfCorr}, with standard normal distributed error terms and additional errors, which cancel out hierarchically, $\eta_t~N(0,10)$ and $\omega_t~N(0,6)$.
Note that the representation in terms of added errors ($\eta$ and $\omega$) can equally be achieved through the representation with a single, appropriately correlated error term. In the above case, this can be simplified to 
\begin{align*}
    \begin{bmatrix}
           y_{AA, t} \\
           y_{AB, t} \\
           y_{BA, t} \\
           y_{BB, t}
    \end{bmatrix} 
    & = 
    \begin{bmatrix}
           z_{AA, t} \\
           z_{AB, t} \\
           z_{BA, t} \\
           z_{BB, t}
    \end{bmatrix} 
    + \mathbf{\zeta}_t,
\end{align*}
where $\mathbf{\zeta}_t$ are multivariate normal errors with zero mean and covariance 
{\footnotesize
\begin{align*}%todo
    \begin{bmatrix}
           \phantom{-}11.5 & -8.5 & \phantom{-}8.5 & -11.5 \\
            -8.5 & \phantom{-}11.5 & -11.5 & \phantom{-}8.5 \\
            \phantom{-}8.5 & -11.5 & \phantom{-}11.5 & -8.5 \\
            -11.5 & \phantom{-}8.5 & -8.5 & \phantom{-}11.5 
    \end{bmatrix}.
\end{align*}}
This results in an additional error variance of $6$ on level 1 and $11.5$ on level 2.

The time series for the 3-level time series are simulated analogously with multivariate normal errors with covariance
{\footnotesize
\begin{align*}
 \setlength{\arraycolsep}{3pt}
    \begin{bmatrix}
        \phantom{-}11.75 & -8.25 & \phantom{-}8.75 & -11.25 & \phantom{-}11.25 & -8.75 & \phantom{-}8.25 & -11.75 \\
        -8.25 & \phantom{-}11.75 & -11.25 & \phantom{-}8.75 & -8.75 & \phantom{-}11.25 & -11.75 & \phantom{-}8.25 \\
        \phantom{-}8.75 & -11.25 & \phantom{-}11.75 & -8.25 & \phantom{-}8.25 & -11.75 & \phantom{-}11.25 & -8.75 \\
        -11.25 & \phantom{-}8.75 & -8.25 & \phantom{-}11.75 & -11.75 & \phantom{-}8.25 & -8.75 & \phantom{-}11.25 \\
        \phantom{-}11.25 & -8.75 & \phantom{-}8.25 & -11.75 & \phantom{-}11.75 & -8.25 & \phantom{-}8.75 & -11.25 \\
        -8.75 & \phantom{-}11.25 & -11.75 & \phantom{-}8.25 & -8.25 & \phantom{-}11.75 & -11.25 & \phantom{-}8.75 \\
        \phantom{-}8.25 & -11.75 & \phantom{-}11.25 & -8.75 & \phantom{-}8.75 & -11.25 & \phantom{-}11.75 & -8.25 \\
        -11.75 & \phantom{-}8.25 & -8.75 & \phantom{-}11.25 & -11.25 & \phantom{-}8.75 & -8.25 & \phantom{-}11.75
    \end{bmatrix} .
\end{align*}}
resulting in added error variances of $4, 7,$ and $11.75$ on levels 1 to 3, respectively. For the time series of the totals, these errors cancel out.  

Similarly to \cref{sec:sim_impactOfCorr}, the results are presented in three tables for the three respective base forecast methods: \cref{tab:sim_smoo_ETS} contains the changes in RMSE for the ETS generated forecasts, \cref{tab:sim_smoo_ARIMA} those for the ARIMA generated forecasts, and \cref{tab:sim_smoo_GPR} those for the GPR generated forecasts. 

\begin{figure}[ht]
%\vskip 0.2in
\begin{center}
\centerline{\includegraphics[width=\columnwidth]{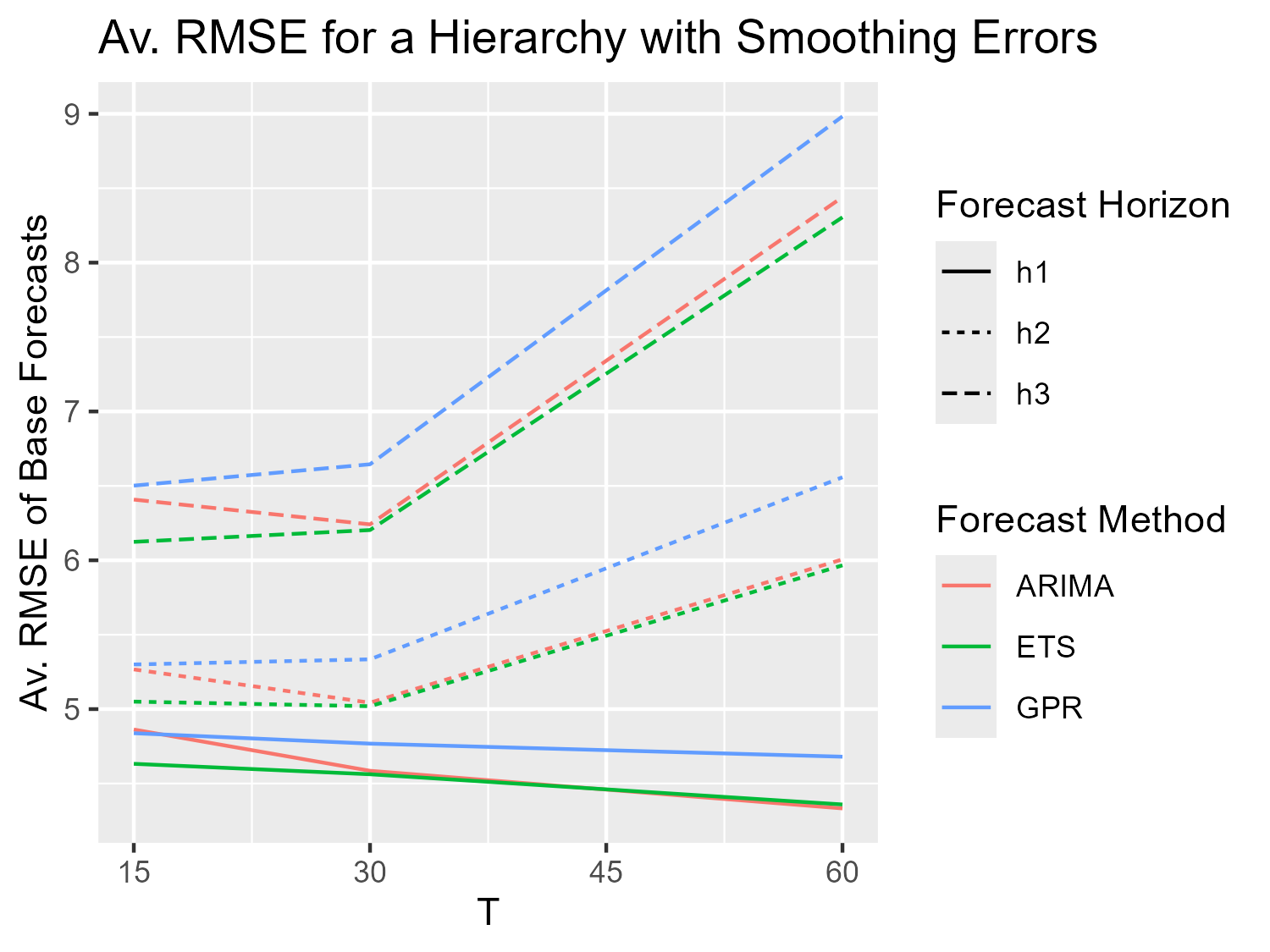}}
\caption{Average RMSE for the various base forecasts in \cref{sec:sim_effOfAggr} (ARIMA in red, ETS in green, and GPR in blue) by total length of the time series. A forecast horizon of $h1$ corresponds to $h = 1$, $h2$ to $h = 1:2$ in the case of $T = 15, 30$ and $h = 1:4$ for $T = 60$. Similarly, $h3$ corresponds to the longer forecasting horizons of $h = 1:4$ and $h = 1:8$, respectively.}
\label{fig:sim_cor_smoo}
\end{center}
\vskip -0.3in
\end{figure}

\textit{Results for Arima and ETS forecasts.} In the case of the ETS and ARIMA forecasts, the MinTit methods and MinT achieved a comparable average decrease in RMSE leading the other methods. MinTit performed better for time series lengths of $T = 15$ and $T = 30$ observations and lower hierarchical levels while MinT showed stronger results for time series of length $T = 60$ and for higher levels of aggregation. The ARIMA and ETS forecasts were comparable in RMSE with ETS having a slight edge, especially for the very short time series ($T = 15$), as shown in \cref{fig:sim_cor_smoo}.

Similarly to \cref{sec:sim_impactOfCorr}, MinT excelled with forecasts generated by GRP. However, these in turn had a much higher base RMSE than ETS and ARIMA -- particularly for $T = 30$ and $T = 60$, as can be seen in \cref{fig:sim_cor_smoo}.

\begin{table}[t]
  \centering
    \footnotesize\tabcolsep2.5pt
  \renewcommand{\arraystretch}{.9}
  \caption{\small Out-of-sample forecast performance for the hierarchy with hierarchically smoothing time series
  in \cref{sec:sim_effOfAggr} for ETS-generated forecasts relative to the ETS-generated base forecasts.}
    % Table generated by Excel2LaTeX from sheet 'paper'
% Table generated by Excel2LaTeX from sheet 'Sheet3'
\begin{tabular}{lrrrlrrrlrrrr|r}
      & \multicolumn{13}{c}{ETS} \bigstrut[b]\\
\cline{2-14}      & \multicolumn{3}{c}{T = 15} &       & \multicolumn{3}{c}{T = 30} &       & \multicolumn{3}{c}{T = 60} & \multicolumn{1}{c}{} &  \bigstrut\\
\cline{2-4}\cline{6-8}\cline{10-12}      & \multicolumn{1}{c}{h=1} & \multicolumn{1}{c}{1:2} & \multicolumn{1}{c}{1:4} &       & \multicolumn{1}{c}{h=1} & \multicolumn{1}{c}{1:2} & \multicolumn{1}{c}{1:4} &       & \multicolumn{1}{c}{h=1} & \multicolumn{1}{c}{1:4} & \multicolumn{1}{c}{1:8} & \multicolumn{1}{c}{} & \multicolumn{1}{c}{Average} \bigstrut\\
\cline{2-12}\cline{14-14}      & \multicolumn{13}{c}{Top-Level} \bigstrut\\
\cline{2-14}MinT  & \textbf{-1.9} & \textbf{-1.2} & \textbf{-1.5} &       & \textbf{-3.1} & \textbf{-3.2} & -4.9  &       & \textbf{-2.3} & \textbf{-4.3} & -5.7  &       & \textbf{-3.1} \bigstrut[t]\\
WLS$_s$ & 2.8   & 2.7   & -0.4  &       & 2.5   & 1.7   & -3.9  &       & 3.6   & -2.6  & \textbf{-6.8} &       & 0.0 \\
WLS$_v$ & 2.8   & 2.7   & -0.4  &       & 2.5   & 1.7   & -3.9  &       & 3.6   & -2.6  & \textbf{-6.8} &       & 0.0 \\
BU    & 39.3  & 32.5  & 18.4  &       & 45.1  & 37.9  & 20.3  &       & 44.6  & 19.3  & 7.1   &       & 29.4 \\
MinTit$_g$ & -1.3  & -0.5  & -1.1  &       & -2.7  & -2.7  & -4.9  &       & -2.1  & \textbf{-4.3} & -6.1  &       & -2.9 \\
MinTit$_l$ & -1.4  & -0.6  & -1.3  &       & -2.8  & -2.8  & \textbf{-5} &       & -2.1  & \textbf{-4.3} & -6.1  &       & -2.9 \\
      & \multicolumn{13}{c}{Level 1} \bigstrut[b]\\
\cline{2-14}MinT  & \textbf{-3.5} & -2.9  & -2.2  &       & -3.7  & -4.1  & -4.4  &       & \textbf{-4.3} & -4.8  & -4.8  &       & -3.9 \bigstrut[t]\\
WLS$_s$ & -1.8  & -1.6  & \textbf{-2.4} &       & -2.1  & -2.5  & -4.8  &       & -2.1  & -4.2  & -5.9  &       & -3.0 \\
WLS$_v$ & -1.8  & -1.6  & \textbf{-2.4} &       & -2.1  & -2.6  & -4.9  &       & -2.1  & -4.3  & \textbf{-6} &       & -3.1 \\
BU    & 20    & 17.6  & 11.4  &       & 22    & 19.6  & 12.3  &       & 20.5  & 11    & 4.9   &       & 15.5 \\
MinTit$_g$ & \textbf{-3.5} & -2.9  & -2.2  &       & \textbf{-4.1} & -4.3  & -5    &       & \textbf{-4.3} & \textbf{-4.9} & -5.2  &       & -4.0 \\
MinTit$_l$ & \textbf{-3.5} & \textbf{-3} & \textbf{-2.4} &       & \textbf{-4.1} & \textbf{-4.4} & \textbf{-5.1} &       & -4.2  & -4.8  & -5.1  &       & \textbf{-4.1} \\
      & \multicolumn{13}{c}{Level 2} \bigstrut[b]\\
\cline{2-14}MinT  & -3.3  & -3.3  & -2.4  &       & -3.8  & -4.1  & -3.7  &       & \textbf{-4.3} & \textbf{-4} & -3.5  &       & -3.6 \bigstrut[t]\\
WLS$_s$ & -3.2  & -3.1  & \textbf{-3.1} &       & -3.8  & -4.2  & -4.7  &       & -3    & -3.6  & -4.3  &       & \textbf{-3.7} \\
WLS$_v$ & -3.2  & -3.1  & \textbf{-3.1} &       & -3.9  & -4.2  & \textbf{-4.8} &       & -3.1  & -3.7  & \textbf{-4.4} &       & \textbf{-3.7} \\
BU    & 6.8   & 6.4   & 4.9   &       & 6.4   & 6.2   & 4.6   &       & 6.1   & 4.4   & 2.4   &       & 5.4 \\
MinTit$_g$ & \textbf{-3.5} & \textbf{-3.4} & -2.7  &       & \textbf{-4.1} & \textbf{-4.4} & -4.3  &       & -3.7  & -3.8  & -3.8  &       & \textbf{-3.7} \\
MinTit$_l$ & \textbf{-3.5} & \textbf{-3.4} & -2.8  &       & \textbf{-4.1} & \textbf{-4.4} & -4.3  &       & -3.6  & -3.8  & -3.8  &       & \textbf{-3.7} \\
      & \multicolumn{13}{c}{Bottom-Level} \bigstrut[b]\\
\cline{2-14}MinT  & -2.4  & -2.4  & -1.9  &       & -2.2  & -2.3  & -1.9  &       & \textbf{-3.5} & \textbf{-3.1} & -2    &       & -2.4 \bigstrut[t]\\
WLS$_s$ & -2.2  & -2.3  & -2.2  &       & -2.3  & -2.5  & -2.7  &       & -2    & -2.2  & -2.3  &       & -2.3 \\
WLS$_v$ & -2.4  & -2.5  & \textbf{-2.4} &       & \textbf{-2.5} & \textbf{-2.7} & \textbf{-2.9} &       & -2.3  & -2.6  & \textbf{-2.8} &       & -2.6 \\
BU    & 0     & 0     & 0     &       & 0     & 0     & 0     &       & 0     & 0     & 0     &       & 0.0 \\
MinTit$_g$ & \textbf{-2.6} & \textbf{-2.6} & -2.3  &       & \textbf{-2.5} & \textbf{-2.7} & -2.8  &       & -2.6  & -2.9  & \textbf{-2.8} &       & -2.6 \\
MinTit$_l$ & \textbf{-2.6} & \textbf{-2.6} & \textbf{-2.4} &       & \textbf{-2.5} & \textbf{-2.7} & -2.8  &       & -2.6  & -2.9  & \textbf{-2.8} &       & \textbf{-2.7} \\
      & \multicolumn{13}{c}{Average} \bigstrut[b]\\
\cline{2-14}MinT  & -2.7  & -2.6  & -2    &       & -2.9  & -3.1  & -3.2  &       & \textbf{-3.7} & \textbf{-3.8} & -3.6  &       & -3.1 \bigstrut[t]\\
WLS$_s$ & -2    & -1.9  & -2.2  &       & -2.3  & -2.5  & -3.7  &       & -1.9  & -3    & -4.4  &       & -2.7 \\
WLS$_v$ & -2.1  & -2    & \textbf{-2.3} &       & -2.4  & -2.7  & \textbf{-3.9} &       & -2    & -3.2  & \textbf{-4.6} &       & -2.8 \\
BU    & 7.8   & 7.7   & 5.8   &       & 8.6   & 8.6   & 6.4   &       & 7.9   & 5.8   & 3     &       & 6.8 \\
MinTit$_g$ & -2.8  & -2.6  & -2.2  &       & \textbf{-3.1} & \textbf{-3.4} & -3.8  &       & -3    & -3.7  & -4.2  &       & \textbf{-3.2} \\
MinTit$_l$ & \textbf{-2.9} & \textbf{-2.7} & \textbf{-2.3} &       & \textbf{-3.1} & \textbf{-3.4} & \textbf{-3.9} &       & -3    & -3.6  & -4.1  &       & \textbf{-3.2} \\
\end{tabular}%

  \label{tab:sim_smoo_ETS}%
\end{table}%

\begin{table}[t]
  \centering
    \footnotesize\tabcolsep2.5pt
  \renewcommand{\arraystretch}{.9}
  \caption{\small Out-of-sample forecast performance for the hierarchy with hierarchically smoothing time series
  in \cref{sec:sim_effOfAggr} for ARIMA-generated forecasts relative to the ARIMA-generated base forecasts.}
% Table generated by Excel2LaTeX from sheet 'Sheet3'
\begin{tabular}{lrrrlrrrlrrrr|r}
      & \multicolumn{13}{c}{ARIMA} \bigstrut[b]\\
\cline{2-14}      & \multicolumn{3}{c}{T = 15} &       & \multicolumn{3}{c}{T = 30} &       & \multicolumn{3}{c}{T = 60} & \multicolumn{1}{c}{} &  \bigstrut\\
\cline{2-4}\cline{6-8}\cline{10-12}      & \multicolumn{1}{c}{h=1} & \multicolumn{1}{c}{1:2} & \multicolumn{1}{c}{1:4} &       & \multicolumn{1}{c}{h=1} & \multicolumn{1}{c}{1:2} & \multicolumn{1}{c}{1:4} &       & \multicolumn{1}{c}{h=1} & \multicolumn{1}{c}{1:4} & \multicolumn{1}{c}{1:8} & \multicolumn{1}{c}{} & \multicolumn{1}{c}{Average} \bigstrut\\
\cline{2-12}\cline{14-14}      & \multicolumn{13}{c}{Top-Level} \bigstrut\\
\cline{2-14}MinT  & \textbf{-7} & \textbf{-6.4} & \textbf{-6.5} &       & \textbf{-5} & \textbf{-5} & \textbf{-5.7} &       & \textbf{-3.7} & \textbf{-4.9} & -5.9  &       & \textbf{-5.6} \bigstrut[t]\\
WLS$_s$ & 0.4   & -0.8  & -5.8  &       & 4.4   & 2.3   & -3.3  &       & 4.4   & -2.6  & \textbf{-6.9} &       & -0.9 \\
WLS$_v$ & 0.4   & -0.8  & -5.8  &       & 4.4   & 2.3   & -3.3  &       & 4.4   & -2.6  & \textbf{-6.9} &       & -0.9 \\
BU    & 51.1  & 41    & 21    &       & 63.2  & 48.4  & 24.3  &       & 55.8  & 21.1  & 5.7   &       & 36.8 \\
MinTit$_g$ & -6    & -5.4  & -6.1  &       & -4.9  & -4.7  & -5.6  &       & -3.3  & -4.7  & -6.1  &       & -5.2 \\
MinTit$_l$ & -6.1  & -5.5  & -6.2  &       & -4.8  & -4.7  & -5.6  &       & -3.3  & -4.7  & -6.1  &       & -5.2 \\
      & \multicolumn{13}{c}{Level 1} \bigstrut[b]\\
\cline{2-14}MinT  & \textbf{-8.2} & \textbf{-7.8} & -7    &       & -5.9  & \textbf{-6} & \textbf{-5.8} &       & -5.3  & -4.9  & -4.8  &       & \textbf{-6.2} \bigstrut[t]\\
WLS$_s$ & -5.4  & -5.7  & -7.3  &       & -2.5  & -2.9  & -4.7  &       & -2.9  & -4.1  & \textbf{-5.8} &       & -4.6 \\
WLS$_v$ & -5.4  & -5.7  & \textbf{-7.4} &       & -2.5  & -2.9  & -4.7  &       & -3    & -4.2  & \textbf{-5.8} &       & -4.6 \\
BU    & 24.8  & 20.7  & 12    &       & 29.8  & 24.9  & 15    &       & 24.7  & 12.8  & 4.5   &       & 18.8 \\
MinTit$_g$ & -8    & -7.6  & -6.9  &       & \textbf{-6.2} & \textbf{-6} & \textbf{-5.8} &       & \textbf{-5.5} & \textbf{-5} & -5    &       & \textbf{-6.2} \\
MinTit$_l$ & -8    & -7.6  & -6.9  &       & \textbf{-6.2} & \textbf{-6} & \textbf{-5.8} &       & -5.4  & \textbf{-5} & -4.9  &       & \textbf{-6.2} \\
      & \multicolumn{13}{c}{Level 2} \bigstrut[b]\\
\cline{2-14}MinT  & \textbf{-6.3} & \textbf{-6.1} & -4.9  &       & \textbf{-5.8} & \textbf{-5.7} & \textbf{-5} &       & \textbf{-5.1} & \textbf{-4.5} & -3.5  &       & \textbf{-5.2} \bigstrut[t]\\
WLS$_s$ & -5.7  & -5.5  & -5.6  &       & -4.5  & -4.3  & -4.4  &       & -3.9  & -3.9  & -4.1  &       & -4.7 \\
WLS$_v$ & -5.7  & -5.5  & \textbf{-5.7} &       & -4.6  & -4.4  & -4.5  &       & -4    & -4.1  & \textbf{-4.2} &       & -4.7 \\
BU    & 9     & 8.2   & 5.9   &       & 9.9   & 9.2   & 6.6   &       & 7.3   & 4.8   & 2.3   &       & 7.0 \\
MinTit$_g$ & \textbf{-6.3} & -6    & -5    &       & \textbf{-5.8} & -5.6  & \textbf{-5} &       & -4.5  & -4.3  & -3.6  &       & -5.1 \\
MinTit$_l$ & \textbf{-6.3} & -6    & -5.1  &       & \textbf{-5.8} & -5.6  & \textbf{-5} &       & -4.5  & -4.2  & -3.6  &       & -5.1 \\
      & \multicolumn{13}{c}{Bottom-Level} \bigstrut[b]\\
\cline{2-14}MinT  & -3.7  & -3.6  & -2.8  &       & \textbf{-4} & -3.9  & -3.4  &       & \textbf{-3.7} & \textbf{-3.3} & -2    &       & -3.4 \bigstrut[t]\\
WLS$_s$ & -3.4  & -3.4  & -3.3  &       & -3.3  & -3.3  & -3.1  &       & -2.5  & -2.5  & -2.2  &       & -3.0 \\
WLS$_v$ & -3.8  & -3.7  & \textbf{-3.6} &       & -3.6  & -3.6  & -3.5  &       & -2.7  & -2.9  & \textbf{-2.6} &       & -3.3 \\
BU    & 0     & 0     & 0     &       & 0     & 0     & 0     &       & 0     & 0     & 0     &       & 0.0 \\
MinTit$_g$ & \textbf{-4} & \textbf{-3.9} & -3.5  &       & -3.9  & \textbf{-4} & \textbf{-3.8} &       & -3    & -3.1  & \textbf{-2.6} &       & \textbf{-3.5} \\
MinTit$_l$ & -3.9  & \textbf{-3.9} & -3.5  &       & -3.8  & \textbf{-4} & \textbf{-3.8} &       & -3    & -3.1  & \textbf{-2.6} &       & \textbf{-3.5} \\
      & \multicolumn{13}{c}{Average} \bigstrut[b]\\
\cline{2-14}MinT  & \textbf{-5.2} & \textbf{-5.1} & -4.5  &       & \textbf{-4.8} & \textbf{-4.7} & -4.5  &       & \textbf{-4.2} & \textbf{-4.1} & -3.7  &       & \textbf{-4.5} \bigstrut[t]\\
WLS$_s$ & -3.9  & -4    & -4.9  &       & -3    & -3    & -3.7  &       & -2.4  & -3.1  & -4.3  &       & -3.6 \\
WLS$_v$ & -4.1  & -4.2  & \textbf{-5.1} &       & -3.1  & -3.2  & -3.9  &       & -2.6  & -3.3  & \textbf{-4.5} &       & -3.8 \\
BU    & 10.5  & 9.7   & 6.7   &       & 11.8  & 11    & 7.9   &       & 9.4   & 6.5   & 2.6   &       & 8.5 \\
MinTit$_g$ & \textbf{-5.2} & \textbf{-5.1} & -4.8  &       & -4.7  & \textbf{-4.7} & \textbf{-4.7} &       & -3.7  & -3.9  & -4    &       & \textbf{-4.5} \\
MinTit$_l$ & \textbf{-5.2} & \textbf{-5.1} & -4.8  &       & -4.7  & \textbf{-4.7} & \textbf{-4.7} &       & -3.7  & -3.9  & -4    &       & \textbf{-4.5} \\
\end{tabular}%

  \label{tab:sim_smoo_ARIMA}%
\end{table}%

\begin{table}[t]
  \centering
    \footnotesize\tabcolsep2.5pt
  \renewcommand{\arraystretch}{.9}
  \caption{\small Out-of-sample forecast performance for the hierarchy with hierarchically smoothing time series
  in \cref{sec:sim_effOfAggr} for GPR-generated forecasts relative to the GPR-generated base forecasts.}
% Table generated by Excel2LaTeX from sheet 'Sheet3'
\begin{tabular}{lrrrlrrrlrrrr|r}
      & \multicolumn{13}{c}{GPR} \bigstrut[b]\\
\cline{2-14}      & \multicolumn{3}{c}{T = 15} &       & \multicolumn{3}{c}{T = 30} &       & \multicolumn{3}{c}{T = 60} & \multicolumn{1}{c}{} &  \bigstrut\\
\cline{2-4}\cline{6-8}\cline{10-12}      & \multicolumn{1}{c}{h=1} & \multicolumn{1}{c}{1:2} & \multicolumn{1}{c}{1:4} &       & \multicolumn{1}{c}{h=1} & \multicolumn{1}{c}{1:2} & \multicolumn{1}{c}{1:4} &       & \multicolumn{1}{c}{h=1} & \multicolumn{1}{c}{1:4} & \multicolumn{1}{c}{1:8} & \multicolumn{1}{c}{} & \multicolumn{1}{c}{Average} \bigstrut\\
\cline{2-12}\cline{14-14}      & \multicolumn{13}{c}{Top-Level} \bigstrut\\
\cline{2-14}MinT  & \textbf{-3.5} & \textbf{-3.6} & \textbf{-3.9} &       & \textbf{-4.2} & \textbf{-3.1} & \textbf{-3.5} &       & \textbf{-3.8} & \textbf{-1.6} & \textbf{-1.9} &       & \textbf{-3.2} \bigstrut[t]\\
WLS$_s$ & 6.4   & 6.2   & 2     &       & 7.6   & 9.8   & 5.5   &       & 6.6   & 7.9   & 3.7   &       & 6.2 \\
WLS$_v$ & 6.4   & 6.2   & 2     &       & 7.6   & 9.8   & 5.5   &       & 6.6   & 7.9   & 3.7   &       & 6.2 \\
BU    & 48.4  & 39.4  & 25.7  &       & 56.3  & 52.7  & 36.6  &       & 56.7  & 42.1  & 26.6  &       & 42.7 \\
MinTit$_g$ & -1.8  & -1.8  & -2.9  &       & -2.5  & -1.4  & -2.4  &       & -2.1  & -0.4  & -1.4  &       & -1.9 \\
MinTit$_l$ & -1.9  & -1.9  & -3    &       & -2.6  & -1.6  & -2.6  &       & -2.2  & -0.6  & -1.5  &       & -2.0 \\
      & \multicolumn{13}{c}{Level 1} \bigstrut[b]\\
\cline{2-14}MinT  & \textbf{-5.3} & \textbf{-6.1} & \textbf{-6.2} &       & \textbf{-5.2} & \textbf{-4.9} & \textbf{-4.5} &       & \textbf{-5.5} & \textbf{-3.8} & \textbf{-3.1} &       & \textbf{-5.0} \bigstrut[t]\\
WLS$_s$ & -0.4  & -0.5  & -2.1  &       & 0.7   & 2.7   & 2.1   &       & 0.2   & 3.2   & 1.9   &       & 0.9 \\
WLS$_v$ & -0.4  & -0.5  & -2.2  &       & 0.7   & 2.8   & 2.1   &       & 0.2   & 3.2   & 1.9   &       & 0.9 \\
BU    & 25.8  & 21.4  & 15.3  &       & 29.1  & 29.9  & 24.3  &       & 28.9  & 27.4  & 19.8  &       & 24.7 \\
MinTit$_g$ & -4.9  & -5.3  & -5.6  &       & -4.4  & -3.8  & -3.5  &       & -4.1  & -2.7  & -2.4  &       & -4.1 \\
MinTit$_l$ & -4.9  & -5.3  & -5.6  &       & -4.4  & -3.8  & -3.6  &       & -4.1  & -2.8  & -2.5  &       & -4.1 \\
      & \multicolumn{13}{c}{Level 2} \bigstrut[b]\\
\cline{2-14}MinT  & \textbf{-4.8} & \textbf{-5.6} & \textbf{-5.4} &       & \textbf{-5.2} & \textbf{-6.2} & \textbf{-6.1} &       & \textbf{-6.8} & \textbf{-6.4} & \textbf{-5.4} &       & \textbf{-5.8} \bigstrut[t]\\
WLS$_s$ & -3    & -2.7  & -2.7  &       & -2.4  & -1.9  & -1.6  &       & -2.2  & -1    & -0.9  &       & -2.0 \\
WLS$_v$ & -2.9  & -2.8  & -2.9  &       & -2.4  & -2    & -1.7  &       & -2.3  & -1.1  & -1    &       & -2.1 \\
BU    & 10.1  & 8.4   & 7.3   &       & 10.6  & 11.5  & 10.7  &       & 10    & 12.2  & 10.1  &       & 10.1 \\
MinTit$_g$ & -4.6  & -5    & -4.9  &       & -4.3  & -5    & -5    &       & -4    & -4.4  & -4    &       & -4.6 \\
MinTit$_l$ & -4.6  & -5    & -4.9  &       & -4.3  & -5    & -5    &       & -3.9  & -4.4  & -4    &       & -4.6 \\
      & \multicolumn{13}{c}{Bottom-Level} \bigstrut[b]\\
\cline{2-14}MinT  & -3.1  & -3.5  & -3.8  &       & \textbf{-5.1} & \textbf{-6.1} & \textbf{-6.9} &       & \textbf{-7.4} & \textbf{-8.4} & \textbf{-7.9} &       & \textbf{-5.8} \bigstrut[t]\\
WLS$_s$ & -3.1  & -2.8  & -2.9  &       & -3.1  & -3.4  & -3.6  &       & -2.8  & -3.9  & -3.8  &       & -3.3 \\
WLS$_v$ & -3.2  & -3.1  & -3.4  &       & -3.4  & -3.9  & -4.3  &       & -3.3  & -4.5  & -4.5  &       & -3.7 \\
BU    & 0     & 0     & 0     &       & 0     & 0     & 0     &       & 0     & 0     & 0     &       & 0.0 \\
MinTit$_g$ & \textbf{-3.5} & \textbf{-3.8} & \textbf{-4.1} &       & -3.9  & -5    & -5.8  &       & -4.1  & -6.1  & -6.4  &       & -4.7 \\
MinTit$_l$ & \textbf{-3.5} & \textbf{-3.8} & \textbf{-4.1} &       & -3.9  & -5    & -5.8  &       & -4.1  & -6.1  & -6.4  &       & -4.7 \\
      & \multicolumn{13}{c}{Average} \bigstrut[b]\\
\cline{2-14}MinT  & \textbf{-3.9} & \textbf{-4.4} & \textbf{-4.7} &       & \textbf{-5} & \textbf{-5.7} & \textbf{-5.8} &       & \textbf{-6.8} & \textbf{-6.2} & \textbf{-5.3} &       & \textbf{-5.3} \bigstrut[t]\\
WLS$_s$ & -1.9  & -1.5  & -2    &       & -1.5  & -0.8  & -0.9  &       & -1.6  & -0.4  & -0.7  &       & -1.3 \\
WLS$_v$ & -2    & -1.7  & -2.3  &       & -1.7  & -1.1  & -1.2  &       & -1.9  & -0.7  & -0.9  &       & -1.5 \\
BU    & 10.6  & 9.8   & 8.3   &       & 12.1  & 13.5  & 12.5  &       & 11.4  & 13.9  & 11.4  &       & 11.5 \\
MinTit$_g$ & -3.8  & -4.1  & -4.4  &       & -4    & -4.5  & -4.7  &       & -3.9  & -4.4  & -4.1  &       & -4.2 \\
MinTit$_l$ & -3.8  & -4.1  & -4.4  &       & -4    & -4.5  & -4.8  &       & -3.9  & -4.4  & -4.2  &       & -4.2 \\
\end{tabular}%

  \label{tab:sim_smoo_GPR}%
\end{table}%

\subsubsection{Impact of Seasonality}
\label{sec:sim_season}

This subsection studies the effect seasonality has on the choice of the reconciliation method. For comparability, the bottom time series $\mathbf{b}_t$ are simulated analogous to the structural time series model consisting of a trend component $\mathbf{\mu}_t$, a quarterly seasonal component $\mathbf{\gamma}_t$, and an error component $\mathbf{\eta}_t$ as used in \citet{Wickramasuriya2019}:
\begin{align*}
    \mathbf{b}_t &= \mathbf{\mu}_t+\mathbf{\gamma}_t+\mathbf{\eta}_t,\\
    \mathbf{\mu}_t &= \mathbf{\mu}_{t-1}+\mathbf{\nu}_t+\mathbf{\varrho}_t, & \varrho_t \sim MN(\mathbf{0}, 2\mathbf{I_8})\\
    \mathbf{\nu}_t &= \mathbf{\nu}_{t-1} + \mathbf{\zeta}_t, & \mathbf{\zeta}_t \sim MN(\mathbf{0}, 0.007 \mathbf{I}_8)\\
    \mathbf{\gamma}_t &= -\sum_{i=1}^3\mathbf{\gamma}_{t-i}+\mathbf{\omega}_t, & \mathbf{\omega}_t \sim MN(\mathbf{0},7\mathbf{I}_8),
\end{align*}
with independently distributed initial values $\mathbf{\mu}_0$, $\mathbf{\nu}_0$, $\mathbf{\gamma}_0$, $\mathbf{\gamma}_1$, and $\mathbf{\gamma}_2 \sim MN(\mathbf{0}, \mathbf{I}_8)$, where $MN$ denotes the multivariate normal distribution. $\mathbf{\eta}_t$ is simulated as an ARIMA process similar to \cref{sec:sim_impactOfCorr} and as detailed in \cref{tab:ARIMA_coef} but with $d$ held at zero and $p$ and $q$ only ranging from zero to one. The resulting time series are summed according to the hierarchical structure as illustrated in \cref{fig:sim_corr_hier}.

\begin{figure}[ht]
%\vskip 0.2in
\begin{center}
\centerline{\includegraphics[width=\columnwidth]{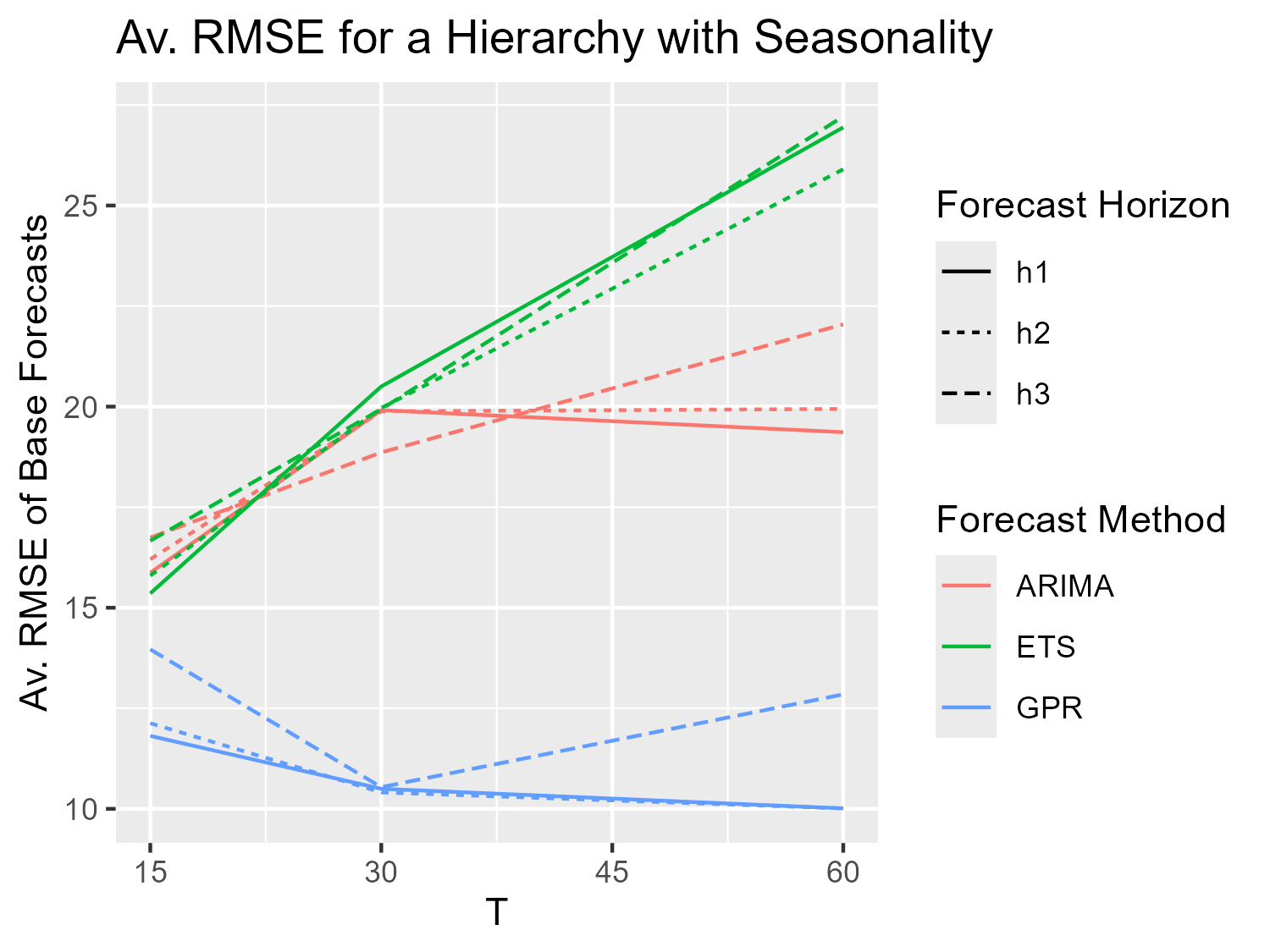}}
\caption{Average RMSE for the various base forecasts in \cref{sec:sim_season} (ARIMA in red, ETS in green, and GPR in blue) by total length of the time series. A forecast horizon of $h1$ corresponds to $h = 1$, $h2$ to $h = 1:2$ in the case of $T = 15, 30$ and $h = 1:4$ for $T = 60$. Similarly, $h3$ corresponds to the longer forecasting horizons of $h = 1:4$ and $h = 1:8$, respectively.}
\label{fig:sim_seas}
\end{center}
\vskip -0.3in
\end{figure}

\textit{Results for ETS forecasts.} The performance of the base forecasts can be obtained from \cref{fig:sim_seas}. In contrast to \cref{sec:sim_impactOfCorr} and \cref{sec:sim_effOfAggr}, it is noticeable that ETS performed poorly in this scenario. This is likely due to an inability of the model to capture and identify both the seasonal and trend components with few observations. Especially a failure to capture trend components could explain the worsening performance as the number of observations increases.

In terms of improvements through reconciliation, MinT achieves superior performance (average 2.6\% improvements for MinT vs. a 2.2\% improvement for MinTiT) followed by the WLS approaches (1.9\% and 1.8\% improvements). BU fares worst with only a marginal improvement of 0.1\%.
%However, \citet{Wickramasuriya2019} note that MinT may be able to offset severe model mis-specification. 

The \textit{results for the ARIMA forecasts} are similar. ARIMA also struggled with the identification of an appropriate model (\cref{fig:sim_seas}). However, the RMSE for the ARIMA model stabilizes at $T = 60$ observations. With regard to the reconciliation performance, for which \cref{tab:sim_seas_ARIMA} provides an overview, the results mirror those of the ETS forecasts but with much larger improvements (except in the case of BU which leads to an increase in RMSE).

\textit{Results for GPR forecasts.} The GPR forecasts fared significantly better than its ETS and ARIMA counterparts in this scenario and showcase a steady improvement in accuracy as the time series lengths increase (\cref{fig:sim_seas}).
WLS reconciliation achieved by far the best results with an RMSE improvement of 6\% (MinTit$_g$: 4.2\%, MinTit$_l$, MinT: 3.2\%).

\begin{table}[t]
  \centering
    \footnotesize\tabcolsep2.5pt
  \renewcommand{\arraystretch}{.9}
  \caption{\small Out-of-sample forecast performance for the hierarchy with seasonal time series
  in \cref{sec:sim_season} for ETS-generated forecasts relative to the ETS-generated base forecasts.}
% Table generated by Excel2LaTeX from sheet 'Sheet3'
\begin{tabular}{lrrrlrrrlrrrr|r}
      & \multicolumn{13}{c}{ETS} \bigstrut[b]\\
\cline{2-14}      & \multicolumn{3}{c}{T = 15} &       & \multicolumn{3}{c}{T = 30} &       & \multicolumn{3}{c}{T = 60} & \multicolumn{1}{c}{} &  \bigstrut\\
\cline{2-4}\cline{6-8}\cline{10-12}      & \multicolumn{1}{c}{h=1} & \multicolumn{1}{c}{1:2} & \multicolumn{1}{c}{1:4} &       & \multicolumn{1}{c}{h=1} & \multicolumn{1}{c}{1:2} & \multicolumn{1}{c}{1:4} &       & \multicolumn{1}{c}{h=1} & \multicolumn{1}{c}{1:4} & \multicolumn{1}{c}{1:8} & \multicolumn{1}{c}{} & \multicolumn{1}{c}{Average} \bigstrut\\
\cline{2-12}\cline{14-14}      & \multicolumn{13}{c}{Top-Level} \bigstrut\\
\cline{2-14}MinT  & \textbf{-4.1} & \textbf{-3.7} & \textbf{-4.5} &       & \textbf{-2.4} & \textbf{-2.9} & \textbf{-3.8} &       & -1.3  & \textbf{-2.4} & \textbf{-3.1} &       & \textbf{-3.1} \bigstrut[t]\\
WLS$_s$ & -3    & -2.6  & -2.6  &       & -2    & -2.5  & -3.1  &       & \textbf{-1.5} & -2.1  & -2.9  &       & -2.5 \\
WLS$_v$ & -3    & -2.6  & -2.6  &       & -2    & -2.5  & -3.1  &       & \textbf{-1.5} & -2.1  & -2.9  &       & -2.5 \\
BU    & -0.8  & -0.3  & -0.2  &       & 0     & -0.2  & -0.4  &       & -0.3  & 0     & 0     &       & -0.2 \\
MinTit$_g$ & -3.5  & -3.1  & -3.4  &       & -2.3  & -2.8  & -3.5  &       & \textbf{-1.5} & -2.2  & -3    &       & -2.8 \\
MinTit$_l$ & -3.4  & -3    & -3.3  &       & -2.3  & -2.7  & -3.4  &       & \textbf{-1.5} & -2.2  & -3    &       & -2.8 \\
      & \multicolumn{13}{c}{Level 1} \bigstrut[b]\\
\cline{2-14}MinT  & \textbf{-3.5} & \textbf{-3.6} & \textbf{-4.4} &       & -2.1  & \textbf{-2.5} & \textbf{-3.4} &       & -1.3  & \textbf{-2.2} & \textbf{-3} &       & \textbf{-2.9} \bigstrut[t]\\
WLS$_s$ & -2.3  & -2.5  & -2.5  &       & -1.7  & -2.1  & -2.6  &       & \textbf{-1.5} & -1.9  & -2.7  &       & -2.2 \\
WLS$_v$ & -2.3  & -2.5  & -2.5  &       & -1.7  & -2.1  & -2.6  &       & -1.4  & -1.9  & -2.7  &       & -2.2 \\
BU    & -0.2  & -0.3  & -0.2  &       & 0.1   & 0.2   & 0.1   &       & -0.2  & 0     & 0     &       & -0.1 \\
MinTit$_g$ & -2.8  & -3    & -3.3  &       & -2    & -2.4  & -3.1  &       & \textbf{-1.5} & -2.1  & -2.9  &       & -2.6 \\
MinTit$_l$ & -2.8  & -3    & -3.2  &       & \textbf{-2.2} & \textbf{-2.5} & -3.2  &       & \textbf{-1.5} & -2.1  & -2.9  &       & -2.6 \\
      & \multicolumn{13}{c}{Level 2} \bigstrut[b]\\
\cline{2-14}MinT  & \textbf{-3.2} & \textbf{-3.2} & \textbf{-3.8} &       & \textbf{-2} & \textbf{-2.2} & \textbf{-3.2} &       & -0.8  & \textbf{-2} & \textbf{-2.6} &       & \textbf{-2.6} \bigstrut[t]\\
WLS$_s$ & -2.1  & -2    & -1.9  &       & -1.5  & -1.6  & -2.2  &       & -0.9  & -1.6  & -2.2  &       & -1.8 \\
WLS$_v$ & -2.1  & -2    & -2    &       & -1.5  & -1.6  & -2.3  &       & -0.9  & -1.6  & -2.2  &       & -1.8 \\
BU    & -0.2  & -0.1  & 0     &       & 0     & 0.4   & 0.1   &       & 0.2   & 0     & 0     &       & 0.0 \\
MinTit$_g$ & -2.4  & -2.4  & -2.6  &       & -1.8  & -2    & -2.7  &       & \textbf{-1} & -1.8  & -2.4  &       & -2.1 \\
MinTit$_l$ & -2.4  & -2.4  & -2.5  &       & \textbf{-2} & -2.1  & -2.8  &       & \textbf{-1} & -1.8  & -2.4  &       & -2.2 \\
      & \multicolumn{13}{c}{Bottom-Level} \bigstrut[b]\\
\cline{2-14}MinT  & \textbf{-2.3} & \textbf{-2.3} & \textbf{-3} &       & \textbf{-1.6} & \textbf{-2} & \textbf{-2.6} &       & -0.6  & \textbf{-1.5} & \textbf{-1.8} &       & \textbf{-2.0} \bigstrut[t]\\
WLS$_s$ & -0.9  & -0.9  & -1    &       & -0.7  & -1    & -1.1  &       & -0.6  & -0.8  & -1.1  &       & -0.9 \\
WLS$_v$ & -0.9  & -0.9  & -1    &       & -0.9  & -1.1  & -1.4  &       & -0.6  & -0.9  & -1.2  &       & -1.0 \\
BU    & 0     & 0     & 0     &       & 0     & 0     & 0     &       & 0     & 0     & 0     &       & 0.0 \\
MinTit$_g$ & -1.1  & -1.2  & -1.4  &       & -1.2  & -1.5  & -1.8  &       & \textbf{-0.7} & -1.1  & -1.4  &       & -1.3 \\
MinTit$_l$ & -1.1  & -1.2  & -1.4  &       & -1.3  & -1.6  & -1.9  &       & -0.6  & -1.1  & -1.4  &       & -1.3 \\
      & \multicolumn{13}{c}{Average} \bigstrut[b]\\
\cline{2-14}MinT  & \textbf{-3.3} & \textbf{-3.2} & \textbf{-3.9} &       & \textbf{-2} & \textbf{-2.4} & \textbf{-3.2} &       & -1    & \textbf{-2} & \textbf{-2.6} &       & \textbf{-2.6} \bigstrut[t]\\
WLS$_s$ & -2.1  & -2    & -2    &       & -1.5  & -1.8  & -2.3  &       & -1.1  & -1.6  & -2.2  &       & -1.8 \\
WLS$_v$ & -2.1  & -2    & -2    &       & -1.5  & -1.8  & -2.4  &       & -1.1  & -1.6  & -2.3  &       & -1.9 \\
BU    & -0.3  & -0.2  & -0.1  &       & 0     & 0.1   & 0     &       & -0.1  & 0     & 0     &       & -0.1 \\
MinTit$_g$ & -2.4  & -2.4  & -2.7  &       & -1.8  & -2.2  & -2.8  &       & \textbf{-1.2} & -1.8  & -2.4  &       & -2.2 \\
MinTit$_l$ & -2.4  & -2.4  & -2.6  &       & -1.9  & -2.2  & -2.8  &       & \textbf{-1.2} & -1.8  & -2.4  &       & -2.2 \\
\end{tabular}%
  \label{tab:sim_seas_ETS}%
\end{table}%

\begin{table}[t]
  \centering
    \tabcolsep1.75pt\footnotesize
  \renewcommand{\arraystretch}{.9}
  \caption{\small Out-of-sample forecast performance for the hierarchy with seasonal time series
  in \cref{sec:sim_season} for ARIMA-generated forecasts relative to the ARIMA-generated base forecasts.}
% Table generated by Excel2LaTeX from sheet 'Sheet3'
\begin{tabular}{lrrrlrrrlrrrr|r}
      & \multicolumn{13}{c}{ARIMA} \bigstrut[b]\\
\cline{2-14}      & \multicolumn{3}{c}{T = 15} &       & \multicolumn{3}{c}{T = 30} &       & \multicolumn{3}{c}{T = 60} & \multicolumn{1}{c}{} &  \bigstrut\\
\cline{2-4}\cline{6-8}\cline{10-12}      & \multicolumn{1}{c}{h=1} & \multicolumn{1}{c}{1:2} & \multicolumn{1}{c}{1:4} &       & \multicolumn{1}{c}{h=1} & \multicolumn{1}{c}{1:2} & \multicolumn{1}{c}{1:4} &       & \multicolumn{1}{c}{h=1} & \multicolumn{1}{c}{1:4} & \multicolumn{1}{c}{1:8} & \multicolumn{1}{c}{} & \multicolumn{1}{c}{Average} \bigstrut\\
\cline{2-12}\cline{14-14}      & \multicolumn{13}{c}{Top-Level} \bigstrut\\
\cline{2-14}MinT  & \textbf{-12.7} & \textbf{-10.3} & \textbf{-10.9} &       & \textbf{-30.3} & \textbf{-27.4} & \textbf{-23.6} &       & \textbf{-36.8} & \textbf{-32.4} & \textbf{-26.2} &       & \textbf{-23.4} \bigstrut[t]\\
WLS$_s$ & -8.1  & -6.7  & -6.6  &       & -12   & -10.8 & -10.5 &       & -16.4 & -12.6 & -9.6  &       & -10.4 \\
WLS$_v$ & -8.1  & -6.7  & -6.6  &       & -12   & -10.8 & -10.5 &       & -16.4 & -12.6 & -9.6  &       & -10.4 \\
BU    & -0.2  & 0.1   & -0.1  &       & 1.8   & 0.9   & 0.9   &       & 0.3   & -0.1  & -0.1  &       & 0.4 \\
MinTit$_g$ & -10.2 & -8.6  & -8.9  &       & -21.2 & -20.4 & -19.3 &       & -28.4 & -25.3 & -20.1 &       & -18.0 \\
MinTit$_l$ & -10.4 & -8.7  & -9    &       & -21.5 & -20.7 & -19.2 &       & -28.8 & -25.7 & -20.4 &       & -18.3 \\
      & \multicolumn{13}{c}{Level 1} \bigstrut[b]\\
\cline{2-14}MinT  & \textbf{-11.7} & \textbf{-9.9} & \textbf{-10.2} &       & \textbf{-30} & \textbf{-26.2} & \textbf{-22.8} &       & \textbf{-35.1} & \textbf{-31.3} & \textbf{-25.4} &       & \textbf{-22.5} \bigstrut[t]\\
WLS$_s$ & -6.9  & -6.4  & -6    &       & -12   & -10   & -10.1 &       & -14.2 & -11.7 & -9    &       & -9.6 \\
WLS$_v$ & -7    & -6.4  & -6    &       & -12.1 & -10.1 & -10.2 &       & -14.4 & -11.8 & -9.1  &       & -9.7 \\
BU    & 0.4   & -0.2  & 0     &       & 0.5   & 0.6   & 0.1   &       & 1.3   & -0.3  & -0.2  &       & 0.2 \\
MinTit$_g$ & -9.2  & -8.2  & -8.3  &       & -21.6 & -19.8 & -19.2 &       & -27   & -25.1 & -20.3 &       & -17.6 \\
MinTit$_l$ & -9.2  & -8.2  & -8.3  &       & -22.2 & -20.4 & -19.3 &       & -28   & -25.7 & -20.9 &       & -18.0 \\
      & \multicolumn{13}{c}{Level 2} \bigstrut[b]\\
\cline{2-14}MinT  & \textbf{-10.8} & \textbf{-8.7} & \textbf{-9} &       & \textbf{-27.8} & \textbf{-24} & \textbf{-20.1} &       & \textbf{-33.2} & \textbf{-28.9} & \textbf{-23.5} &       & \textbf{-20.7} \bigstrut[t]\\
WLS$_s$ & -6.4  & -5.4  & -5.1  &       & -10   & -8.5  & -8.2  &       & -11.9 & -9.8  & -7.6  &       & -8.1 \\
WLS$_v$ & -6.5  & -5.5  & -5.2  &       & -10.5 & -8.9  & -8.7  &       & -12.8 & -10.6 & -8.3  &       & -8.6 \\
BU    & -0.5  & -0.3  & -0.1  &       & 0.3   & 0.1   & 0.2   &       & 0.3   & -0.4  & -0.3  &       & -0.1 \\
MinTit$_g$ & -8    & -6.8  & -6.9  &       & -19   & -17.4 & -16.3 &       & -24.7 & -22.8 & -18.6 &       & -15.6 \\
MinTit$_l$ & -8.1  & -6.9  & -6.9  &       & -19.1 & -17.3 & -16.1 &       & -25.7 & -23.3 & -19.1 &       & -15.8 \\
      & \multicolumn{13}{c}{Bottom-Level} \bigstrut[b]\\
\cline{2-14}MinT  & \textbf{-7.8} & \textbf{-6.1} & \textbf{-6.4} &       & \textbf{-23.4} & \textbf{-19.3} & \textbf{-15.7} &       & \textbf{-28.2} & \textbf{-23.4} & \textbf{-19} &       & \textbf{-16.6} \bigstrut[t]\\
WLS$_s$ & -2.9  & -2.5  & -2.5  &       & -5    & -4.2  & -4.1  &       & -5.9  & -4.6  & -3.6  &       & -3.9 \\
WLS$_v$ & -3.3  & -2.9  & -2.8  &       & -7.5  & -6.3  & -6.3  &       & -10   & -8    & -6.4  &       & -5.9 \\
BU    & 0     & 0     & 0     &       & 0     & 0     & 0     &       & 0     & 0     & 0     &       & 0.0 \\
MinTit$_g$ & -4.4  & -3.7  & -3.8  &       & -13.5 & -12   & -11.2 &       & -18.2 & -16.3 & -13.4 &       & -10.7 \\
MinTit$_l$ & -4.4  & -3.7  & -3.8  &       & -13.4 & -11.7 & -10.8 &       & -18.2 & -16.1 & -13.2 &       & -10.6 \\
      & \multicolumn{13}{c}{Average} \bigstrut[b]\\
\cline{2-14}MinT  & \textbf{-10.7} & \textbf{-8.7} & \textbf{-9.1} &       & \textbf{-27.8} & \textbf{-24.2} & \textbf{-20.5} &       & \textbf{-33.2} & \textbf{-28.9} & \textbf{-23.5} &       & \textbf{-20.7} \bigstrut[t]\\
WLS$_s$ & -6.1  & -5.2  & -5    &       & -9.7  & -8.3  & -8.2  &       & -12   & -9.6  & -7.4  &       & -7.9 \\
WLS$_v$ & -6.2  & -5.3  & -5.1  &       & -10.5 & -9    & -8.9  &       & -13.3 & -10.7 & -8.3  &       & -8.6 \\
BU    & -0.1  & -0.1  & -0.1  &       & 0.7   & 0.4   & 0.3   &       & 0.5   & -0.2  & -0.2  &       & 0.1 \\
MinTit$_g$ & -7.9  & -6.8  & -7    &       & -18.8 & -17.3 & -16.5 &       & -24.4 & -22.3 & -18.1 &       & -15.5 \\
MinTit$_l$ & -8    & -6.9  & -7    &       & -19   & -17.5 & -16.3 &       & -25   & -22.6 & -18.3 &       & -15.6 
\end{tabular}%
  \label{tab:sim_seas_ARIMA}%
\end{table}%

\begin{table}[t]
  \centering
    \footnotesize\tabcolsep2.5pt
  \renewcommand{\arraystretch}{.9}
  \caption{\small Out-of-sample forecast performance for the hierarchy with seasonal time series
  in \cref{sec:sim_season} for GPR-generated forecasts relative to the GPR-generated base forecasts.}
% Table generated by Excel2LaTeX from sheet 'Sheet3'
\begin{tabular}{lrrrlrrrlrrrr|r}
      & \multicolumn{13}{c}{GPR} \bigstrut[b]\\
\cline{2-14}      & \multicolumn{3}{c}{T = 15} &       & \multicolumn{3}{c}{T = 30} &       & \multicolumn{3}{c}{T = 60} & \multicolumn{1}{c}{} &  \bigstrut\\
\cline{2-4}\cline{6-8}\cline{10-12}      & \multicolumn{1}{c}{h=1} & \multicolumn{1}{c}{1:2} & \multicolumn{1}{c}{1:4} &       & \multicolumn{1}{c}{h=1} & \multicolumn{1}{c}{1:2} & \multicolumn{1}{c}{1:4} &       & \multicolumn{1}{c}{h=1} & \multicolumn{1}{c}{1:4} & \multicolumn{1}{c}{1:8} & \multicolumn{1}{c}{} & \multicolumn{1}{c}{Average} \bigstrut\\
\cline{2-12}\cline{14-14}      & \multicolumn{13}{c}{Top-Level} \bigstrut\\
\cline{2-14}MinT  & -6.3  & -5    & -4.3  &       & -6.2  & -6    & -5.8  &       & \textbf{-5} & \textbf{-4.2} & \textbf{-5} &       & -5.3 \bigstrut[t]\\
WLS$_s$ & \textbf{-12.8} & \textbf{-13.3} & \textbf{-18} &       & \textbf{-6.5} & \textbf{-6.4} & \textbf{-6.4} &       & -4.5  & -3.8  & -4.7  &       & \textbf{-8.5} \\
WLS$_v$ & \textbf{-12.8} & \textbf{-13.3} & \textbf{-18} &       & \textbf{-6.5} & \textbf{-6.4} & \textbf{-6.4} &       & -4.5  & -3.8  & -4.7  &       & \textbf{-8.5} \\
BU    & -1.6  & -0.3  & 0.9   &       & 0.4   & -0.3  & -0.4  &       & -0.3  & 0     & 0     &       & -0.2 \\
MinTit$_g$ & -9.7  & -9    & -10.7 &       & -6.2  & -6.1  & -6    &       & -4.5  & -3.8  & -4.7  &       & -6.7 \\
MinTit$_l$ & -7.8  & -6.7  & -8.2  &       & -5.7  & -5.6  & -5.5  &       & -4.4  & -3.8  & -4.6  &       & -5.8 \\
      & \multicolumn{13}{c}{Level 1} \bigstrut[b]\\
\cline{2-14}MinT  & -3.5  & -3.2  & -0.5  &       & -5.7  & -5.4  & -4.9  &       & \textbf{-4.3} & \textbf{-3.7} & \textbf{-4.4} &       & -4.0 \bigstrut[t]\\
WLS$_s$ & \textbf{-9.8} & \textbf{-11.3} & \textbf{-14} &       & \textbf{-6.1} & \textbf{-5.8} & \textbf{-5.5} &       & -3.7  & -3.3  & -4.2  &       & \textbf{-7.1} \\
WLS$_v$ & \textbf{-9.8} & \textbf{-11.3} & \textbf{-14} &       & \textbf{-6.1} & \textbf{-5.8} & -5.4  &       & -3.7  & -3.3  & -4.2  &       & \textbf{-7.1} \\
BU    & 1.1   & 0.9   & 4.2   &       & 0.1   & -0.3  & 0     &       & 0.6   & 0.3   & 0.3   &       & 0.8 \\
MinTit$_g$ & -6.2  & -6.6  & -6.4  &       & -5.7  & -5.4  & -5    &       & -3.7  & -3.4  & -4.2  &       & -5.2 \\
MinTit$_l$ & -4.1  & -4.3  & -3.4  &       & -5.2  & -4.9  & -4.5  &       & -3.6  & -3.3  & -4    &       & -4.1 \\
      & \multicolumn{13}{c}{Level 2} \bigstrut[b]\\
\cline{2-14}MinT  & -1.8  & -1.6  & 0.8   &       & -3.8  & -3.9  & -3.9  &       & \textbf{-3.7} & \textbf{-3.4} & \textbf{-3.9} &       & -2.8 \bigstrut[t]\\
WLS$_s$ & \textbf{-8} & \textbf{-9.4} & \textbf{-11.7} &       & \textbf{-4.2} & \textbf{-4.2} & \textbf{-4.3} &       & -3.1  & -2.8  & -3.5  &       & \textbf{-5.7} \\
WLS$_v$ & -7.9  & -9.3  & \textbf{-11.7} &       & \textbf{-4.2} & \textbf{-4.2} & \textbf{-4.3} &       & -3.1  & -2.8  & -3.5  &       & \textbf{-5.7} \\
BU    & 0.3   & 0.4   & 3.4   &       & 1     & 0.5   & 0.2   &       & 0.2   & 0     & 0     &       & 0.7 \\
MinTit$_g$ & -4    & -4.2  & -3.5  &       & -3.7  & -3.8  & -3.9  &       & -3.1  & -2.9  & -3.5  &       & -3.6 \\
MinTit$_l$ & -1.6  & -1.5  & 0     &       & -3.1  & -3.2  & -3.4  &       & -3.1  & -2.8  & -3.4  &       & -2.5 \\
      & \multicolumn{13}{c}{Bottom-Level} \bigstrut[b]\\
\cline{2-14}MinT  & 1.8   & 1.9   & 2.7   &       & \textbf{-2.7} & \textbf{-2.5} & \textbf{-2.3} &       & \textbf{-2.6} & \textbf{-2.2} & \textbf{-2.5} &       & -0.9 \bigstrut[t]\\
WLS$_s$ & \textbf{-4.1} & \textbf{-4.8} & -7    &       & -2.5  & -2.3  & -2.2  &       & -1.6  & -1.4  & -1.7  &       & \textbf{-3.1} \\
WLS$_v$ & -3.7  & -4.5  & \textbf{-7.1} &       & -2.4  & -2.2  & -2.2  &       & -1.6  & -1.3  & -1.7  &       & -3.0 \\
BU    & 0     & 0     & 0     &       & 0     & 0     & 0     &       & 0     & 0     & 0     &       & 0.0 \\
MinTit$_g$ & 0     & -0.2  & -0.5  &       & -2.2  & -2    & -2    &       & -1.7  & -1.5  & -1.8  &       & -1.3 \\
MinTit$_l$ & 1.9   & 2     & 2.3   &       & -1.8  & -1.7  & -1.6  &       & -1.7  & -1.5  & -1.7  &       & -0.4 \\
      & \multicolumn{13}{c}{Average} \bigstrut[b]\\
\cline{2-14}MinT  & -2.4  & -1.9  & -0.3  &       & -4.6  & -4.4  & -4.2  &       & \textbf{-3.9} & \textbf{-3.4} & \textbf{-3.9} &       & -3.2 \bigstrut[t]\\
WLS$_s$ & \textbf{-8.7} & \textbf{-9.6} & -12.5 &       & \textbf{-4.8} & \textbf{-4.7} & \textbf{-4.6} &       & -3.2  & -2.8  & -3.5  &       & \textbf{-6.0} \\
WLS$_v$ & -8.5  & -9.5  & \textbf{-12.6} &       & \textbf{-4.8} & \textbf{-4.7} & \textbf{-4.6} &       & -3.2  & -2.8  & -3.5  &       & \textbf{-6.0} \\
BU    & -0.1  & 0.2   & 2.1   &       & 0.4   & 0     & -0.1  &       & 0.1   & 0.1   & 0.1   &       & 0.3 \\
MinTit$_g$ & -5    & -4.9  & -5.2  &       & -4.5  & -4.3  & -4.2  &       & -3.2  & -2.9  & -3.5  &       & -4.2 \\
MinTit$_l$ & -2.9  & -2.6  & -2.2  &       & -4    & -3.8  & -3.8  &       & -3.2  & -2.8  & -3.4  &       & -3.2 \\
\end{tabular}%

  \label{tab:sim_seas_GPR}%
\end{table}%

\subsubsection{Impact of Differing Time Series Lengths}
\label{sec:sim_len}
This section investigates the impact of differing time series lengths throughout the hierarchical structure. Differing time series lengths in a hierarchical context often occur in practice due to several reasons: Product categories may get more refined over time as markets grow and products become more differentiated or the organizational structure becomes more refined as a company grows, leading to more differentiated reporting of revenues over time. Incorporating the available information can allow for more precise forecasts, especially when the time series are already very short.

As in \cref{sec:sim_effOfAggr}, time series according to the hierarchical structure illustrated in \cref{fig:sim_corr_hier} are simulated to produce higher SNRs on higher hierarchical levels. The resulting time series were then cut to the respective lengths as detailed in \cref{tab:sim_lens}, which ensures that hierarchically higher time series have at least as many observations as their constituents.

\begin{table}[t]
  \footnotesize\centering
  \caption{\small Time series lengths for the simulation in \cref{sec:sim_len}.}
    \begin{tabular}{lrrrr}
    \multicolumn{1}{c}{\textbf{Time Series}} & \multicolumn{1}{c}{\textbf{Obs.}} &       & \multicolumn{1}{c}{\textbf{Time Series}} & \multicolumn{1}{c}{\textbf{Obs.}} \\
    \multicolumn{2}{c}{Top-Level} &       & \multicolumn{2}{c}{Bottom-Level} \bigstrut[b]\\
\cline{1-2}\cline{4-5}    T     & 120   &       & \multicolumn{1}{l}{AAA} & 60 \bigstrut[t]\\
    \multicolumn{2}{c}{Level 1} &       & \multicolumn{1}{l}{AAB} & 60 \bigstrut[b]\\
\cline{1-2}    A     & 120   &       & \multicolumn{1}{l}{ABA} & 60 \bigstrut[t]\\
    B     & 90    &       & \multicolumn{1}{l}{ABB} & 60 \\
    \multicolumn{2}{c}{Level 2} &       & \multicolumn{1}{l}{BAA} & 45 \bigstrut[b]\\
\cline{1-2}    AA    & 120   &       & \multicolumn{1}{l}{BAB} & 45 \bigstrut[t]\\
    AB    & 90    &       & \multicolumn{1}{l}{BBA} & 15 \\
    BA    & 90    &       & \multicolumn{1}{l}{BBB} & 15 \\
    BB    & 90    &       &       &  \\
    \end{tabular}%
  \label{tab:sim_lens}%
\end{table}%

In this scenario, MinTit$_l$ should have an advantage: Because the covariance matrices are calculated separately for each sub-hierarchy, more observations can be used in some cases potentially leading to more accurate estimates. For example, overall, there are only 15 observations without missing data. But the local algorithm can utilize 90 observations to estimate the covariance matrix encompassing the top-level and level 1 time series. 

The \textit{results} of this simulation are presented in \cref{tab:sim_len_res}. As in \cref{sec:sim_effOfAggr}, it is not surprising that the BU approach performed poorly, leading to an increase in average RMSE of a factor between $9\%$ and over $90\%$ compared to the base forecasts. The WLS approaches achieved a slight error reduction, which was always exceeded by MinT. As expected, the best performing reconciliation method for ARIMA and ETS was MinTit$_l$ with reductions in average RMSE up to over $20\%$. It performed slightly better than its global variant in this context. When the forecasts were generated with the GPR model, MinT again performed on-par or slightly better than the iterative approaches.

%todo convergence rate???

\begin{table}[t]
  \centering
      \footnotesize\tabcolsep2.5pt
  \renewcommand{\arraystretch}{.9}
  \caption{\small Average out-of-sample forecast performance for the hierarchy with differing time series lengths in \cref{sec:sim_len}.}
    \begin{tabular}{lrrrlrrrlrrr}
          & \multicolumn{3}{c}{ARIMA} &       & \multicolumn{3}{c}{ETS} &       & \multicolumn{3}{c}{GPR} \bigstrut[b]\\
\cline{2-4}\cline{6-8}\cline{10-12}          & \multicolumn{1}{c}{h=1} & \multicolumn{1}{c}{1:2} & \multicolumn{1}{c}{1:4} &       & \multicolumn{1}{c}{h=1} & \multicolumn{1}{c}{1:2} & \multicolumn{1}{c}{1:4} &       & \multicolumn{1}{c}{h=1} & \multicolumn{1}{c}{1:3} & \multicolumn{1}{c}{1:6} \bigstrut\\
\cline{2-12}    MinT  & -19.6 & -19.4 & -19.1 &       & -2.8  & -3    & -3.2  &       & \textbf{-4.2} & \textbf{-4.3} & \textbf{-4.4} \bigstrut[t]\\
    WLS$_s$ & -3.5  & -3.5  & -3.3  &       & -1.5  & -1.6  & -2.1  &       & -2.2  & -1.4  & -1 \\
    WLS$_v$ & -2.6  & -2.6  & -2.5  &       & -1.6  & -1.7  & -2.2  &       & -2.5  & -1.7  & -1.4 \\
    BU    & 90.4  & 90    & 88.8  &       & 10.1  & 10.4  & 9.1   &       & 12.8  & 14.6  & 14.6 \\
    MinTit$_g$ & -19.9 & -19.8 & -19.4 &       & -2.7  & -3    & -3.2  &       & -4    & -4.1  & -4.2 \\
    MinTit$_l$ & \textbf{-20.3} & \textbf{-20.2} & \textbf{-19.8} &       & \textbf{-2.9} & \textbf{-3.1} & \textbf{-3.5} &       & \textbf{-4.2} & -4.2  & -4.3 \\
    \end{tabular}%
  \label{tab:sim_len_res}%
\end{table}%

\subsubsection{Impact of Degeneracy}
\label{sec:sim_deg}
\begin{figure}[ht]
%\vskip 0.2in
\begin{center}
\centerline{\includegraphics[width=1.0\linewidth]{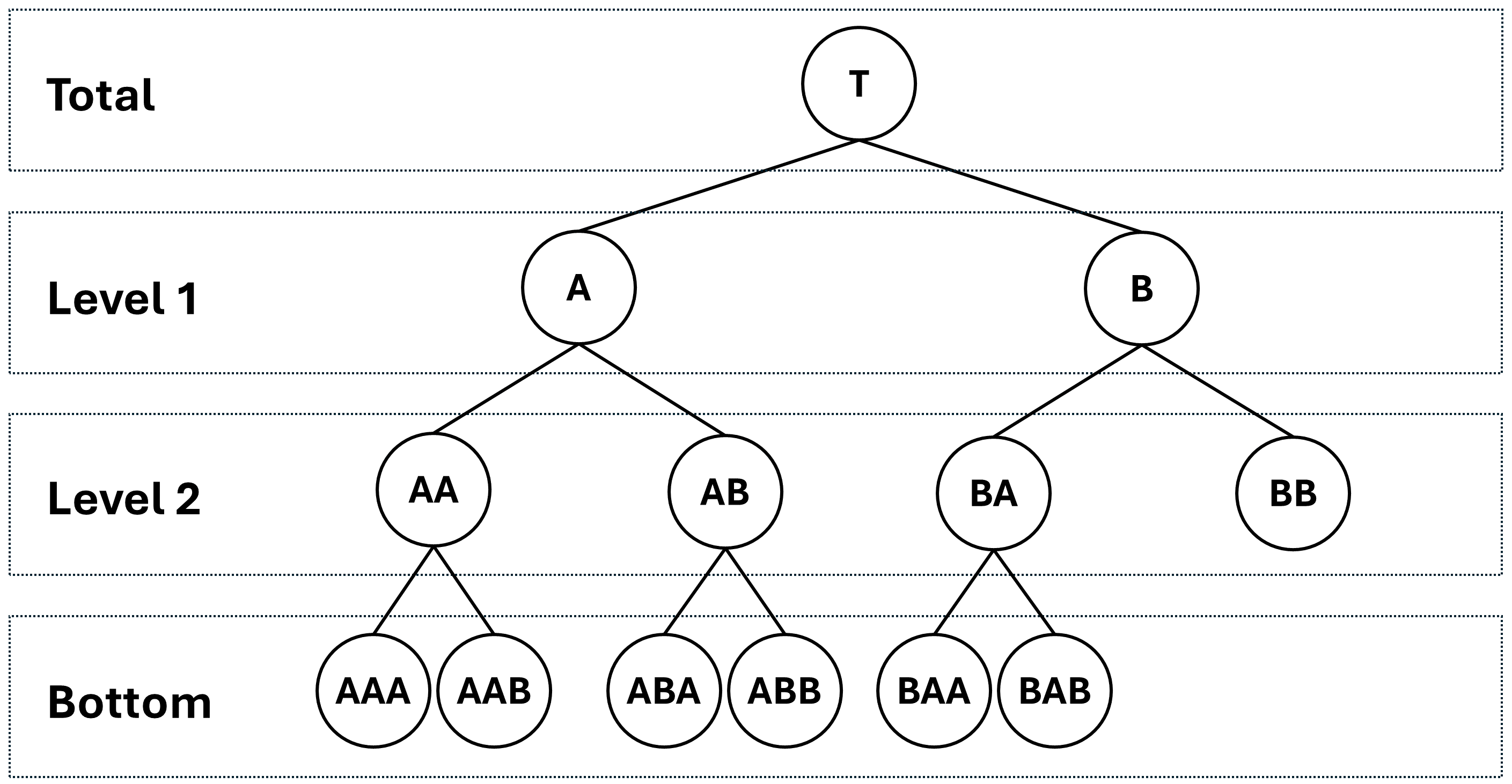}}
\caption{Hierarchical structure for the simulations in \cref{sec:sim_deg}.}
\label{fig:sim_deg_hier}
\end{center}
\vskip -0.3in
\end{figure}

This section examines the performance of the reconciliation methods under a degenerate hierarchical structure, such as illustrated in \cref{fig:sim_deg_hier}. The time series are generated in the same way as in \cref{sec:sim_impactOfCorr} but the bottom time series $y_{BBA}$ and $y_{BBB}$ are deleted after aggregation whereby the illustrated hierarchical structure is created. It should be noted that the convenient interface provided by the R packages \verb|hts| \citep{Hyndman2021htsLib} and \verb|fable| is unable to handle this case. However, an adapted implementation of the algorithm found in \citet{Hyndman2021htsLib} is capable of handling degenerate hierarchies and is provided in \citet{Steinmeister2024}.

\begin{figure}[ht]
%\vskip 0.2in
\begin{center}
\centerline{\includegraphics[width=\columnwidth]{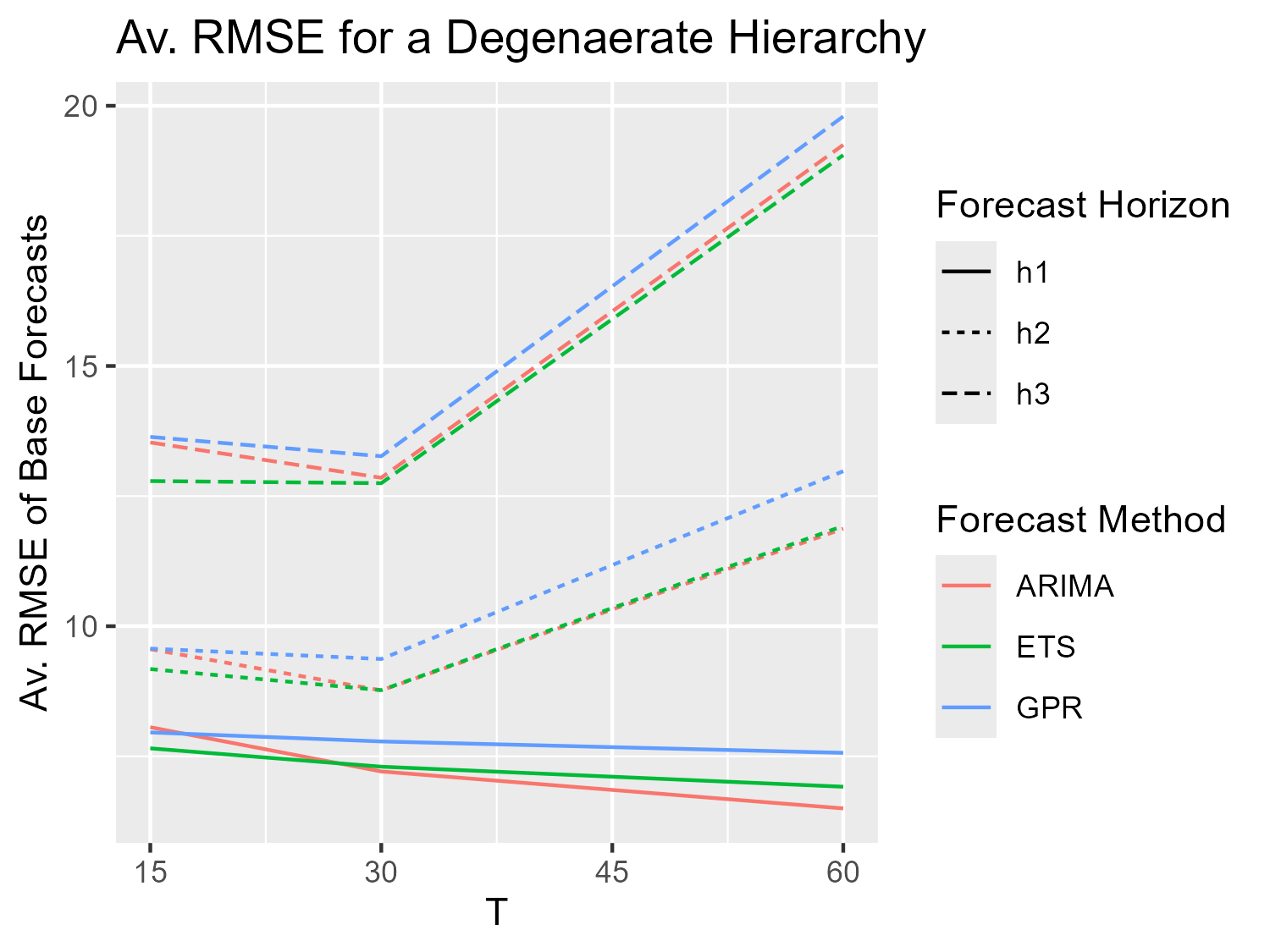}}
\caption{Average RMSE for the various base forecasts in \cref{sec:sim_deg} (ARIMA in red, ETS in green, and GPR in blue) by total length of the time series. A forecast horizon of $h1$ corresponds to $h = 1$, $h2$ to $h = 1:2$ in the case of $T = 15, 30$ and $h = 1:4$ for $T = 60$. Similarly, $h3$ corresponds to the longer forecasting horizons of $h = 1:4$ and $h = 1:8$, respectively.}
\label{fig:sim_deg}
\end{center}
\vskip -0.3in
\end{figure}

\textit{Results.}
In contrast to \cref{sec:sim_season} and in line with the other sections, ARIMA and ETS forecasts proved more accurate in this setting than GPR forecasts, see \cref{fig:sim_deg}. Similarly, MinT performed best for the GPR forecasts with an average 6.8\% RMSE reduction (compared to 6.1\% for WLS and 4.6\% for MinTit) but MinTit excelled for the ETS and ARIMA forecasts. In the case of ETS base forecasts, MinTit achieved a slightly higher reduction in average RMSE than WLS and MinT: 5.6\% against 5.2\% and 4.8\%, respectively. For the ARIMA base forecasts, the case was a bit more clear cut: MinTit achieved an average RMSE reduction of 9.4\% against 8.8\% for MinT and 7.5\% for WLS.

% Table generated by Excel2LaTeX from sheet 'Paper'
\begin{table}[t]
  \centering
      \footnotesize\tabcolsep2.5pt
  \renewcommand{\arraystretch}{.9}
  \caption{\small Out-of-sample forecast performance for the degenerate hierarchical structure
  in \cref{sec:sim_deg} for ETS-generated forecasts relative to the ETS-generated base forecasts.}
% Table generated by Excel2LaTeX from sheet 'Sheet3'
\begin{tabular}{lrrrlrrrlrrrr|r}
      & \multicolumn{13}{c}{ETS} \bigstrut[b]\\
\cline{2-14}      & \multicolumn{3}{c}{T = 15} &       & \multicolumn{3}{c}{T = 30} &       & \multicolumn{3}{c}{T = 60} & \multicolumn{1}{c}{} &  \bigstrut\\
\cline{2-4}\cline{6-8}\cline{10-12}      & \multicolumn{1}{c}{h=1} & \multicolumn{1}{c}{1:2} & \multicolumn{1}{c}{1:4} &       & \multicolumn{1}{c}{h=1} & \multicolumn{1}{c}{1:2} & \multicolumn{1}{c}{1:4} &       & \multicolumn{1}{c}{h=1} & \multicolumn{1}{c}{1:4} & \multicolumn{1}{c}{1:8} & \multicolumn{1}{c}{} & \multicolumn{1}{c}{Average} \bigstrut\\
\cline{2-12}\cline{14-14}      & \multicolumn{13}{c}{Top-Level} \bigstrut\\
\cline{2-14}MinT  & -6.2  & -4.8  & -3    &       & -8.1  & -7.4  & -6.5  &       & -8.2  & -6.2  & -4.2  &       & -6.1 \bigstrut[t]\\
WLS$_s$ & -7.1  & \textbf{-6.5} & \textbf{-6.3} &       & -8.3  & -8.8  & \textbf{-8.9} &       & -6.8  & -6.9  & \textbf{-6.5} &       & -7.3 \\
WLS$_v$ & -7    & -6.4  & -6.2  &       & -8.2  & -8.6  & -8.8  &       & -6.6  & -6.8  & -6.4  &       & -7.2 \\
BU    & -7    & -5.4  & -4.6  &       & \textbf{-9.9} & -8.9  & -8    &       & \textbf{-8.7} & -6.2  & -3.6  &       & -6.9 \\
MinTit$_g$ & \textbf{-7.2} & -6.1  & -5.2  &       & -9.2  & \textbf{-9} & -8.6  &       & -8.5  & \textbf{-7.5} & -6.2  &       & \textbf{-7.5} \\
MinTit$_l$ & -7.1  & -5.9  & -5    &       & -9.2  & -8.8  & -8.3  &       & \textbf{-8.7} & \textbf{-7.5} & -6.1  &       & -7.4 \\
      & \multicolumn{13}{c}{Level 1} \bigstrut[b]\\
\cline{2-14}MinT  & -5.9  & -4.9  & -3.2  &       & -7.7  & -7    & -6.1  &       & -7.1  & -5.9  & -4.4  &       & -5.8 \bigstrut[t]\\
WLS$_s$ & \textbf{-6.4} & \textbf{-6} & \textbf{-5.8} &       & -7.3  & -7.7  & \textbf{-7.9} &       & -5.5  & -6.1  & \textbf{-5.8} &       & -6.5 \\
WLS$_v$ & -6.1  & -5.8  & -5.6  &       & -7    & -7.4  & -7.6  &       & -5.3  & -6    & \textbf{-5.8} &       & -6.3 \\
BU    & -6.2  & -4.9  & -4    &       & \textbf{-9} & \textbf{-7.9} & -7    &       & \textbf{-7.5} & -5.6  & -3.2  &       & -6.1 \\
MinTit$_g$ & -6.3  & -5.5  & -4.5  &       & -8.2  & \textbf{-7.9} & -7.6  &       & -7.3  & \textbf{-6.8} & \textbf{-5.8} &       & \textbf{-6.7} \\
MinTit$_l$ & -6.2  & -5.2  & -4.2  &       & -8.2  & -7.7  & -7.3  &       & \textbf{-7.5} & \textbf{-6.8} & -5.6  &       & -6.5 \\
      & \multicolumn{13}{c}{Level 2} \bigstrut[b]\\
\cline{2-14}MinT  & \textbf{-3.2} & -2.7  & -1.7  &       & -3.9  & -3.8  & -3.3  &       & -3.3  & -2.6  & -2.3  &       & -3.0 \bigstrut[t]\\
WLS$_s$ & -3    & \textbf{-3} & \textbf{-3.1} &       & -3    & -3.7  & \textbf{-4.1} &       & -1.8  & -2.5  & -3    &       & -3.0 \\
WLS$_v$ & -2.6  & -2.6  & -2.7  &       & -2.7  & -3.4  & -3.7  &       & -1.5  & -2.3  & -2.9  &       & -2.7 \\
BU    & -2.9  & -2    & -1.6  &       & \textbf{-4.5} & \textbf{-4} & -3.3  &       & \textbf{-3.4} & -2    & -0.8  &       & -2.7 \\
MinTit$_g$ & -3    & -2.6  & -2.1  &       & -3.9  & \textbf{-4} & -4    &       & -3.3  & \textbf{-3.1} & \textbf{-3.1} &       & \textbf{-3.2} \\
MinTit$_l$ & -2.9  & -2.3  & -1.7  &       & -3.7  & -3.7  & -3.6  &       & -3.3  & -2.9  & -2.8  &       & -3.0 \\
      & \multicolumn{13}{c}{Bottom-Level} \bigstrut[b]\\
\cline{2-14}MinT  & \textbf{-0.8} & \textbf{-1.2} & -0.7  &       & 0.1   & -0.4  & -0.7  &       & \textbf{-0.8} & -1    & -1.8  &       & \textbf{-0.8} \bigstrut[t]\\
WLS$_s$ & 0.1   & -0.7  & \textbf{-1.1} &       & 1.6   & 0.5   & -0.5  &       & 1.5   & -0.3  & -1.6  &       & -0.1 \\
WLS$_v$ & 0.8   & -0.2  & -0.5  &       & 1.9   & 0.5   & -0.3  &       & 1.8   & -0.3  & -1.5  &       & 0.2 \\
BU    & 0     & 0     & 0     &       & \textbf{0} & 0     & 0     &       & 0     & 0     & 0     &       & 0.0 \\
MinTit$_g$ & -0.4  & -1    & -1    &       & 0.2   & \textbf{-0.5} & \textbf{-1.1} &       & -0.3  & \textbf{-1.2} & \textbf{-2.2} &       & \textbf{-0.8} \\
MinTit$_l$ & -0.4  & -0.9  & -0.8  &       & 0.4   & -0.2  & -0.8  &       & -0.4  & -1.1  & -1.9  &       & -0.7 \\
      & \multicolumn{13}{c}{Average} \bigstrut[b]\\
\cline{2-14}MinT  & -4.9  & -4    & -2.5  &       & -6.3  & -5.8  & -5.1  &       & -6.2  & -4.8  & -3.5  &       & -4.8 \bigstrut[t]\\
WLS$_s$ & -5.2  & \textbf{-4.9} & \textbf{-4.8} &       & -6    & -6.4  & \textbf{-6.7} &       & -4.6  & -5    & \textbf{-4.9} &       & -5.4 \\
WLS$_v$ & -5    & -4.7  & -4.6  &       & -5.7  & -6.2  & -6.4  &       & -4.4  & -4.9  & -4.8  &       & -5.2 \\
BU    & -5.2  & -3.9  & -3.2  &       & \textbf{-7.6} & \textbf{-6.7} & -5.8  &       & \textbf{-6.4} & -4.5  & -2.4  &       & -5.1 \\
MinTit$_g$ & \textbf{-5.3} & -4.6  & -3.8  &       & -6.9  & \textbf{-6.7} & -6.5  &       & -6.3  & \textbf{-5.7} & \textbf{-4.9} &       & \textbf{-5.6} \\
MinTit$_l$ & -5.2  & -4.4  & -3.5  &       & -6.8  & -6.5  & -6.2  &       & \textbf{-6.4} & -5.6  & -4.7  &       & -5.5 \\
\end{tabular}%

  \label{tab:sim_deg_ETS}%
\end{table}%

% Table generated by Excel2LaTeX from sheet 'Paper'
\begin{table}[t]
  \centering
      \footnotesize\tabcolsep2.5pt
  \renewcommand{\arraystretch}{.9}
  \caption{\small Out-of-sample forecast performance for the degenerate hierarchical structure
  in \cref{sec:sim_deg} for ARIMA-generated forecasts relative to the ARIMA-generated base forecasts.}
% Table generated by Excel2LaTeX from sheet 'Sheet3'
\begin{tabular}{lrrrlrrrlrrrl|r}
      & \multicolumn{13}{c}{ARIMA} \bigstrut[b]\\
\cline{2-14}      & \multicolumn{3}{c}{T = 15} &       & \multicolumn{3}{c}{T = 30} &       & \multicolumn{3}{c}{T = 60} & \multicolumn{1}{c}{} &  \bigstrut\\
\cline{2-4}\cline{6-8}\cline{10-12}      & \multicolumn{1}{c}{h=1} & \multicolumn{1}{c}{1:2} & \multicolumn{1}{c}{1:4} &       & \multicolumn{1}{c}{h=1} & \multicolumn{1}{c}{1:2} & \multicolumn{1}{c}{1:4} &       & \multicolumn{1}{c}{h=1} & \multicolumn{1}{c}{1:4} & \multicolumn{1}{c}{1:8} & \multicolumn{1}{c}{} & \multicolumn{1}{c}{Average} \bigstrut\\
\cline{2-12}\cline{14-14}      & \multicolumn{13}{c}{Top-Level} \bigstrut\\
\cline{2-14}MinT  & -15   & -12.9 & -9.5  &       & -11.1 & -10.7 & -10.1 &       & -8.5  & -7.2  & -6.3  &       & -10.1 \bigstrut[t]\\
WLS$_s$ & -13.3 & -12   & \textbf{-10.7} &       & -9.7  & -9.8  & -10   &       & -8    & -7.7  & -7.7  &       & -9.9 \\
WLS$_v$ & -13.1 & -11.9 & -10.5 &       & -9.4  & -9.6  & -9.8  &       & -7.9  & -7.5  & -7.6  &       & -9.7 \\
BU    & -12.8 & -10.7 & -8.4  &       & -8.6  & -8.8  & -9.4  &       & -9    & -8.2  & -7    &       & -9.2 \\
MinTit$_g$ & \textbf{-15.3} & \textbf{-13.2} & -10.5 &       & \textbf{-12.4} & \textbf{-12} & \textbf{-11.4} &       & \textbf{-10.1} & \textbf{-8.8} & \textbf{-8.2} &       & \textbf{-11.3} \\
MinTit$_l$ & -15.2 & -13   & -10.3 &       & -12.2 & -11.7 & -11.1 &       & -9.9  & -8.6  & -8    &       & -11.1 \\
      & \multicolumn{13}{c}{Level 1} \bigstrut[b]\\
\cline{2-14}MinT  & \textbf{-13.1} & \textbf{-11} & -7.8  &       & -12.2 & -11.1 & -9.3  &       & -9.1  & -7.9  & -6.8  &       & -9.8 \bigstrut[t]\\
WLS$_s$ & -10.6 & -9.4  & \textbf{-8.3} &       & -9.6  & -9.2  & -8.5  &       & -7.6  & -7.5  & -7.3  &       & -8.7 \\
WLS$_v$ & -10.5 & -9.2  & -8    &       & -9.5  & -9    & -8.3  &       & -7.7  & -7.4  & -7.2  &       & -8.5 \\
BU    & -10.1 & -8.2  & -6.2  &       & -7.6  & -7.2  & -7.2  &       & -8.8  & -8.2  & -6.7  &       & -7.8 \\
MinTit$_g$ & -12.9 & -10.8 & -8.2  &       & \textbf{-12.9} & \textbf{-11.8} & \textbf{-10.1} &       & \textbf{-10.4} & \textbf{-8.9} & \textbf{-7.9} &       & \textbf{-10.4} \\
MinTit$_l$ & -12.8 & -10.6 & -8    &       & -12.7 & -11.4 & -9.6  &       & -10.2 & -8.8  & -7.7  &       & -10.2 \\
      & \multicolumn{13}{c}{Level 2} \bigstrut[b]\\
\cline{2-14}MinT  & \textbf{-8.8} & \textbf{-7.2} & \textbf{-4.8} &       & -10.3 & -8.8  & -6.6  &       & -6.3  & -4.8  & -3.7  &       & -6.8 \bigstrut[t]\\
WLS$_s$ & -5.4  & -4.8  & -4.4  &       & -6.2  & -5.5  & -5    &       & -3.8  & -3.7  & -3.6  &       & -4.7 \\
WLS$_v$ & -5.3  & -4.6  & -4.1  &       & -6.6  & -5.7  & -4.9  &       & -4.2  & -3.7  & -3.5  &       & -4.7 \\
BU    & -4.7  & -3.5  & -2.4  &       & -3.6  & -3.3  & -3.6  &       & -4.7  & -4.2  & -3    &       & -3.7 \\
MinTit$_g$ & -8    & -6.4  & -4.6  &       & \textbf{-10.4} & \textbf{-8.9} & \textbf{-6.8} &       & \textbf{-7.1} & \textbf{-5.4} & \textbf{-4.4} &       & \textbf{-6.9} \\
MinTit$_l$ & -7.8  & -6.1  & -4.2  &       & -10   & -8.4  & -6.1  &       & -6.8  & -5.2  & -4.1  &       & -6.5 \\
      & \multicolumn{13}{c}{Bottom-Level} \bigstrut[b]\\
\cline{2-14}MinT  & \textbf{-5.1} & \textbf{-4.3} & \textbf{-2.6} &       & \textbf{-12.6} & \textbf{-9.9} & \textbf{-5.7} &       & -2.6  & -1.1  & -0.9  &       & \textbf{-5.0} \bigstrut[t]\\
WLS$_s$ & -0.3  & -0.9  & -1.4  &       & -2.3  & -1.9  & -1.1  &       & 1.1   & 0.8   & -0.2  &       & -0.7 \\
WLS$_v$ & -1.1  & -1.4  & -1.4  &       & -8.1  & -6.3  & -3.5  &       & -0.6  & 0.1   & -0.6  &       & -2.5 \\
BU    & 0     & 0     & 0     &       & 0     & 0     & 0     &       & 0     & 0     & 0     &       & 0.0 \\
MinTit$_g$ & -3.9  & -3.3  & -2.1  &       & -12.5 & -9.8  & \textbf{-5.7} &       & \textbf{-3.3} & \textbf{-1.5} & \textbf{-1.4} &       & -4.8 \\
MinTit$_l$ & -3.9  & -3.1  & -1.8  &       & -12.3 & -9.5  & -5.2  &       & -3.1  & -1.3  & -1.2  &       & -4.6 \\
      & \multicolumn{13}{c}{Average} \bigstrut[b]\\
\cline{2-14}MinT  & \textbf{-12.2} & \textbf{-10.3} & -7.2  &       & -11.4 & -10.3 & -8.6  &       & -7.7  & -6.2  & -5.2  &       & -8.8 \bigstrut[t]\\
WLS$_s$ & -9.6  & -8.6  & \textbf{-7.6} &       & -8.1  & -7.9  & -7.5  &       & -6.2  & -5.9  & -5.8  &       & -7.5 \\
WLS$_v$ & -9.5  & -8.5  & -7.4  &       & -8.7  & -8.2  & -7.6  &       & -6.4  & -5.9  & -5.7  &       & -7.5 \\
BU    & -9.1  & -7.3  & -5.5  &       & -6.3  & -6.2  & -6.4  &       & -7.2  & -6.5  & -5.2  &       & -6.6 \\
MinTit$_g$ & -11.9 & -10   & \textbf{-7.6} &       & \textbf{-12.1} & \textbf{-11} & \textbf{-9.4} &       & \textbf{-8.9} & \textbf{-7.3} & \textbf{-6.5} &       & \textbf{-9.4} \\
MinTit$_l$ & -11.9 & -9.8  & -7.3  &       & -11.9 & -10.7 & -9    &       & -8.8  & -7.2  & -6.2  &       & -9.2 \\
\end{tabular}%

  \label{tab:sim_deg_ARIMA}%
\end{table}%

% Table generated by Excel2LaTeX from sheet 'Paper'
\begin{table}[t]
  \centering
      \footnotesize\tabcolsep2.5pt
  \renewcommand{\arraystretch}{.9}
  \caption{\small Out-of-sample forecast performance for the degenerate hierarchical structure
  in \cref{sec:sim_deg} for GPR-generated forecasts relative to the GPR-generated base forecasts.}
% Table generated by Excel2LaTeX from sheet 'Sheet3'
\begin{tabular}{lrrrlrrrlrrrr|r}
      & \multicolumn{13}{c}{GPR} \bigstrut[b]\\
\cline{2-14}      & \multicolumn{3}{c}{T = 15} &       & \multicolumn{3}{c}{T = 30} &       & \multicolumn{3}{c}{T = 60} & \multicolumn{1}{c}{} &  \bigstrut\\
\cline{2-4}\cline{6-8}\cline{10-12}      & \multicolumn{1}{c}{h=1} & \multicolumn{1}{c}{1:2} & \multicolumn{1}{c}{1:4} &       & \multicolumn{1}{c}{h=1} & \multicolumn{1}{c}{1:2} & \multicolumn{1}{c}{1:4} &       & \multicolumn{1}{c}{h=1} & \multicolumn{1}{c}{1:4} & \multicolumn{1}{c}{1:8} & \multicolumn{1}{c}{} & \multicolumn{1}{c}{Average} \bigstrut\\
\cline{2-12}\cline{14-14}      & \multicolumn{13}{c}{Top-Level} \bigstrut\\
\cline{2-14}MinT  & -12.1 & -11   & -7.8  &       & \textbf{-9.9} & \textbf{-8.9} & \textbf{-7.7} &       & \textbf{-8.7} & \textbf{-5.7} & \textbf{-4.8} &       & -8.5 \bigstrut[t]\\
WLS$_s$ & \textbf{-12.8} & \textbf{-13.3} & \textbf{-18} &       & -6.5  & -6.4  & -6.4  &       & -5.7  & -4.3  & -4.3  &       & \textbf{-8.6} \\
WLS$_v$ & \textbf{-12.8} & \textbf{-13.3} & \textbf{-18} &       & -6.5  & -6.4  & -6.4  &       & -5.6  & -4.2  & -4.3  &       & \textbf{-8.6} \\
BU    & -1.6  & -0.3  & 0.9   &       & 0.4   & -0.3  & -0.4  &       & -7.1  & -3.9  & -4    &       & -1.8 \\
MinTit$_g$ & -9.7  & -9    & -10.7 &       & -6.2  & -6.1  & -6    &       & -8.1  & -5.3  & \textbf{-4.8} &       & -7.3 \\
MinTit$_l$ & -7.8  & -6.7  & -8.2  &       & -5.7  & -5.6  & -5.5  &       & -8.2  & -5.4  & -4.7  &       & -6.4 \\
      & \multicolumn{13}{c}{Level 1} \bigstrut[b]\\
\cline{2-14}MinT  & \textbf{-11.3} & -9.7  & -6.2  &       & \textbf{-9.6} & \textbf{-8.1} & \textbf{-6.7} &       & \textbf{-8.7} & \textbf{-5.1} & \textbf{-3.7} &       & \textbf{-7.7} \bigstrut[t]\\
WLS$_s$ & -9.8  & \textbf{-11.3} & \textbf{-14} &       & -6.1  & -5.8  & -5.5  &       & -5.4  & -3.4  & -3    &       & -7.1 \\
WLS$_v$ & -9.8  & \textbf{-11.3} & \textbf{-14} &       & -6.1  & -5.8  & -5.4  &       & -5.2  & -3.3  & -2.9  &       & -7.1 \\
BU    & 1.1   & 0.9   & 4.2   &       & 0.1   & -0.3  & 0     &       & -7.3  & -3.5  & -2.9  &       & -0.9 \\
MinTit$_g$ & -6.2  & -6.6  & -6.4  &       & -5.7  & -5.4  & -5    &       & -8    & -4.5  & -3.5  &       & -5.7 \\
MinTit$_l$ & -4.1  & -4.3  & -3.4  &       & -5.2  & -4.9  & -4.5  &       & -8.1  & -4.7  & -3.5  &       & -4.7 \\
      & \multicolumn{13}{c}{Level 2} \bigstrut[b]\\
\cline{2-14}MinT  & -7.1  & -5.8  & -2.4  &       & \textbf{-6.4} & \textbf{-5.3} & -3.8  &       & \textbf{-5.7} & \textbf{-3.3} & \textbf{-2.2} &       & -4.7 \bigstrut[t]\\
WLS$_s$ & \textbf{-8} & \textbf{-9.4} & \textbf{-11.7} &       & -4.2  & -4.2  & \textbf{-4.3} &       & -2.4  & -1.4  & -1.3  &       & \textbf{-5.2} \\
WLS$_v$ & -7.9  & -9.3  & \textbf{-11.7} &       & -4.2  & -4.2  & \textbf{-4.3} &       & -2.1  & -1.2  & -1.1  &       & -5.1 \\
BU    & 0.3   & 0.4   & 3.4   &       & 1     & 0.5   & 0.2   &       & -4    & -1.6  & -1.3  &       & -0.1 \\
MinTit$_g$ & -4    & -4.2  & -3.5  &       & -3.7  & -3.8  & -3.9  &       & -4.6  & -2.3  & -1.6  &       & -3.5 \\
MinTit$_l$ & -1.6  & -1.5  & 0     &       & -3.1  & -3.2  & -3.4  &       & -4.6  & -2.3  & -1.4  &       & -2.3 \\
      & \multicolumn{13}{c}{Bottom-Level} \bigstrut[b]\\
\cline{2-14}MinT  & -2.5  & -2.7  & -0.7  &       & -2.3  & \textbf{-2.6} & -1.9  &       & \textbf{-2.3} & \textbf{-2.1} & \textbf{-1.3} &       & -2.0 \bigstrut[t]\\
WLS$_s$ & \textbf{-4.1} & \textbf{-4.8} & -7    &       & \textbf{-2.5} & -2.3  & \textbf{-2.2} &       & 1.6   & 0.2   & 0     &       & \textbf{-2.3} \\
WLS$_v$ & -3.7  & -4.5  & \textbf{-7.1} &       & -2.4  & -2.2  & \textbf{-2.2} &       & 1.7   & 0     & -0.1  &       & \textbf{-2.3} \\
BU    & 0     & 0     & 0     &       & 0     & 0     & 0     &       & 0     & 0     & 0     &       & 0.0 \\
MinTit$_g$ & 0     & -0.2  & -0.5  &       & -2.2  & -2    & -2    &       & -0.9  & -1    & -0.6  &       & -1.0 \\
MinTit$_l$ & 1.9   & 2     & 2.3   &       & -1.8  & -1.7  & -1.6  &       & -1    & -1.2  & -0.7  &       & -0.2 \\
      & \multicolumn{13}{c}{Average} \bigstrut[b]\\
\cline{2-14}MinT  & \textbf{-9.9} & -8.6  & -5.4  &       & \textbf{-8.3} & \textbf{-7.2} & \textbf{-5.9} &       & \textbf{-7.4} & \textbf{-4.6} & \textbf{-3.5} &       & \textbf{-6.8} \bigstrut[t]\\
WLS$_s$ & -8.7  & \textbf{-9.6} & -12.5 &       & -4.8  & -4.7  & -4.6  &       & -4.2  & -2.9  & -2.7  &       & -6.1 \\
WLS$_v$ & -8.5  & -9.5  & \textbf{-12.6} &       & -4.8  & -4.7  & -4.6  &       & -4    & -2.8  & -2.7  &       & -6.0 \\
BU    & -0.1  & 0.2   & 2.1   &       & 0.4   & 0     & -0.1  &       & -5.8  & -2.8  & -2.6  &       & -1.0 \\
MinTit$_g$ & -5    & -4.9  & -5.2  &       & -4.5  & -4.3  & -4.2  &       & -6.6  & -3.9  & -3.2  &       & -4.6 \\
MinTit$_l$ & -2.9  & -2.6  & -2.2  &       & -4    & -3.8  & -3.8  &       & -6.7  & -4    & -3.1  &       & -3.7 \\
\end{tabular}%

  \label{tab:sim_deg_GPR}%
\end{table}%

\subsubsection{A ``large'' Hierarchy}
\label{sec:sim_largeHier}
Finally, this section explores the performance of the reconciliation methods on a much larger hierarchy -- a scenario which motivated the formulation of the iterative trace minimization algorithm. This hierarchy is made up of 6 times series on the first hierarchical level, followed by 4 component time series each on the following three levels respectively. This results in a total of 6, 24, 96, 384 time series on levels 1 through 4, respectively. The total number of time series is 510. Here, only 500 Monte Carlo iterations were used (i.e. 500 such time series hierarchies were simulated) due to the high number of time series already contained in this setup. Similarly, \citet{Wickramasuriya2019} only used 200 Monte Carlo iterations in their simulation of a large hierarchy.

The 384 bottom time series are simulated similarly to \cref{sec:sim_effOfAggr}. Error terms with variances of $0.4$, $1/40$, and $1/640$ which cancel out levels three, two, and one, respectively. This results in additional variances of about $0.427$ on the bottom level, $0.425$ on level 3, and $0.4$ on level 2.

The \textit{results for ETS forecasts} can be assessed in \cref{tab:sim_large_ETS}.
On average, the iterative MinT methods, particularly MinTit$_l$, produced the largest reduction in RMSE (3.6\%). The second highest RMSE-reduction was achieved by variance scaling and structural scaling (WLS$_v$ and WLS$_s$). MinT led to a slight increase in forecasting error (av. 0.7 \%), especially on lower hierarchical levels (7.5\%).

\textit{Results for ARIMA forecasts} (\cref{tab:sim_large_ARIMA}). The reconciliation results for the forecasts based on ARIMA are similar to those based on ETS. MinTit$_l$ and MinTit$_g$ achieved the largest accuracy gains (4.2\% and 4\% respectively), followed by WLS (3.4\% and 3.3\%). Again, MinT led to an increase in forecasting error (av. 2.2 \%, bottom-level: 10.5\%). BU generated an average increase in RMSE of 11\%.

\begin{figure}[ht]
%\vskip 0.8in
\begin{center}
\centerline{\includegraphics[width=\columnwidth]{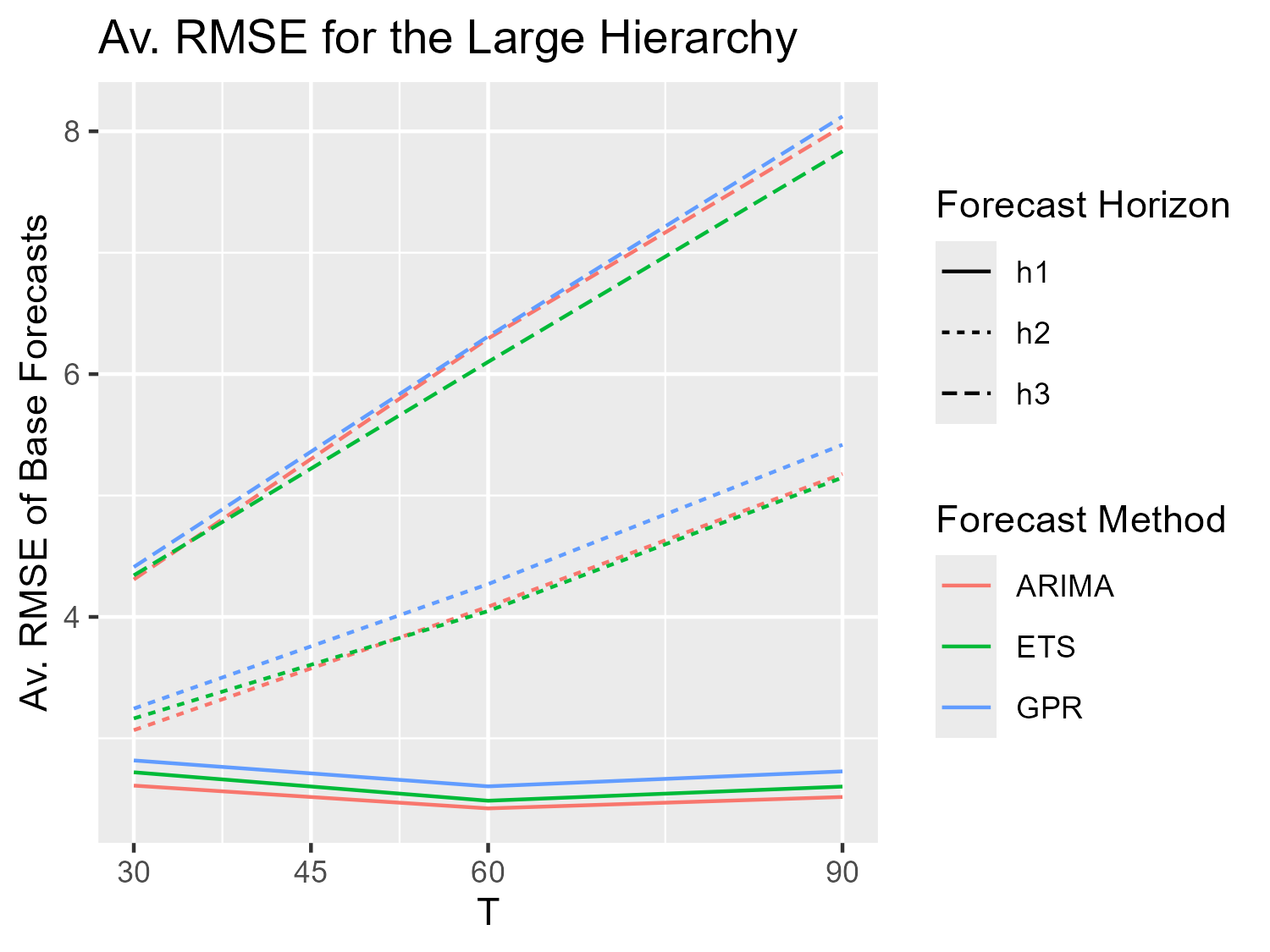}}
\caption{Average RMSE for the various base forecasts in \cref{sec:sim_largeHier} (ARIMA in red, ETS in green, and GPR in blue) by total length of the time series. A forecast horizon of $h1$ corresponds to $h = 1$, $h2$ to $h = 1:2$ in the case of $T = 15, 30$ and $h = 1:3$ for $T = 60$. Similarly, $h3$ corresponds to the longer forecasting horizons of $h = 1:4$ and $h = 1:6$, respectively.}
\label{fig:sim_largeHier}
\end{center}
\vskip -0.3in
\end{figure}

\textit{Result for GPR forecasts} (\cref{tab:sim_large_GPR}). In contrast to the ETS and ARIMA forecasts, and similar to other subsections, MinT achieved the largest error reduction as measured in RMSE for forecasts based on GPR (5.3\% vs. MinTit: 4.3\%, WLS: 1.5\% and 1.2\%). However, GPR had a worse average base-forecast performance than ETS and ARIMA as illustrated in \cref{fig:sim_largeHier}. Mirroring the results in \cref{sec:sim_effOfAggr}, the BU approach was the worst performing reconciliation method (with an 11.6\% RMSE increase) due to its inability to incorporate the more accurate forecasts of time series on higher hierarchical levels.

\begin{table}[t]
  \centering
    \footnotesize\tabcolsep2.5pt
  \renewcommand{\arraystretch}{.9}
  \vskip -0.8in
  \caption{\small Out-of-sample forecast performance for the large hierarchy
  in \cref{sec:sim_largeHier} for ETS-generated forecasts relative to the ETS-generated base forecasts.}
% Table generated by Excel2LaTeX from sheet 'Sheet3'
\begin{tabular}{lrrrlrrrrrrrr|r}
      & \multicolumn{13}{c}{ETS} \bigstrut[b]\\
\cline{2-14}      & \multicolumn{3}{c}{T = 30} &       & \multicolumn{3}{c}{T = 60} &       & \multicolumn{3}{c}{T = 90} & \multicolumn{1}{c}{} &  \bigstrut\\
\cline{2-4}\cline{6-8}\cline{10-12}      & \multicolumn{1}{c}{h=1} & \multicolumn{1}{c}{1:2} & \multicolumn{1}{c}{1:4} &       & \multicolumn{1}{c}{h=1} & \multicolumn{1}{c}{1:3} & \multicolumn{1}{c}{1:6} &       & \multicolumn{1}{c}{h=1} & \multicolumn{1}{c}{1:4} & \multicolumn{1}{c}{1:8} & \multicolumn{1}{c}{} & \multicolumn{1}{c}{Average} \bigstrut\\
\cline{2-12}\cline{14-14}      & \multicolumn{13}{c}{Top-Level} \bigstrut\\
\cline{2-14}MinT  & -1.8  & -5.7  & -5.7  &       & -2.5  & -3.6  & -1.3  &       & -1.6  & -2.7  & -1.9  &       & -3.0 \bigstrut[t]\\
WLS$_s$ & -5.1  & -6.9  & \textbf{-9.1} &       & -1.6  & -6.2  & \textbf{-8.5} &       & -4.4  & \textbf{-6.5} & \textbf{-6.6} &       & -6.1 \\
WLS$_v$ & -5.1  & -6.9  & \textbf{-9.1} &       & -1.6  & -6.2  & \textbf{-8.5} &       & -4.4  & \textbf{-6.5} & \textbf{-6.6} &       & -6.1 \\
BU    & 33.9  & 25.2  & 11.2  &       & 48    & 19.2  & 6     &       & 25.2  & 6.6   & 1.4   &       & 19.6 \\
MinTit$_g$ & -5.2  & -6.5  & -7.4  &       & \textbf{-6} & -7.6  & -8.1  &       & -4.5  & -6.3  & -6.1  &       & -6.4 \\
MinTit$_l$ & \textbf{-6} & \textbf{-7} & -7.8  &       & -5.9  & \textbf{-7.8} & -8.3  &       & \textbf{-4.7} & -6.2  & -6    &       & \textbf{-6.6} \\
      & \multicolumn{13}{c}{Level 1} \bigstrut[b]\\
\cline{2-14}MinT  & -2.4  & \textbf{-5.8} & -5.9  &       & 0.6   & -2.3  & -2.3  &       & -0.9  & -1.8  & -1.5  &       & -2.5 \bigstrut[t]\\
WLS$_s$ & -2.7  & -3.8  & \textbf{-7.6} &       & 1.2   & -4.3  & \textbf{-8.6} &       & 0     & -4.7  & \textbf{-6.3} &       & -4.1 \\
WLS$_v$ & -2.7  & -3.8  & \textbf{-7.6} &       & 1.2   & -4.3  & \textbf{-8.6} &       & 0     & -4.7  & \textbf{-6.3} &       & -4.1 \\
BU    & 32.8  & 27    & 12.3  &       & 42.1  & 18.1  & 3.9   &       & 37.5  & 10.4  & 2.8   &       & 20.8 \\
MinTit$_g$ & -3.9  & -4.7  & -6.9  &       & -2.6  & -5.3  & -7.5  &       & -3.1  & -4.8  & -5.6  &       & -4.9 \\
MinTit$_l$ & \textbf{-4.5} & -5.3  & -7.4  &       & \textbf{-2.8} & \textbf{-5.6} & -7.8  &       & \textbf{-3.3} & \textbf{-5} & -5.7  &       & \textbf{-5.3} \\
      & \multicolumn{13}{c}{Level 2} \bigstrut[b]\\
\cline{2-14}MinT  & -1.5  & \textbf{-3.9} & -3.2  &       & -2    & -1.9  & 0     &       & -1.9  & -1.2  & 0.7   &       & -1.7 \bigstrut[t]\\
WLS$_s$ & -1.4  & -1.6  & \textbf{-4.5} &       & 0.5   & -2.8  & \textbf{-6} &       & 0.4   & -3.5  & \textbf{-4.8} &       & -2.6 \\
WLS$_v$ & -1.4  & -1.6  & \textbf{-4.5} &       & 0.5   & -2.8  & \textbf{-6} &       & 0.4   & -3.5  & \textbf{-4.8} &       & -2.6 \\
BU    & 24.3  & 21.4  & 11.5  &       & 28    & 14.5  & 4.7   &       & 27.4  & 8.6   & 3     &       & 15.9 \\
MinTit$_g$ & -2.6  & -2.6  & -4    &       & -2.2  & -3.9  & -5.4  &       & -2.4  & -3.9  & -4.3  &       & -3.5 \\
MinTit$_l$ & \textbf{-3.1} & -3    & \textbf{-4.5} &       & \textbf{-2.4} & \textbf{-4.1} & -5.6  &       & \textbf{-2.6} & \textbf{-4.1} & -4.5  &       & \textbf{-3.8} \\
      & \multicolumn{13}{c}{Level 3} \bigstrut[b]\\
\cline{2-14}MinT  & -4.4  & \textbf{-5.3} & -2.2  &       & -5.9  & -1.9  & 3.1   &       & -4    & 2     & 5.9   &       & -1.4 \bigstrut[t]\\
WLS$_s$ & -1.3  & -1.1  & \textbf{-2.4} &       & -0.6  & -1.6  & \textbf{-3.4} &       & 0.9   & -1    & \textbf{-2.2} &       & -1.4 \\
WLS$_v$ & -1.3  & -1.1  & \textbf{-2.4} &       & -0.6  & -1.7  & \textbf{-3.4} &       & 0.9   & -1.1  & \textbf{-2.2} &       & -1.4 \\
BU    & 10.7  & 10.8  & 6.9   &       & 11    & 7.3   & 2.8   &       & 13.7  & 6.6   & 3     &       & 8.1 \\
MinTit$_g$ & -1.7  & -1.5  & -2    &       & -1.7  & -1.9  & -2.5  &       & -0.8  & -1.4  & -1.7  &       & -1.7 \\
MinTit$_l$ & \textbf{-1.9} & -1.8  & \textbf{-2.3} &       & \textbf{-1.8} & \textbf{-2.2} & -2.7  &       & \textbf{-0.9} & \textbf{-1.6} & -1.9  &       & \textbf{-1.9} \\
      & \multicolumn{13}{c}{Bottom-Level} \bigstrut[b]\\
\cline{2-14}MinT  & 3.2   & 3.2   & 8.4   &       & 1.1   & 5.8   & 15    &       & 1.2   & 10.3  & 19.5  &       & 7.5 \bigstrut[t]\\
WLS$_s$ & -1.6  & -1.8  & -1.8  &       & -1.5  & -1.7  & -1.5  &       & -1.7  & -1.7  & -1.5  &       & -1.6 \\
WLS$_v$ & -1.8  & -2    & -2    &       & -1.7  & -2    & -2    &       & -1.9  & -2.1  & -1.9  &       & -1.9 \\
BU    & 0     & 0     & 0     &       & 0     & 0     & 0     &       & 0     & 0     & 0     &       & 0.0 \\
MinTit$_g$ & -1.9  & -2.1  & -2    &       & \textbf{-2.2} & -2.5  & -2    &       & \textbf{-2.4} & -2.7  & -2.2  &       & -2.2 \\
MinTit$_l$ & -1.9  & -2.2  & -2.1  &       & \textbf{-2.2} & \textbf{-2.6} & \textbf{-2.1} &       & \textbf{-2.4} & \textbf{-2.8} & \textbf{-2.3} &       & \textbf{-2.3} \\
      & \multicolumn{13}{c}{Average} \bigstrut[b]\\
\cline{2-14}MinT  & -0.2  & -1.9  & -0.2  &       & -1.2  & 0.2   & 3.5   &       & -0.8  & 2.1   & 4.7   &       & 0.7 \bigstrut[t]\\
WLS$_s$ & -2    & -2.5  & -4.4  &       & -0.8  & -2.9  & -5.3  &       & -1    & -3.3  & -4.2  &       & -2.9 \\
WLS$_v$ & -2.1  & -2.6  & \textbf{-4.5} &       & -0.9  & -3    & \textbf{-5.4} &       & -1.1  & -3.4  & \textbf{-4.3} &       & -3.0 \\
BU    & 13.6  & 12.6  & 7.3   &       & 15.9  & 10    & 3.3   &       & 14    & 6.1   & 2.1   &       & 9.4 \\
MinTit$_g$ & -2.5  & -2.9  & -3.9  &       & -2.4  & -3.8  & -4.9  &       & -2.4  & -3.7  & -4    &       & -3.4 \\
MinTit$_l$ & \textbf{-2.8} & \textbf{-3.2} & -4.3  &       & \textbf{-2.5} & \textbf{-4} & -5.1  &       & \textbf{-2.5} & \textbf{-3.8} & -4.1  &       & \textbf{-3.6} \\
\end{tabular}%

  \label{tab:sim_large_ETS}%
\end{table}%

\begin{table}[t]
  \centering
    \footnotesize\tabcolsep2.5pt
  \renewcommand{\arraystretch}{.9}
  \vskip -0.8in
  \caption{\small Out-of-sample forecast performance for the large hierarchy
  in \cref{sec:sim_largeHier} for ARIMA-generated forecasts relative to the ARIMA-generated base forecasts.}
% Table generated by Excel2LaTeX from sheet 'Sheet3'
\begin{tabular}{lrrrrrrrlrrrr|r}
      & \multicolumn{13}{c}{ARIMA} \bigstrut[b]\\
\cline{2-14}      & \multicolumn{3}{c}{T = 30} &       & \multicolumn{3}{c}{T = 60} &       & \multicolumn{3}{c}{T = 90} & \multicolumn{1}{c}{} &  \bigstrut\\
\cline{2-4}\cline{6-8}\cline{10-12}      & \multicolumn{1}{c}{h=1} & \multicolumn{1}{c}{1:2} & \multicolumn{1}{c}{1:4} &       & \multicolumn{1}{c}{h=1} & \multicolumn{1}{c}{1:3} & \multicolumn{1}{c}{1:6} &       & \multicolumn{1}{c}{h=1} & \multicolumn{1}{c}{1:4} & \multicolumn{1}{c}{1:8} & \multicolumn{1}{c}{} & \multicolumn{1}{c}{Average} \bigstrut\\
\cline{2-12}\cline{14-14}      & \multicolumn{13}{c}{Top-Level} \bigstrut\\
\cline{2-14}MinT  & 1.2   & -3.1  & -5    &       & 3     & -3.8  & -3.9  &       & 3.9   & -0.2  & -1.1  &       & -1.0 \bigstrut[t]\\
WLS$_s$ & -4.2  & -5.8  & \textbf{-8} &       & -3.2  & -8    & \textbf{-10.6} &       & -4.3  & \textbf{-5.9} & \textbf{-6.4} &       & -6.3 \\
WLS$_v$ & -4.2  & -5.8  & \textbf{-8} &       & -3.2  & -8    & \textbf{-10.6} &       & -4.3  & \textbf{-5.9} & \textbf{-6.4} &       & -6.3 \\
BU    & 48.1  & 32.6  & 14.4  &       & 52.8  & 19    & 4.1   &       & 33.4  & 7.6   & 1     &       & 23.7 \\
MinTit$_g$ & -6.5  & -6.6  & -7.3  &       & -7.2  & -9    & -10.1 &       & -4.6  & -5.6  & -5.6  &       & -6.9 \\
MinTit$_l$ & \textbf{-6.8} & \textbf{-6.9} & -7.7  &       & \textbf{-7.4} & \textbf{-9.3} & -10.3 &       & \textbf{-5} & -5.6  & -5.8  &       & \textbf{-7.2} \\
      & \multicolumn{13}{c}{Level 1} \bigstrut[b]\\
\cline{2-14}MinT  & 0.4   & -2.2  & -4.5  &       & 4.4   & -1.7  & -3.7  &       & 3.1   & -2.9  & -3.8  &       & -1.2 \bigstrut[t]\\
WLS$_s$ & -2.9  & -3.1  & -6.7  &       & 0.3   & -5.7  & \textbf{-9.9} &       & 0.5   & -5.6  & \textbf{-7.7} &       & -4.5 \\
WLS$_v$ & -2.9  & -3.1  & -6.7  &       & 0.3   & -5.7  & \textbf{-9.9} &       & 0.5   & -5.6  & \textbf{-7.7} &       & -4.5 \\
BU    & 47.8  & 38    & 17.7  &       & 49.3  & 18    & 2.4   &       & 46.4  & 10.4  & 1.1   &       & 25.7 \\
MinTit$_g$ & -6.2  & -5.4  & -6.7  &       & -4.5  & -6.5  & -8.9  &       & -3.9  & -6.2  & -7    &       & -6.1 \\
MinTit$_l$ & \textbf{-6.6} & \textbf{-5.8} & \textbf{-7.1} &       & \textbf{-4.6} & \textbf{-6.6} & -9    &       & \textbf{-4.2} & \textbf{-6.3} & -7.2  &       & \textbf{-6.4} \\
      & \multicolumn{13}{c}{Level 2} \bigstrut[b]\\
\cline{2-14}MinT  & 0.7   & -2    & -3.8  &       & 0.7   & -0.7  & -0.7  &       & 0.7   & -0.8  & 0     &       & -0.7 \bigstrut[t]\\
WLS$_s$ & -2.2  & -2.1  & -5    &       & -0.2  & -3.2  & \textbf{-6.1} &       & 0.2   & -3.4  & \textbf{-5.3} &       & -3.0 \\
WLS$_v$ & -2.2  & -2.1  & -5    &       & -0.2  & -3.2  & \textbf{-6.1} &       & 0.2   & -3.4  & \textbf{-5.3} &       & -3.0 \\
BU    & 33.1  & 27.4  & 13.9  &       & 32.7  & 15.6  & 4.6   &       & 32.4  & 9.8   & 2.2   &       & 19.1 \\
MinTit$_g$ & -4.8  & -4.3  & -5.4  &       & -3.4  & -4.1  & -5.4  &       & -3    & -4    & -4.8  &       & -4.4 \\
MinTit$_l$ & \textbf{-5.3} & \textbf{-4.7} & \textbf{-5.7} &       & \textbf{-3.7} & \textbf{-4.3} & -5.7  &       & \textbf{-3.2} & \textbf{-4.2} & -4.9  &       & \textbf{-4.6} \\
      & \multicolumn{13}{c}{Level 3} \bigstrut[b]\\
\cline{2-14}MinT  & \textbf{-4.2} & \textbf{-4.9} & -2.9  &       & \textbf{-3.5} & -1.3  & 1.9   &       & 0.6   & 2.4   & 5.5   &       & -0.7 \bigstrut[t]\\
WLS$_s$ & -1.6  & -1.5  & -2.5  &       & -1.2  & -1.9  & -3.3  &       & 0.3   & -1.4  & -2.9  &       & -1.8 \\
WLS$_v$ & -1.6  & -1.5  & -2.6  &       & -1.2  & -2    & \textbf{-3.4} &       & 0.3   & -1.4  & \textbf{-3} &       & -1.8 \\
BU    & 14.6  & 13.2  & 8.2   &       & 12.8  & 7.9   & 2.9   &       & 15.4  & 6.8   & 2     &       & 9.3 \\
MinTit$_g$ & -3.2  & -3    & -2.9  &       & -2.4  & -2.3  & -2.6  &       & -1.5  & -1.9  & -2.4  &       & -2.5 \\
MinTit$_l$ & -3.4  & -3.2  & \textbf{-3.1} &       & -2.5  & \textbf{-2.5} & -2.8  &       & \textbf{-1.6} & \textbf{-2} & -2.6  &       & \textbf{-2.6} \\
      & \multicolumn{13}{c}{Bottom-Level} \bigstrut[b]\\
\cline{2-14}MinT  & 3.4   & 4.3   & 7.7   &       & 2.2   & 11.4  & 20.4  &       & 4.6   & 15.4  & 25.1  &       & 10.5 \bigstrut[t]\\
WLS$_s$ & -2.2  & -2.2  & -2    &       & -1.8  & -1.9  & -1.6  &       & -1.9  & -1.8  & -1.4  &       & -1.9 \\
WLS$_v$ & -2.5  & -2.6  & -2.4  &       & -2    & -2.3  & -2    &       & -2.2  & -2.3  & -1.9  &       & -2.2 \\
BU    & 0     & 0     & 0     &       & 0     & 0     & 0     &       & 0     & 0     & 0     &       & 0.0 \\
MinTit$_g$ & \textbf{-2.9} & \textbf{-3.1} & \textbf{-2.8} &       & -2.4  & -2.7  & -2    &       & -2.5  & -2.8  & -2.1  &       & -2.6 \\
MinTit$_l$ & \textbf{-2.9} & \textbf{-3.1} & \textbf{-2.8} &       & \textbf{-2.5} & \textbf{-2.8} & \textbf{-2.1} &       & \textbf{-2.6} & \textbf{-2.9} & \textbf{-2.2} &       & \textbf{-2.7} \\
      & \multicolumn{13}{c}{Average} \bigstrut[b]\\
\cline{2-14}MinT  & 0.9   & -0.2  & -0.3  &       & 1     & 2.4   & 3.6   &       & 3     & 3.8   & 5.3   &       & 2.2 \bigstrut[t]\\
WLS$_s$ & -2.3  & -2.5  & -4.3  &       & -1.4  & -3.6  & -6    &       & -1.2  & -3.4  & -4.8  &       & -3.3 \\
WLS$_v$ & -2.4  & -2.7  & -4.4  &       & -1.5  & -3.7  & \textbf{-6.1} &       & -1.3  & -3.5  & \textbf{-4.9} &       & -3.4 \\
BU    & 18.4  & 16.1  & 9.3   &       & 17.6  & 10.3  & 2.7   &       & 16.6  & 6.5   & 1.3   &       & 11.0 \\
MinTit$_g$ & -3.9  & -3.9  & -4.5  &       & -3.1  & -4.3  & -5.5  &       & -2.7  & -3.9  & -4.4  &       & -4.0 \\
MinTit$_l$ & \textbf{-4.1} & \textbf{-4.1} & \textbf{-4.8} &       & \textbf{-3.2} & \textbf{-4.5} & -5.7  &       & \textbf{-2.8} & \textbf{-4} & -4.5  &       & \textbf{-4.2} \\
\end{tabular}%

  \label{tab:sim_large_ARIMA}%
\end{table}%

\begin{table}[t]
  \centering
    \footnotesize\tabcolsep2.5pt
  \renewcommand{\arraystretch}{.9}
  \vskip -0.8in
  \caption{\small Out-of-sample forecast performance for the large hierarchy
  in \cref{sec:sim_largeHier} for GPR-generated forecasts relative to the GPR-generated base forecasts.}
% Table generated by Excel2LaTeX from sheet 'Sheet3'
\begin{tabular}{lrrrlrrrlrrrr|r}
      & \multicolumn{13}{c}{GPR} \bigstrut[b]\\
\cline{2-14}      & \multicolumn{3}{c}{T = 30} &       & \multicolumn{3}{c}{T = 60} &       & \multicolumn{3}{c}{T = 90} & \multicolumn{1}{c}{} &  \bigstrut\\
\cline{2-4}\cline{6-8}\cline{10-12}      & \multicolumn{1}{c}{h=1} & \multicolumn{1}{c}{1:2} & \multicolumn{1}{c}{1:4} &       & \multicolumn{1}{c}{h=1} & \multicolumn{1}{c}{1:3} & \multicolumn{1}{c}{1:6} &       & \multicolumn{1}{c}{h=1} & \multicolumn{1}{c}{1:4} & \multicolumn{1}{c}{1:8} & \multicolumn{1}{c}{} & \multicolumn{1}{c}{Average} \bigstrut\\
\cline{2-12}\cline{14-14}      & \multicolumn{13}{c}{Top-Level} \bigstrut\\
\cline{2-14}MinT  & \textbf{-3.5} & \textbf{-3.6} & \textbf{-3.9} &       & \textbf{-4.2} & \textbf{-3.1} & \textbf{-3.5} &       & \textbf{-4.5} & \textbf{-2.4} & \textbf{-2.2} &       & \textbf{-3.4} \bigstrut[t]\\
WLS$_s$ & 6.4   & 6.2   & 2     &       & 7.6   & 9.8   & 5.5   &       & 6.7   & 9.4   & 6     &       & 6.6 \\
WLS$_v$ & 6.4   & 6.2   & 2     &       & 7.6   & 9.8   & 5.5   &       & 6.7   & 9.4   & 6     &       & 6.6 \\
BU    & 48.4  & 39.4  & 25.7  &       & 56.3  & 52.7  & 36.6  &       & 54    & 47    & 32.8  &       & 43.7 \\
MinTit$_g$ & -1.8  & -1.8  & -2.9  &       & -2.5  & -1.4  & -2.4  &       & -2.6  & -0.6  & -1.1  &       & -1.9 \\
MinTit$_l$ & -1.9  & -1.9  & -3    &       & -2.6  & -1.6  & -2.6  &       & -2.7  & -0.7  & -1.2  &       & -2.0 \\
      & \multicolumn{13}{c}{Level 1} \bigstrut[b]\\
\cline{2-14}MinT  & \textbf{-5.3} & \textbf{-6.1} & \textbf{-6.2} &       & \textbf{-5.2} & \textbf{-4.9} & \textbf{-4.5} &       & \textbf{-5.5} & \textbf{-4.8} & \textbf{-4.1} &       & \textbf{-5.2} \bigstrut[t]\\
WLS$_s$ & -0.4  & -0.5  & -2.1  &       & 0.7   & 2.7   & 2.1   &       & -0.1  & 2.9   & 2.2   &       & 0.8 \\
WLS$_v$ & -0.4  & -0.5  & -2.2  &       & 0.7   & 2.8   & 2.1   &       & -0.1  & 2.9   & 2.2   &       & 0.8 \\
BU    & 25.8  & 21.4  & 15.3  &       & 29.1  & 29.9  & 24.3  &       & 27.6  & 28.5  & 22.5  &       & 24.9 \\
MinTit$_g$ & -4.9  & -5.3  & -5.6  &       & -4.4  & -3.8  & -3.5  &       & -4.4  & -3.6  & -3.1  &       & -4.3 \\
MinTit$_l$ & -4.9  & -5.3  & -5.6  &       & -4.4  & -3.8  & -3.6  &       & -4.4  & -3.7  & -3.2  &       & -4.3 \\
      & \multicolumn{13}{c}{Level 2} \bigstrut[b]\\
\cline{2-14}MinT  & \textbf{-4.8} & \textbf{-5.6} & \textbf{-5.4} &       & \textbf{-5.2} & \textbf{-6.2} & \textbf{-6.1} &       & \textbf{-6.2} & \textbf{-6.7} & \textbf{-6.2} &       & \textbf{-5.8} \bigstrut[t]\\
WLS$_s$ & -3    & -2.7  & -2.7  &       & -2.4  & -1.9  & -1.6  &       & -2    & -1.1  & -0.9  &       & -2.0 \\
WLS$_v$ & -2.9  & -2.8  & -2.9  &       & -2.4  & -2    & -1.7  &       & -2.1  & -1.2  & -1.1  &       & -2.1 \\
BU    & 10.1  & 8.4   & 7.3   &       & 10.6  & 11.5  & 10.7  &       & 10.7  & 12.5  & 11.1  &       & 10.3 \\
MinTit$_g$ & -4.6  & -5    & -4.9  &       & -4.3  & -5    & -5    &       & -3.9  & -4.5  & -4.5  &       & -4.6 \\
MinTit$_l$ & -4.6  & -5    & -4.9  &       & -4.3  & -5    & -5    &       & -3.8  & -4.5  & -4.5  &       & -4.6 \\
      & \multicolumn{13}{c}{Level 3} \bigstrut[b]\\
\cline{2-14}MinT  & \textbf{-4.8} & \textbf{-5.6} & \textbf{-5.4} &       & \textbf{-5.2} & \textbf{-6.2} & \textbf{-6.1} &       & \textbf{-6.2} & \textbf{-6.7} & \textbf{-6.2} &       & \textbf{-5.8} \bigstrut[t]\\
WLS$_s$ & -3    & -2.7  & -2.7  &       & -2.4  & -1.9  & -1.6  &       & -2    & -1.1  & -0.9  &       & -2.0 \\
WLS$_v$ & -2.9  & -2.8  & -2.9  &       & -2.4  & -2    & -1.7  &       & -2.1  & -1.2  & -1.1  &       & -2.1 \\
BU    & 10.1  & 8.4   & 7.3   &       & 10.6  & 11.5  & 10.7  &       & 10.7  & 12.5  & 11.1  &       & 10.3 \\
MinTit$_g$ & -4.6  & -5    & -4.9  &       & -4.3  & -5    & -5    &       & -3.9  & -4.5  & -4.5  &       & -4.6 \\
MinTit$_l$ & -4.6  & -5    & -4.9  &       & -4.3  & -5    & -5    &       & -3.8  & -4.5  & -4.5  &       & -4.6 \\
      & \multicolumn{13}{c}{Bottom-Level} \bigstrut[b]\\
\cline{2-14}MinT  & -3.1  & -3.5  & -3.8  &       & \textbf{-5.1} & \textbf{-6.1} & \textbf{-6.9} &       & \textbf{-6.5} & \textbf{-7.9} & \textbf{-8.2} &       & \textbf{-5.7} \bigstrut[t]\\
WLS$_s$ & -3.1  & -2.8  & -2.9  &       & -3.1  & -3.4  & -3.6  &       & -2.9  & -3.7  & -3.9  &       & -3.3 \\
WLS$_v$ & -3.2  & -3.1  & -3.4  &       & -3.4  & -3.9  & -4.3  &       & -3.4  & -4.3  & -4.6  &       & -3.7 \\
BU    & 0     & 0     & 0     &       & 0     & 0     & 0     &       & 0     & 0     & 0     &       & 0.0 \\
MinTit$_g$ & \textbf{-3.5} & \textbf{-3.8} & \textbf{-4.1} &       & -3.9  & -5    & -5.8  &       & -4    & -5.8  & -6.5  &       & -4.7 \\
MinTit$_l$ & \textbf{-3.5} & \textbf{-3.8} & \textbf{-4.1} &       & -3.9  & -5    & -5.8  &       & -4    & -5.8  & -6.5  &       & -4.7 \\
      & \multicolumn{13}{c}{Average} \bigstrut[b]\\
\cline{2-14}MinT  & \textbf{-3.9} & \textbf{-4.4} & \textbf{-4.7} &       & \textbf{-5} & \textbf{-5.7} & \textbf{-5.8} &       & \textbf{-6.1} & \textbf{-6.5} & \textbf{-6} &       & \textbf{-5.3} \bigstrut[t]\\
WLS$_s$ & -1.9  & -1.5  & -2    &       & -1.5  & -0.8  & -0.9  &       & -1.6  & -0.5  & -0.4  &       & -1.2 \\
WLS$_v$ & -2    & -1.7  & -2.3  &       & -1.7  & -1.1  & -1.2  &       & -1.8  & -0.8  & -0.7  &       & -1.5 \\
BU    & 10.6  & 9.8   & 8.3   &       & 12.1  & 13.5  & 12.5  &       & 11.4  & 14    & 12.6  &       & 11.6 \\
MinTit$_g$ & -3.8  & -4.1  & -4.4  &       & -4    & -4.5  & -4.7  &       & -3.9  & -4.5  & -4.5  &       & -4.3 \\
MinTit$_l$ & -3.8  & -4.1  & -4.4  &       & -4    & -4.5  & -4.8  &       & -3.9  & -4.6  & -4.6  &       & -4.3 \\
\end{tabular}%

  \label{tab:sim_large_GPR}%
\end{table}%

\section{Case Study: The World Semiconductor Trade Statistics}
\label{sec:CaseStudy}

Over the past decades, the semiconductor market has increasingly been identified as an economic driver. Semiconductors have become virtually ubiquitous in everyday life: they power applications from electronic toothbrushes, smart LEDs, computers, and washing machines to cars, power generators, and data centers.
The industry's strategic importance was identified by the US and the European Commission, which allocated \$280 billion and 43 billion Euros in subsidies respectively \citep{Taylor.2023, EuropeanCommission.}. In addition, the CEO of OpenAI, Sam Altman, is seeking an additional \$5-7 trillion in investments to boost fabrication capacity of AI chips \citep{Rajan.2024}. This highlights the importance of the industry and motivates a strong interest in predicting its development.

Ensuring that forecasts be coherent across hierarchical levels allows for consistent planning by executives, investors, and market analysts. Additionally, the simulation results in \cref{sec:sim} suggest that ensuring adherence to the constraint implied by the hierarchical structure has potential to increase forecast accuracy, allowing practitioners to plan more certainty.

The World Semiconductor Trade Statistics (WSTS) is a premier provider of semiconductor market data \citep{WSTS}. Starting from January 1991, they report monthly worldwide market sizes of semiconductor products (here considered in units of thousand USD), which are hierarchically aggregated into 116 product categories. These are detailed on WSTS' website\footnote{\url{wsts.org}}. A 2021 version is also available from semiconductors.org\footnote{\url{www.semiconductors.org/wp-content/uploads/2021/02/Product_Classification_2021.pdf}} without login. The semiconductor market is highly dynamic, meaning that these categorizations are subject to change as the market develops. Some of the newer ones are aggregated into larger categories to have sufficient historical data to fit quantitative models, leaving a total of 110 product categories following the procedure described in \citep{steinmeister2024human}. All data is available from WSTS\footnote{wsts.org/61/subscription}.

%\begin{landscape}
\begin{figure*}[ht]
%\vskip 0.2in
\begin{center}
\centerline{\includegraphics[width=\linewidth]{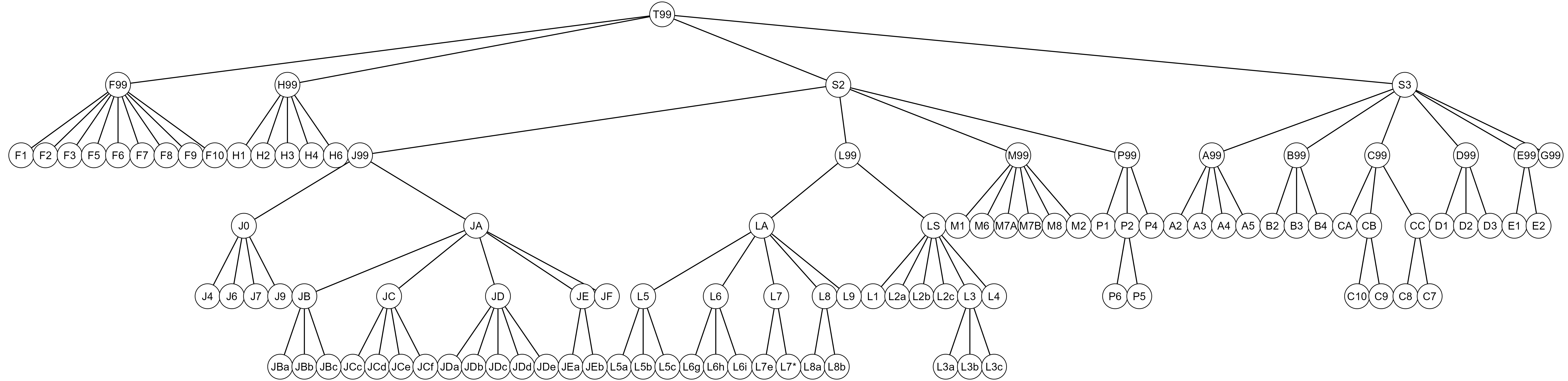}}
\caption{Illustration of the WSTS product classification Hierarchy. L$^\ast$ refers to L7a/b/c/d/f.}
\label{fig:WSTS_hier}
\end{center}
\vskip -0.3in
\end{figure*}
%\end{landscape}

\subsection{Experimental Design}
\label{sec:ExpDesign}
The semiconductor sales data from January 1991 to August 2023 is obtained from WSTS 
and split into a training- and a test-set: the training set comprised all data from January 1991 to December 2017 to allow for a minimum of two years of history for model training. 

The compared models for the base forecasts are ARIMA obtained with the \verb|auto.arima| function of the \verb|forecast| R-package, Error Trend and Seasonality (ETS), which is also implemented in the \verb|forecast| package\citep{Hyndman.2008forecastLibrary}, and Gaussian Process Regression as implemented in \verb|kernlab|\citep{Karatzoglou.2004}. These are the same models which were compared in the simulation in \cref{sec:sim}.

These models are then successively evaluated on the remaining 68 months of data spanning from January 2018 to August 2023 and comprising roughly 17.4\% of the complete data set. A 1-step ahead cross-validation is used to obtain 68 out-of-sample forecasts for each product category. These forecasts are reconciled using the different methods detailed in \cref{sec:sim_rec} and out-of-sample RMSE estimates are computed.

\subsection{Results}
\label{sec:cs_Res}

The results are presented in \cref{tab:cs_wsts}. For each hierarchical level, it details the percentage improvements over the respective base forecasts and the average RMSE of the base forecasts on that level. Note that, while the RMSE numbers look quite small, the semiconductor market size is vast. For example, in the case of the average top level forecast error for ARIMA, an RMSE of 1.719 billion USD only amounts to about 4\% of the total semiconductor market in August 2023. For the base forecasts, ARIMA achieved the lowest forecast error (average RMSE of $308,621$) across all hierarchical levels, closely followed by GPR ($310,638$). 

The average RMSE of the ETS base forecasts was about 11\% higher than that of the ARIMA model. However, the ETS model performed better on lower hierarchical levels. This mirrors the observations of \citet{steinmeister2024human}, as lower hierarchical levels tend to contain shorter time series.

The bottom-up approach generally fared poorly, suggesting that time series on higher hierarchical levels benefit from some smoothing of aggregation as was explored in \cref{sec:sim_effOfAggr}. Though, BU did achieve a small reduction of average RMSE -- even on the highest levels of aggregation -- for the ETS forecasts. 

For the forecasts generated by ARIMA and ETS, MinT achieved the largest reduction in RMSE: 1.9\% and 4.9\%, respectively. MinT also resulted in the most consistent improvements in this case study. However, when it comes to the GPR-generated forecasts, MinTit$_l$ excelled with an average reduction of 6\% when it converged. Overall, this yields the most accurate forecasts, with an estimated average RMSE of $292,000$.

\begin{table}[t]
  \centering
      \footnotesize\tabcolsep2.5pt
  \renewcommand{\arraystretch}{.9}
  \caption{\small Out-of-sample forecast performance for ARIMA, ETS, and GPR forecasts on the WSTS case study data in \cref{sec:CaseStudy}. The numbers for the reconciliation methods are percentage changes relative to the respective base forecasts. In addition, the last row of each section provides the average baseline RMSE of forecasts of the respective model and on the respective level. The last section, which presents the average results, includes a row indicating the average RMSE of forecasts with the respective model and the best average reconciliation method in terms of average RMSE reduction.}
% Table generated by Excel2LaTeX from sheet 'Sheet3'
\begin{tabular}{rrrlrrrr}
      & \multicolumn{1}{c}{ARIMA} & \multicolumn{1}{c}{ETS} & \multicolumn{1}{c}{GPR} &       & \multicolumn{1}{c}{ARIMA} & \multicolumn{1}{c}{ETS} & \multicolumn{1}{c}{GPR} \bigstrut[b]\\
\cline{2-4}\cline{6-8}      & \multicolumn{3}{c}{Top-Level} &       & \multicolumn{3}{c}{Level 3} \bigstrut\\
\cline{2-4}\cline{6-8}\multicolumn{1}{l}{MinT} & \textbf{3.1} & \textbf{-6.5} & \multicolumn{1}{r}{0.5} &       & \textbf{-7} & \textbf{-4.7} & -7.1 \bigstrut[t]\\
\multicolumn{1}{l}{WLS$_s$} & 4.1   & -5.3  & \multicolumn{1}{r}{1.8} &       & -0.6  & -1    & -2.4 \\
\multicolumn{1}{l}{WLS$_v$} & 3.2   & -5.2  & \multicolumn{1}{r}{1.3} &       & -5.8  & -3.7  & -5.8 \\
\multicolumn{1}{l}{BU} & 19.3  & -2.4  & \multicolumn{1}{r}{7.8} &       & 1.4   & 0     & -1.9 \\
\multicolumn{1}{l}{MinTit$_g$} & 6.2   & -5.4  & \multicolumn{1}{r}{1.9} &       & -5.5  & -4    & -6.2 \\
\multicolumn{1}{l}{MinTit$_l$} & 9.3   & -5.2  & \multicolumn{1}{r}{\textbf{-1.3}} &       & -3.6  & -2.6  & \textbf{-11.4} \\
\multicolumn{1}{l}{Base} & \textbf{1,719,047} &          2,062,625  & \multicolumn{1}{r}{         1,754,233 } &       & \textbf{256,173} &          259,774  &          257,230  \bigstrut[b]\\
\cline{2-4}\cline{6-8}      & \multicolumn{3}{c}{Level 1} &       & \multicolumn{3}{c}{Level 4} \bigstrut\\
\cline{2-4}\cline{6-8}\multicolumn{1}{l}{MinT} & 3.8   & \textbf{-5.2} & \multicolumn{1}{r}{1} &       & \textbf{-9.1} & \textbf{-4.8} & -1.7 \bigstrut[t]\\
\multicolumn{1}{l}{WLS$_s$} & 6     & -3.6  & \multicolumn{1}{r}{3.1} &       & -2.8  & -0.2  & 0.8 \\
\multicolumn{1}{l}{WLS$_v$} & \textbf{3.6} & -3.6  & \multicolumn{1}{r}{2.1} &       & -6.2  & -2.1  & 0.8 \\
\multicolumn{1}{l}{BU} & 19.9  & -0.7  & \multicolumn{1}{r}{7.9} &       & -0.1  & 0.5   & 0.9 \\
\multicolumn{1}{l}{MinTit$_g$} & 7.3   & -4    & \multicolumn{1}{r}{2.5} &       & -5.7  & -3.6  & -0.3 \\
\multicolumn{1}{l}{MinTit$_l$} & 11.1  & -3.8  & \multicolumn{1}{r}{\textbf{-4.9}} &       & 0.2   & -2.4  & \textbf{-8.4} \\
\multicolumn{1}{l}{Base} & \textbf{802,795} &              960,035  & \multicolumn{1}{r}{839,987} &       &            88,835  & \textbf{87,637} &            90,870  \bigstrut[b]\\
\cline{2-4}\cline{6-8}      & \multicolumn{3}{c}{Level 2} &       & \multicolumn{3}{c}{Level 5} \bigstrut\\
\cline{2-4}\cline{6-8}\multicolumn{1}{l}{MinT} & \textbf{-8.9} & \textbf{-2.3} & \multicolumn{1}{r}{-2.3} &       & \textbf{-9.7} & \textbf{-4.2} & -2.1 \bigstrut[t]\\
\multicolumn{1}{l}{WLS$_s$} & -3.1  & 0     & \multicolumn{1}{r}{0.6} &       & -2.1  & -0.7  & -0.2 \\
\multicolumn{1}{l}{WLS$_v$} & -7.2  & -1.1  & \multicolumn{1}{r}{-1.3} &       & -7.6  & -1.6  & -0.2 \\
\multicolumn{1}{l}{BU} & 0.8   & 2.9   & \multicolumn{1}{r}{3.6} &       & 0     & 0     & 0 \\
\multicolumn{1}{l}{MinTit$_g$} & -7    & -1.3  & \multicolumn{1}{r}{-1.4} &       & -6.9  & -3.3  & -1.1 \\
\multicolumn{1}{l}{MinTit$_l$} & -5.1  & -0.3  & \multicolumn{1}{r}{\textbf{-8.8}} &       & -1.3  & -2.3  & \textbf{-11.9} \\
\multicolumn{1}{l}{Base} &              338,426  &              334,742  & \multicolumn{1}{r}{\textbf{322,153}} &       &            75,200  & \textbf{73,083} &            76,204  \bigstrut[b]\\
\cline{6-8}      &       &       &       &       & \multicolumn{3}{c}{Average} \bigstrut\\
\cline{6-8}      &       &       & \multicolumn{2}{l}{MinT} & \textbf{-1.9} & \textbf{-4.9} & -1.4 \bigstrut[t]\\
      &       &       & \multicolumn{2}{l}{WLS$_s$} & 1.7   & -2.9  & 1.1 \\
      &       &       & \multicolumn{2}{l}{WLS$_v$} & -1.2  & -3.6  & -0.3 \\
      &       &       & \multicolumn{2}{l}{BU} & 11.2  & -0.4  & 5 \\
      &       &       & \multicolumn{2}{l}{MinTit$_g$} & 0.8   & -3.9  & -0.2 \\
      &       &       & \multicolumn{2}{l}{MinTit$_l$} & 3.6   & -3.3  & \textbf{-6} \\
      &       &       & \multicolumn{2}{l}{Base} & \textbf{308,621} &          341,970  &          310,638  \\
      &       &       & \multicolumn{2}{l}{Best} &          302,757  &          325,213  & \textbf{292,000} \\
\end{tabular}%

  \label{tab:cs_wsts}%
\end{table}%

\section{Discussion}
\label{sec:Discussion}

The Monte Carlo simulation in \cref{sec:sim} showcases the potential for additional reductions in expected RMSE in various scenarios by utilizing the iterative reconciliation method introduced in \cref{sec:ItMinT}. 

MinTit achieved the lowest RMSE values for forecasts based on ARIMA and ETS for correlated time series in \cref{fig:sim_corr_hier}, time series of differing lengths in \cref{sec:sim_len}, in the presence of hierarchical degeneracy (\cref{sec:sim_deg}), and for particularly large hierarchical time series structures in \cref{sec:sim_largeHier}. 

Both, \cref{sec:sim_effOfAggr} and \cref{sec:sim_largeHier} incorporate a dependence structure, which leads to higher signal-to-noise-ratios (SNR) on higher hierarchical levels. This is also referred to as the smoothing effect of aggregation. While MinTit outperformed MinT for the ARIMA and the ETS forecasts in the context of the larger hierarchy (\cref{sec:sim_largeHier}), this effect was less noticeable in the context of \cref{sec:sim_effOfAggr}, where both approaches achieved similar RMSE reductions. 

At the same time, MinT proved remarkably robust in most situations. Considering the challenge of estimating large covariance matrices, MinT with the shrinkage estimator performed surprisingly well even when time series histories were very short. 

Furthermore, MinT appears to excel in severe model mis-specification as its strong performance for the GPR forecasts and the scenario with seasonality in \cref{fig:sim_seas} suggests. Meanwhile, MinTit appears to be more susceptible when forecasts are of poor quality. Possibly, outliers in the residuals are more problematic due to the iterative approach.

As a rule of thumb, MinT may be preferable when the data can be assumed to contain a strong seasonal component -- especially when time series are very short and the models have insufficient data to adequately capture this effect. Otherwise, MinTit may improve upon MinT and other existing reconciliation methods for very short time series, large hierarchical structures, or time series of differing lengths.

\appendix

\section*{Acknowledgements}
This paper was supported by a research cooperation between Infineon Technologies AG and TU Dortmund University through the Graduate School of Logistics.

\pagebreak
\onecolumn



%%       \citet{<label>} ==> Jones et al. [21]
%%       \citep{<label>} ==> [21]
%%

 \bibliographystyle{elsarticle-num-names} 
 \bibliography{MinTit}





% \end{thebibliography}
\end{document}